\documentclass[a4paper,11pt]{article}
\pdfoutput=1 

\usepackage{jheppub} 

\usepackage{graphicx,epsfig,wrapfig,amssymb,color,amsmath,bm,breqn}

\usepackage[T1]{fontenc} 

\allowdisplaybreaks

\usepackage{lineno} 

\usepackage{hyperref} 
\hypersetup{
    colorlinks,
    citecolor=black,
    filecolor=black,
    linkcolor=black,
    urlcolor=black
}

\newcommand{\blue}[1]{{\color{blue} #1}}
\newcommand{\red}[1]{{\color{red} #1}}

\newcommand{\be}{\begin{equation}}
\newcommand{\ee}{\end{equation}}
\newcommand{\ba}{\begin{eqnarray}}
\newcommand{\ea}{\end{eqnarray}}
\newcommand{\la}{\langle}
\newcommand{\ra}{\rangle}

\newcommand{\mh}{ m_h }
\newcommand{\with}[3]{{\Biggl|{\renewcommand*{\arraystretch}{0.8}
	\begin{array}{l}
	\phantom{X}\\
	\mbox{\scriptsize ${#1}$}\\
	\mbox{\scriptsize ${#2}$}\\
	\mbox{\scriptsize #3}\end{array}}}}
\newcommand{\slim}{\mskip 1.5mu}       
\def\T{_{_T}}
\def\C{_{_C}}

\definecolor{darkgreen}{rgb}{0,0.65,0}
\definecolor{orange}{rgb}{1,0.4,0.1}

\newcommand{\asym}[2]{{A_{#1}^{#2}}}
\newcommand{\asympre}[2]{{A_{#1,\langle y\rangle}^{#2}}}

\newcommand{\WWtype}{\ensuremath{\stackrel{\text{\scriptsize WW--type}}{\approx}}}

\newcommand{\Gaussian}{\ensuremath{\stackrel{\text{\scriptsize Gauss}}{=}}}

\def\bflperp{{\bm \ell}_\perp}

\def\bfkperp{{\bm k}_\perp}
\def\bfpperp{{\bm P}_\perp}
\def\bfhp{\hat{\bm h}}
\def\bfPhperp{{\bm P}_{hT}}
\def\Phperp{P_{hT}}

\def\kperp{k_\perp}
\def\pperp{P_\perp}
\def\avkperp{\la \kperp^2 \ra}
\def\avpperp{\la \pperp^2 \ra}

\newcommand*{\FigPath}{./figs}%
\preprint{JLAB-THY-18-2775}

\title{	Semi-inclusive deep-inelastic scattering
	in Wandzura--Wilczek-type approximation}

\author[a]{S.~Bastami}
\author[b]{H.~Avakian}
\author[c]{A.~V.~Efremov}
\author[d,e]{A.~Kotzinian}
\author[f]{B.~U.~Musch}
\author[k,e]{B.~Parsamyan}
\author[g,b]{A.~Prokudin}
\author[h]{M.~Schlegel}
\author[i]{G.~Schnell}
\author[a,j]{P.~Schweitzer}
\author[a]{K.~Tezgin}

\affiliation[a]{Department of Physics, University of Connecticut,
	Storrs, CT 06269, U.S.A.}
\affiliation[b]{Thomas Jefferson National Accelerator Facility,
	Newport News, VA 23606, U.S.A.}
\affiliation[c]{Joint Institute for Nuclear Research, Dubna,
	141980 Russia}
\affiliation[d]{Yerevan Physics Institute,  Alikhanyan Brothers St.,
	375036 Yerevan, Armenia}
\affiliation[e]{INFN, Sezione di Torino,
	10125 Torino, Italy}
\affiliation[f]{Institut f\"ur Theoretische Physik, Universit\"at
  	Regensburg, 93040 Regensburg, Germany}
\affiliation[g]{Division of Science, Penn State Berks, Reading,
	PA 19610, USA}
\affiliation[k]{CERN, 1211 Geneva 23, Switzerland}
\affiliation[h]{Department of Physics, New Mexico State University,
	Las Cruces, NM 88003-001, USA}
\affiliation[i]{Department of Theoretical Physics, University of the Basque
	Country UPV/EHU, 48080 Bilbao, Spain, and
	IKERBASQUE, Basque Foundation for Science, 48013 Bilbao, Spain}
\affiliation[j]{Institute for Theoretical Physics, Universit\"at T\"ubingen,
	D-72076 T\"ubingen, Germany} 
 
\emailAdd{saman.bastami@uconn.edu}
\emailAdd{avakian@jlab.org}
\emailAdd{efremov@theor.jinr.ru}
\emailAdd{aram.kotzinian@cern.ch}
\emailAdd{bmusch@b-mu.de}
\emailAdd{bakur@cern.ch}
\emailAdd{prokudin@jlab.org}
\emailAdd{schlegel@nmsu.edu}
\emailAdd{gunar.schnell@desy.de}
\emailAdd{peter.schweitzer@phys.uconn.edu}
\emailAdd{kemal.tezgin@uconn.edu}

\abstract{
We present the complete cross-section
for the production of unpolarized hadrons in semi-inclusive
deep-inelastic scattering up to power-suppressed ${\cal O}(1/Q^2)$ terms in
the Wandzura--Wilczek-type approximation, which consists in systematically
assuming that $\bar{q}gq$--terms are much smaller than $\bar{q}q$--correlators.
We compute all twist-2 and twist-3 structure functions and the corresponding
asymmetries, and discuss the applicability of the Wandzura--Wilczek-type
approximations on the basis of available data. We make predictions that
can be tested by data from COMPASS, HERMES, Jefferson Lab, and the future
Electron-Ion Collider. The results of this paper can be readily used for
phenomenology and for event generators, and will help
   {to improve the description of semi-inclusive deep-inelastic processes
   in terms of transverse momentum dependent parton distribution functions
   and fragmentation functions beyond the leading twist}.}

\begin{document}


\maketitle

\flushbottom
\newpage

\section{Introduction}
\label{Sec-1:introduction}

A great deal of what is known about the quark-gluon structure of
nucleons is due to studies of parton distribution functions (PDFs)
in deep-inelastic reactions. Leading-twist PDFs  tell us  how likely
it is to find an unpolarized parton
[described by PDF $f_1^a(x)$, $a=q,\,\bar q,\,g$]
or a longitudinally polarized parton
[described by PDF $g_1^a(x)$, $a=q,\,\bar q,\,g$]
in a fast-moving unpolarized or longitudinally polarized nucleon,
which carries the fraction $x$ of the nucleon momentum.
This information depends on the ``resolution (renormalization) scale''
associated with the hard scale $Q$ of the process.
Although the PDFs  $f_1^a(x)$ and $g_1^a(x)$ continue being the
subject of intense research (small-$x$, large-$x$, helicity sea
and gluon distributions) they can be considered as rather
well known, and the frontier has been extended in the last years
to go beyond the one-dimensional picture offered by those PDFs.

One way to do this consists in a systematic inclusion of transverse
parton momenta $\kperp$, whose effects manifest themselves in terms of
transverse momenta of the reaction products in the final state.
If these transverse momenta are much smaller than the hard scale $Q$
of the process, the formal description is given in terms of
transverse momentum dependent distribution functions (TMDs)
and fragmentation functions (FFs),
which are defined in terms of quark-quark correlators
\cite{Kotzinian:1994dv,Mulders:1995dh,Boer:1997nt,Goeke:2005hb,Bacchetta:2006tn}.
Both of them depend on two independent variables: in the case of TMDs,
on the fraction $x$ of nucleon momentum carried by the parton and intrinsic transverse momentum $\kperp$
of the parton, while in the case of FFs, on the fraction $z$ of the parton momentum
transferred to the hadron and the transverse momentum of the hadron
acquired during the fragmentation process.
Being a vector in the plane transverse with respect to
the light-cone direction singled out by the hard-momentum flow in the process,
$\kperp$ allows us to access novel information on the nucleon spin structure
through correlations of $\kperp$ with the nucleon and/or parton spin. The
latter is a well-defined concept for twist-2 TMDs interpreted in
the infinite momentum frame or in the lightcone quantization formalism.

One powerful tool to study TMDs are measurements of the
semi-inclusive deep-inelastic scattering (SIDIS) process.
By exploring various possibilities for the lepton beam and target
polarizations unambiguous information can be accessed on the 8 leading-twist
TMDs \cite{Boer:1997nt} and, if one assumes factorization, on certain
linear combinations of the 16 subleading-twist TMDs
\cite{Goeke:2005hb,Bacchetta:2006tn}.
	It is important to stress that this information could not have
	been obtained without advances in target polarization techniques
	employed in the HERMES, COMPASS and Jefferson Lab (JLab) experiments
	\cite{Stock:1994vv,Airapetian:2004yf,Crabb:1997cy,Goertz:2002vv}.
Complementary information can be obtained
from the Drell--Yan process \cite{Arnold:2008kf},
and $e^+e^-$ annihilation \cite{Metz:2016swz}.

In QCD the TMDs are independent functions. Each TMD contains unique
information on a different aspect of the nucleon structure.
Twist-2 TMDs have partonic interpretations. Twist-3 TMDs
give insights on quark-gluon correlations in the nucleon
\cite{Miller:2007ae,Burkardt:2007rv,Burkardt:2009rf}.
Besides positivity constraints \cite{Bacchetta:1999kz}
there is little model-independent information on TMDs.
An important question with practical applications is:
do useful {\sl approximations} for TMDs exist?
Experience from collinear PDFs encourages to explore this possibility:
the twist-3 $g_T^a(x)$ and $h_L^a(x)$ can be respectively expressed
in terms of contributions from twist-2 $g_1^a(x)$ and $h_1^a(x)$, and
additional quark-gluon-quark ($\bar{q}gq$) correlations or current-quark
mass terms \cite{Wandzura:1977qf,Jaffe:1991ra}
(the index $a=q\,,\bar q$ does not include gluons for
$h_1^a$, $h_L^a$ and other chiral-odd TMDs below).
We shall refer to the latter generically as $\bar{q}gq$--terms, keeping in
mind one deals in each case with matrix elements of different operators.
The $\bar{q}gq$--correlations contain new insights on hadron structure,
which are worthwhile exploring for their own sake,
see for instance Ref.~\cite{Jaffe:1989xx} on $g_T^a(x)$.

The striking observation is that the $\bar{q}gq$--terms in $g_T^a(x)$
and $h_L^a(x)$ are small: theoretical mechanisms predict this
\cite{Balla:1997hf,Dressler:1999hc,Gockeler:2000ja,Gockeler:2005vw},
and in the case of $g_T^a(x)$ data confirm or are compatible with these
predictions \cite{Abe:1998wq,Anthony:2002hy,Airapetian:2011wu}.
This approximation (``neglect of $\bar{q}gq$--terms'') is commonly
known as Wandzura--Wilczek (WW) approximation \cite{Wandzura:1977qf}.
The possibility to apply this type of approximation also to TMDs has
been explored in specific cases in \cite{Kotzinian:1995cz,Kotzinian:1997wt,
Kotzinian:2006dw,Avakian:2007mv,Metz:2008ib,Teckentrup:2009tk,Tangerman:1994bb}.
In both cases, PDFs and TMDs, one basically assumes that the
contributions from $\bar{q}gq$--terms can be neglected with respect to
$\bar{q}q$--terms. But the nature of the omitted matrix elements is
different, and in the context of TMDs one often prefers to speak
about WW-type approximations.

The WW-type approximation is not preserved under $Q^2$ evolution.
Some intuition can be obtained from the collinear case. However,
much less is known about the $k_\perp$-evolution especially at 
subleading twist. More theoretical work is required here.

The present work is the first 
study of all SIDIS structure functions up to twist-3 evaluated within one
common systematic theoretical guideline.
Our results are of importance for measurements
performed or in preparation at COMPASS, HERMES, and JLab with
12$\,$GeV beam-energy upgrade, or proposed in the long-term
(Electron-Ion Collider), and provide helpful input for the
development of Monte Carlo event generators \cite{Avakian:2015vha}.

On the theoretical side it is also important to note that the
theory for subleading-twist TMD observables is only poorly developed
as compared to the current state-of-the-art of leading-twist observables.
In order to address subleading-twist TMD observables one has
to restrict oneself to the tree-level formalism
\cite{Kotzinian:1994dv,Mulders:1995dh,Boer:1997nt,Goeke:2005hb,
Bacchetta:2006tn}, which may not be free of doubts
\cite{Metz:2004je,Gamberg:2006ru}.

Our predictions, whether confirmed or not supported by current and
future experimental data, will in any case provide a useful benchmark,
and call for dedicated theoretical studies to explain (i) why the
pertinent $\bar{q}gq$--terms are small or (ii) why they are sizable,
depending on the outcome of the experiments.
In either case our results will deepen the understanding of
$\bar{q}gq$--correlations, pave the
way towards testing the validity of the TMD factorization approach
at subleading twist, and help us to guide further developments.

In this work, after introducing the SIDIS process and defining TMDs and
FFs (Sec.~\ref{Sec-2:SIDIS+TMDs+FF}), we shall introduce the WW(-type)
approximations, and review what is presently known about them
from experiment and theory (Sec.~\ref{Sec-3:WW}).
We will show that under the assumption of the validity of these approximations
all leading and subleading SIDIS structure functions are described in terms of
a basis of 6 TMDs and 2 FFs (Sec.~\ref{Sec-4:SIDIS-in-WW-approximation}),
and review how these basis functions describe available data
(Sec.~\ref{Sec-5:twist-2+basis}).
We will systematically apply the WW and/or WW-type approximations
to SIDIS structure functions at leading (Sec.~\ref{Sec-6:twist-2-and-WW})
and subleading (Sec.~\ref{Sec-7:twist-3-and-WW}) twist, and
conclude with a critical discussion (Sec.~\ref{Sec-8:conclusions}).
The Appendices \ref{App:basis} and \ref{App:factor} contain
technical details.
An open-source package is available which allows one to visualize 
	and reproduce the results presented in this work, and may easily 
	be adapted by interested colleagues for their purposes~\footnote{Open-source packages with implementations of SIDIS structure functions
in the WW-type approximation are publicly available on github.com: in
Mathematica, Version 11.3 on https://github.com/prokudin/WW-SIDIS, in Python on https://jeffersonlab.github.io/jam3d.}.

\section{The SIDIS process in terms of TMDs and FFs}
\label{Sec-2:SIDIS+TMDs+FF}

In this section we review the description of the SIDIS process,
define structure functions, PDFs, TMDs, FFs and recall how they
describe the SIDIS structure functions.

\subsection{The SIDIS process}
\label{Sec-2.1:SIDIS+structure-functions}

\begin{wrapfigure}[9]{RD}{8cm}
\vspace{-7mm}
\centering
	\includegraphics[width=7.2cm]{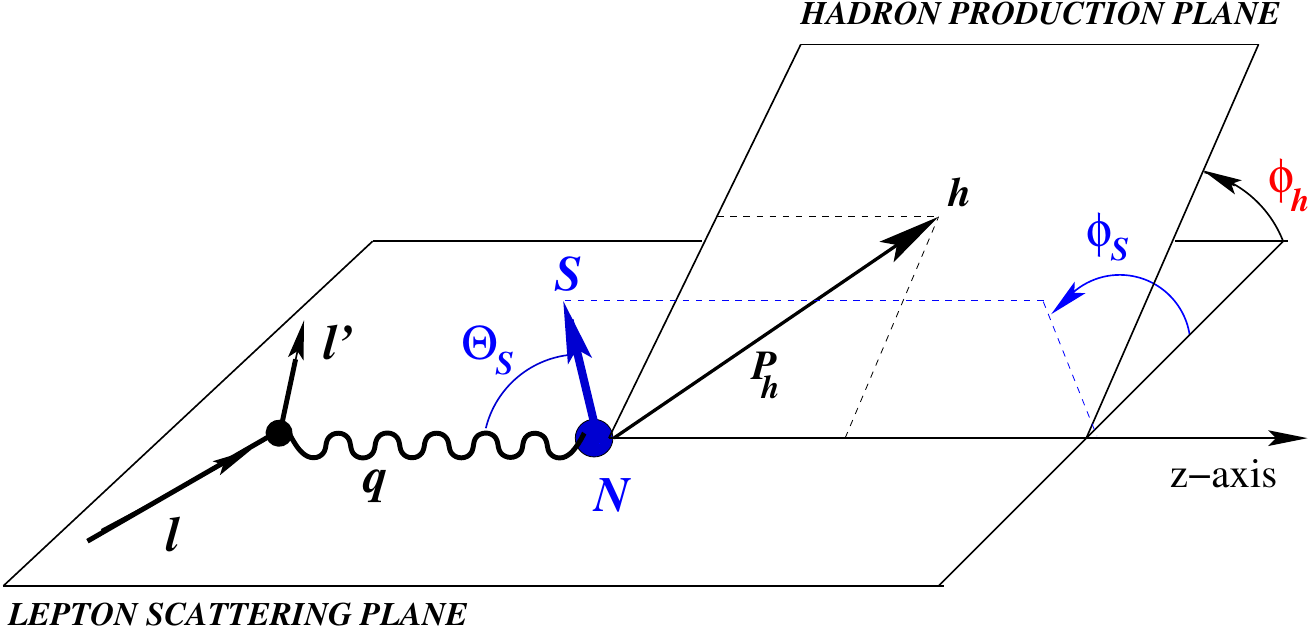}
        \caption{\label{fig-kin-SIDIS}
    	Kinematics of the SIDIS process $lN\to l^\prime h X$
	in the 1-photon exchange approximation.}
\vspace{-5mm}
\end{wrapfigure}

The SIDIS process  $lN\to l^\prime h X$ is sketched in
Fig.~\ref{fig-kin-SIDIS}. Here, $l$ and $P$ are the momenta of the incoming
lepton and nucleon, $l^\prime$ and $P_h$ are the momenta of the outgoing
lepton and produced hadron. The virtual-photon momentum $q=l-l^\prime$
defines the z-axis, and $l^\prime$ points in the direction of the x-axis
from which azimuthal angles are counted. The relevant kinematic invariants
are
\ba
   x  = \frac{Q^2}{2\,P\cdot  q}, \;\;
   y = \frac{P \cdot  q}{P \cdot  l}, \;\;
   z = \frac{P \cdot  P_h}{P\cdot  q}, \;\;
   Q^2=-q^2.
\label{eq:xyz}\;\;\;\;\ea
Note that we consider the production of unpolarized hadrons in DIS of
charged leptons (electrons, positrons, muons) at $Q^2 \ll M_Z^2$
in the single-photon exchange approximation,
where $M_Z$ denotes the mass of the $Z^0$ electroweak gauge boson.
In addition to $x$, $y$, and $z$, the cross section is also differential
in the azimuthal angle $\phi_h$ of the produced hadron and in the square
of the hadron's momentum component $\Phperp$ perpendicular with respect to the
virtual-photon momentum.  The cross section is also
differential with respect to the azimuthal angle $\psi_l$
 characterizing the overall orientation of the lepton scattering plane
around the incoming lepton direction. The angle is calculated with respect
to an arbitrary reference axis, which in case of transversely polarized
targets is chosen to be the transverse component $S_T$ of the target-spin direction.
In the DIS limit, $\psi_l \approx \phi_S$, where the latter is the azimuthal angle of
the spin-vector defined as in Fig.~\ref{fig-kin-SIDIS}.

To leading order in $1/Q$ the SIDIS cross-section is given by
\begin{subequations}\ba\hspace{-1cm}
   &&  \hspace{-5mm}
	\frac{d^6\sigma_{\rm leading}}{dx\,dy\,dz\,d\psi_l\,d\phi_h\,d \Phperp^2}
   =	
	\frac{\alpha_{em}^2}{x\,y\,Q^2}\biggl(1-y+\frac12y^2\biggr)
	\;F_{UU}(x,z,\Phperp^2) 
	\nonumber\\
   && \times     \Biggl\{\;1
        + \cos(2\phi_h)\,   p_1\,A_{UU}^{\cos(2\phi_h)} 
  	+ S_L\sin(2\phi_h)\,p_1\,A_{UL}^{\sin(2\phi_h)}
	+ \lambda\,S_L\,    p_2\,A_{LL}  \nonumber\\
   && \hspace{1cm}
       	+ \; 	S_T\sin( \phi_h-\phi_S)\, A_{UT}^{\sin( \phi_h-\phi_S)}
	\; + \;	S_T\sin( \phi_h+\phi_S)\,p_1\,A_{UT}^{\sin( \phi_h+\phi_S)}
	 \phantom{\frac11}
	\nonumber\\
   && \hspace{1cm}
        + \;	   S_T\sin(3\phi_h-\phi_S)\,p_1\,A_{UT}^{\sin(3\phi_h-\phi_S)}
	+ \lambda\,S_T\cos(\phi_h-\phi_S)\,p_2\,A_{LT}^{\cos( \phi_h-\phi_S)}\Biggr\}
	\,. \hspace{12mm} \label{Eq:SIDIS-leading}
\ea
Here $F_{UU}$ is the structure function due to transverse
polarization of the virtual photon (sometimes denoted as $F_{UU,T}$),
and we 
neglect $1/Q^2$ corrections in kinematic factors
and a structure function (sometimes denoted as $F_{UU,L}$) arising from
longitudinal polarization of the virtual photon (and another
structure function $\propto S_T\,\sin( \phi_h-\phi_S)$, see below).
The structure functions
(and asymmetries) also depend on $Q^2$ via the scale dependence of
TMDs and FFs, which we do not show in formulas throughout this work.

At subleading order in the $1/Q$ expansion one has
\ba\hspace{-1cm}
   &&   \frac{d^6\sigma_{\rm subleading}}{dx\,dy\,dz\,d\psi_l\,d\phi_h\,d \Phperp^2}
   =	
	\frac{\alpha_{em}^2}{x\,y\,Q^2}\biggl(1-y+\frac12y^2\biggr)
	\; F_{UU}(x,z,\Phperp^2)
        \Biggl\{
          \cos(\phi_h)\,p_3\,A_{UU}^{\cos(\phi_h)}
	\nonumber\\
   && \hspace{2cm}
	+ \lambda\sin(\phi_h)\,p_4\,A_{LU}^{\sin(\phi_h)}
	+ S_L\sin(\phi_h)\,p_3\,A_{UL}^{\sin(\phi_h)}
	+ \lambda\,S_L\cos(\phi_h)\,p_4\,A_{LL}^{\cos(\phi_h)}\phantom{\frac11}
	\nonumber\\
   && \hspace{2cm}
	+ S_T\sin(2\phi_h-\phi_S)\,p_3\,A_{UT}^{\sin(2\phi_h-\phi_S)}
        + S_T\sin(\phi_S)\,p_3\,A_{UT}^{\sin(\phi_S)} \phantom{\frac11}\nonumber\\
   && \hspace{2cm}
  	+ \lambda\,S_T\cos(\phi_S)\,p_4\,A_{LT}^{\cos(\phi_S)}
        + \lambda\,S_T\cos(2\phi_h-\phi_S)\,p_4\,A_{LT}^{\cos(2\phi_h-\phi_S)}
	\Biggr\} \, .
   \hspace{1cm} \label{Eq:SIDIS-subleading}
\ea\end{subequations}
Neglecting $1/Q^2$ corrections, the kinematic prefactors $p_i$ are given by
\be\label{Eq:y-prefactors}
	p_1 = \frac{1-y}{1-y+\frac12\,y^2} 		\, , \;\;\;
	p_2 = \frac{y(1-\frac12\,y)}{1-y+\frac12\,y^2}	\, , \;\;\;
	p_3 = \frac{(2-y)\sqrt{1-y}}{1-y+\frac12\,y^2} 	\, , \;\;\;
	p_4 = \frac{y\sqrt{1-y}}{1-y+\frac12\,y^2}     	\, ,
\ee
and the asymmetries {$A_{XY}^{\rm weight}$,  are defined in terms of
structure functions $F_{XY}^{\rm weight}$, as follows}
\be
	A_{XY}^{\rm weight}\equiv A_{XY}^{\rm weight}(x,z,\Phperp)=
	\frac{F_{XY}^{\rm weight}(x,z,\Phperp)}{F_{UU}(x,z,\Phperp)}.
\ee
Here, the first subscript $X=U(L)$ denotes the unpolarized beam
(longitudinally polarized beam with helicity $\lambda$). The second
subscript $Y=U(L\text{ or }T)$ refers to the target, which can be unpolarized
(longitudinally or transversely polarized with respect to the virtual photon).
The superscript ``weight'' indicates the azimuthal dependence with no index
indicating a $\phi_h$-independent asymmetry or structure function. 

In the partonic description the structure functions in
(\ref{Eq:SIDIS-leading})    are ``twist-2.'' Those in
(\ref{Eq:SIDIS-subleading}) are ``twist-3'' and contain a
factor $M_N/Q$ in their definitions, see below,
where $M_N$ is the nucleon mass. In our treatment to $1/Q^2$
accuracy we neglect two structure functions due to longitudinal virtual-photon
polarization, which contribute at order ${\cal O}(M_N^2/Q^2)$ in the
partonic description of the process, one being $F_{UU,L}$ and the other
contributing to the $\sin(\phi_h-\phi_S)$ angular distribution
\cite{Bacchetta:2006tn}.

Experimental collaborations often define asymmetries in terms of counts
$N(\phi_h)$. This means the kinematic prefactors $p_i$ and $1/(x\,y\,Q^2)$
are included in the numerators or denominators of the asymmetries which
are averaged over $y$ within experimental kinematics. We will call the
corresponding asymmetries $\asympre{XY}{\rm weight}$.
For instance, in the unpolarized case one has
\be
	N(x,\dots,\phi_h) = \frac{N_0(x,\dots)}{2\pi} \biggl(1
		+ \cos\phi_h\;\asympre{UU}{\cos\phi_h}(x,\dots)
		+ \cos2\phi_h\;\asympre{UU}{\cos2\phi_h}(x,\dots)\Biggr)
\ee
where $N_0$ denotes the total ($\phi_h$--averaged) number of counts
and the dots indicate further kinematic variables in the kinematic
bin of interest (which may also be averaged over).
It would be preferable if asymmetries were analyzed with known kinematic
prefactors divided out on event-by-event basis. One could then directly
compare asymmetries $\asym{XY}{\rm weight}$ measured in different
experiments and kinematics, and focus on effects of evolution
or power suppression for twist-3. In practice, often the kinematic
factors were included. We will define and comment on the explicit
expressions as needed.

For completeness we remark that after integrating the cross section
over transverse hadron momenta one obtains
\begin{subequations}\ba
     	\frac{d^4\sigma_{\rm leading}}{dx\,dy\,dz\,d\psi_l}
   &=&	 \frac{1}{2\pi}\; \frac{4\pi \alpha_{em}^2}{xyQ^2}\biggl(1-y+\frac12y^2\biggr) \; F_{UU}(x,z)
        \Biggl\{\;1 + \lambda\,S_L\,    p_2\,A_{LL} \Biggr\}
    	\label{Eq:SIDIS-leading-integrated} \\
	\frac{d^4\sigma_{\rm subleading}}{dx\,dy\,dz\,d\psi_l}
   &=&	 \frac{1}{2\pi}\; \frac{4\pi \alpha_{em}^2}{xyQ^2}\biggl(1-y+\frac12y^2\biggr) \; F_{UU}(x,z)
        \Biggl\{ S_T\sin(\phi_S)\,p_3\,A_{UT}^{\sin(\phi_S)} +
  \nonumber \\
  & &	 \lambda\,S_T\cos(\phi_S)\,p_4\,A_{LT}^{\cos(\phi_S)}
          \Biggr\} \, ,
     \hspace{1cm} \label{Eq:SIDIS-subleading-integrated}
\ea\end{subequations}
where (and analogous for the other structure functions)
\be\label{Eq:FUU-integrated}
	F_{UU}(x,z) = \int d^2\Phperp\;F_{UU}(x,z,\Phperp)
\ee
and the asymmetries are defined as
\be
	A_{XY}^{\rm weight}(x,z) = \frac{F_{XY}^{\rm weight}(x,z)}{F_{UU}(x,z)}\,.
\ee

The connection of ``collinear'' SIDIS structure functions
in (\ref{Eq:SIDIS-leading-integrated},~\ref{Eq:SIDIS-subleading-integrated})
to those known from inclusive DIS is established by integrating over $z$
and summing over hadrons as
\begin{subequations}\begin{alignat}{4}
	&\sum\limits_h\int d z\;z\;F_{UU}(x,z)
	&\equiv	&&	& 2\,x\,F_1(x) \;,
	\label{Eq:DIS-F1}\\ 
	&\sum\limits_h\int d z\;z\;F_{LL}(x,z)
	&\equiv && 	& 2\,x\,g_1(x) \;,
	\label{Eq:DIS-g1}\\ 
	&\sum\limits_h\int d z\;z\;F_{LT}^{\cos\phi_S}(x,z) \;\;
	&\equiv && \;\; -\,\gamma\; & 2\,x\biggl(g_1(x)+g_2(x)\biggr) \;,
	\label{Eq:DIS-gT}\\ 
	&\sum\limits_h\int d z\;z\;F_{UT}^{\sin\phi_S}(x,z)
	&=      && 	    & \;\; 0 \, ,
	\label{Eq:DIS-zero}
\end{alignat}\end{subequations}
where $\gamma=2M_Nx/Q$ signals the twist-3 character of $F_{LT}^{\cos\phi_S}(x,z)$.
Notice that $F_{UT}^{\sin\phi_S}(x,z)$ has no DIS counterpart due to time-reversal
symmetry of strong interactions, and terms suppressed by $1/Q^2$ are
consequently neglected throughout this work including the twist-4 DIS
structure function $F_L(x)$.

\subsection{TMDs, FFs and structure functions}
\label{Sec-2.2:def-TMD-FF}

TMDs are defined in terms of light-front correlators
\be\label{Eq:correlator}
    	\Phi(x,\bfkperp)_{ij} = \int\frac{ d \xi^- d^2{\bm \xi}_\perp}{(2\pi)^3}
	\;e^{ik\xi}\;\la N(P,S)|\bar\psi_j(0)\,{\cal W}_{(0,\,\infty)}
	{\cal W}_{(\infty,\,\xi)}\,\psi_i(\xi)|N(P,S)\ra
    	\with{ }{\xi^+\!=\!0}{$k^+ = xP^+$}  \, ,
	\ee
where the Wilson lines  {${\cal W}_{(0,\,\infty)}{\cal W}_{(\infty,\,\xi)}$}
refer to the SIDIS process
\cite{Collins:2002kn}. For a generic four-vector $a^\mu$ we define
the light-cone coordinates $a^\mu=(a^+,a^-,a_\perp)$ with
$a^\pm=(a^0\pm a^3)/\sqrt{2}$.
The light-cone direction is singled out by the virtual-photon momentum
and transverse vectors like $\bfkperp$ are perpendicular to it. In the
virtual-photon--nucleon center-of-mass frame, the nucleon and the partons
inside it move in the $(+)$--lightcone direction, while the struck
quark and the produced hadron move in the $(-)$--light-cone direction.
In the nucleon rest frame the polarization vector is given by
$S=(0,{\bm S}_T,S_L)$ with ${\bm S}_T^2+S_L^2=1$.

The 8 leading-twist TMDs \cite{Boer:1997nt} are projected out from
the correlator (\ref{Eq:correlator}) as follows (\blue{blue: T-even} TMDs,
\red{red: T-odd} TMDs; all TMDs depend on $x$, $k_\perp$, renormalization
scale and carry a flavor index which we do not indicate for brevity):
\begin{subequations}\ba
    \frac12\;{\rm Tr}\biggl[\gamma^+ \;\Phi(x,\bfkperp)\biggr]
    &=& \hspace{5mm}
    \blue{f_1}-\frac{\varepsilon^{jk}\kperp^j S_T^k}{M_N}\,\red{f_{1T}^\perp}\;,
    \label{Eq:TMD-pdfs-I}\\
    \frac12\;{\rm Tr}\biggl[\gamma^+\gamma_5 \;\Phi(x,\bfkperp)\biggr] &=&
    S_L\,\blue{g_1} + \frac{\bfkperp \cdot{\bm S}_T}{M_N}\,\blue{g_{1T}^\perp}\;,
    \label{Eq:TMD-pdfs-II}\\
    \frac12\;{\rm Tr}\biggl[i\sigma^{j+}\gamma_5 \;\Phi(x,\bfkperp)\biggr] &=&
    S_T^j\,\blue{h_1}  + S_L\,\frac{\kperp^j}{M_N}\,\blue{h_{1L}^\perp} +
    \frac{\kappa^{jk}S_T^k}{M_N^2}\,
    \blue{h_{1T}^\perp} + \frac{\varepsilon^{jk}\kperp^k}{M_N}\,\red{h_1^\perp}\;,
    \label{Eq:TMD-pdfs-III} \hspace{15mm}
\ea
and the 16 subleading-twist TMDs \cite{Mulders:1995dh,Bacchetta:2006tn}
are given by
\ba
\hspace{-5mm}
	\frac12{\rm Tr}\biggl[\,1\;\Phi(x,\bfkperp)\biggr]         &=&
    	\frac{M_N}{P^+}\biggl[
	\hspace{5mm}\blue{e}
	-\frac{\varepsilon^{jk}\kperp^j S_T^k}{M_N}\,\red{e_T^\perp}
    	\biggr], \label{Eq:sub-TMD-pdfs-I}\\
\hspace{-5mm}
	\frac12{\rm Tr}\biggl[i\gamma_5\Phi(x,\bfkperp)\biggr]        &=&
        \frac{M_N}{P^+}\biggl[
    	S_L\red{e_L} +\frac{\bfkperp \cdot {\bm S}_T}{M_N}\,\red{e_T}
    	\biggr], \label{Eq:sub-TMD-pdfs-II}\\
\hspace{-5mm}
	\frac12{\rm Tr}\biggl[\,\gamma^j\,\Phi(x,\bfkperp)\biggr]        &=&
        \frac{M_N}{P^+}\biggl[
    	\frac{\kperp^j}{M_N}\blue{f^\perp}\!+\varepsilon^{jk}S_T^k\red{f_T}
	\!+\!S_L\frac{\varepsilon^{jk}\kperp^k}{M_N}\red{f_L^\perp}
	\!-\!\frac{\kappa^{jk}\varepsilon^{kl}S_T^l}{M_N^2}\red{f_T^\perp}\!
	\biggr], \label{Eq:sub-TMD-pdfs-III}\\
\hspace{-5mm}
	\frac12{\rm Tr}\biggl[\,\gamma^j\gamma_5\Phi(x,\bfkperp)\biggr] &=&
    	\frac{M_N}{P^+}\biggl[
    	S_T^j\,\blue{g_T}
	+ S_L\,\frac{\kperp^j}{M_N}\blue{g_L^\perp} +
	\frac{\kappa^{jk}S_T^k}{M_N^2}
    	\,\blue{g_T^\perp}
	+\frac{\varepsilon^{jk}\kperp^k}{M_N}\,\red{g^\perp}
	\biggr], \label{Eq:sub-TMD-pdfs-IV}\\
\hspace{-5mm}
	\frac12{\rm Tr}\biggl[i\,\sigma^{jk}\gamma_5\Phi(x,\bfkperp)\biggr] &=&
    	\frac{M_N}{P^+}\biggl[
    	\frac{S_T^j \kperp^k-S_T^k \kperp^j}{M_N}\,\blue{h_T^\perp}
    	-\varepsilon^{jk}\,\red{h}
	\biggr], \label{Eq:TMD-pdfs-V} \\
\hspace{-5mm}
	\frac12{\rm Tr}\biggl[i\,\sigma^{+-}\,\gamma_5\,\Phi(x,\bfkperp)\biggr]
	&=& \frac{M_N}{P^+}\biggl[
    	S_L\,\blue{h_L} + \frac{\bfkperp\cdot{\bm S}_T}{M_N}\,\blue{h_T}
    	\biggr], \label{Eq:TMD-pdfs-VI}
\ea\end{subequations}
where $\kappa^{jk}\equiv (\kperp^j \kperp^k-\frac12\,\bfkperp^{\:2}\delta^{jk})$.
The indices $j,k,l$ refer to the plane transverse with respect to the
light cone, $\epsilon^{ij}\equiv\epsilon^{-+ij}$ and $\epsilon^{0123}=+1$.
Dirac structures not listed in (\ref{Eq:TMD-pdfs-I}--\ref{Eq:TMD-pdfs-VI})
are twist-4 \cite{Goeke:2005hb}.
Integrating out transverse momenta in the correlator (\ref{Eq:correlator})
leads to the ``usual'' PDFs known from collinear kinematics
\cite{Ralston:1979ys,Jaffe:1991ra}, namely at twist-2 level

\begin{subequations}\ba
    \frac12\;{\rm Tr}\biggl[\gamma^+ \;\Phi(x)\biggr]
    &=& \hspace{5mm}
    \blue{f_1}\;, 	\label{Eq:pdf-I}\\
    \frac12\;{\rm Tr}\biggl[\gamma^+\gamma_5 \;\Phi(x)\biggr] &=&
    S_L\,\blue{g_1}\;, 	\label{Eq:pdf-II}\\
    \frac12\;{\rm Tr}\biggl[i\sigma^{j+}\gamma_5 \;\Phi(x)\biggr] &=&
    S_T^j\,\blue{h_1}\;, \label{Eq:pdf-III} \hspace{75mm}
\ea
and at twist-3 level
\ba
    \frac12\;{\rm Tr}\biggl[\,1\;\Phi(x)\biggr] &=&
    \frac{M_N}{P^+}\;\blue{e}\;,  \label{Eq:sub-pdf-I}\\
    \frac12\;{\rm Tr}\biggl[\gamma^j\gamma_5 \;\Phi(x)\biggr] &=&
    \frac{M_N}{P^+}\;S_T^j\,\blue{g_T} \;, \label{Eq:sub-pdf-II}\\ \hspace{6mm}
    \frac12\;{\rm Tr}\biggl[\,i\,\sigma^{+-}\gamma_5 \;\Phi(x)\biggr]
    &=& \frac{M_N}{P^+}\;S_L\,\blue{h_L}\,. \label{Eq:sub-pdf-III}\hspace{75mm}
\ea\end{subequations}
Other structures drop out either due to explicit $\kperp$--dependence,
or due to the sum rules \cite{Bacchetta:2006tn}
\be\label{Eq:sum-rules-T-odd}
	\int d^2\bfkperp\;f_T^a(x,\kperp^2)=
	\int d^2\bfkperp\;e_L^a(x,\kperp^2)=
	\int d^2\bfkperp\;h^a(x,\kperp^2)=0
\ee
imposed by time reversal constraints.

{Fragmentation functions are defined through the following correlator~\cite{Metz:2016swz}
(where $\bfpperp$ denotes the transverse momentum of the produced hadrons
acquired during the fragmentation process with respect to the quark)}:
\be\label{Eq:correlator-FF}
    \Delta(z,\bfpperp)_{ij}
    = \sum\limits_X\!\int\!
    \frac{ d \xi^+ d^2 {\bm \xi}_\perp}{2z(2\pi)^3}\,e^{ip\xi}
    \, \la 0  |{\cal W}_{(\infty,\xi)}\psi_i(\xi)\,|h,X\ra\,
    \la h,X|\bar{\psi}_j(0){\cal W}_{(0,\infty)}|0\ra
    \with{\xi^-\!=\!0}
	 {p^- \!=\! P_h^-/z}
	 {${\bm p}_\perp \!=\! -\bfpperp/z$} \, .
    \ee
In this work we will consider only unpolarized final-state hadrons.
If the produced hadron moves fast in the $(-)$ light-cone direction,
the twist-2 FFs are projected out as
\begin{subequations}\ba
	\frac{1}{2}{\rm Tr}\big[\gamma^-\Delta(z,\bfpperp)\big]
	&=& \blue{D_1}\, , \label{eq:DeltaTr-twist-2a}\\
	\frac{1}{2}{\rm Tr}\big[i\sigma^{j-}\gamma_5\Delta(z,\bfpperp)\big]
	&=& \epsilon^{jk}\,\frac{\pperp^k}{z\mh}\red{H_1^\perp}\;,
	\label{eq:DeltaTr-twist-2b}
\ea
and at twist-3 level
\ba
    \frac12\;{\rm Tr}\biggl[\,1\;\Delta(z,\bfpperp)\biggr]         &=&
    \phantom{-}\frac{\mh}{P^-_h}\;\blue{E}\;,  \label{eq:DeltaTr-twist-3a}\\
    \frac12\;{\rm Tr}\biggl[\;\,\gamma^j\;\Delta(z,\bfpperp)\biggr]  &=&
    -\frac{\pperp^j}{zP_h^-}\;\blue{D^\perp}\;, \label{eq:DeltaTr-twist-3b}\\
    \frac12\;{\rm Tr}\biggl[\gamma^j\gamma_5 \,\Delta(z,\bfpperp)\biggr] &=&
    \varepsilon^{jk}\,\frac{\pperp^k}{zP_h^-}\,\red{G^\perp}\;,
	\label{eq:DeltaTr-twist-3c}\\
    \frac12\;{\rm Tr}\biggl[i\,\sigma^{jk}\gamma_5\,\Delta(z,\bfpperp)
	\biggr] &=&
    -\varepsilon^{jk}\,\frac{\mh}{P_h^-}\;\red{H}\;.  \label{eq:DeltaTr-twist-3d}
\ea\end{subequations}


\noindent
The FFs depend on $z$, $P_\perp$, renormalization scale, quark flavor and
type of hadron which we do not indicate for brevity.
Integration over transverse hadron momenta leaves us with $D_1(z)$, $E(z)$,
$H(z)$ while the other structures drop out due to their $\pperp$ dependence.


The structure functions in
Eqs.~(\ref{Eq:SIDIS-leading},~\ref{Eq:SIDIS-subleading}) are described
in the Bjorken limit at tree level in terms of convolutions of TMDs
and FFs. We define the unit vector $\bfhp   = \bfPhperp/\Phperp$
and use the following convolution integrals
(see Appendix \ref{ApendixB1} for details)
\be
 \label{Eq:def-convolution-integral}
 {\cal C}\biggl[\omega\;f\;D\biggr]
	= x \sum_a e_a^2\int d^2\bfkperp^{ } d^2\bfpperp
 	\; \delta^{(2)}(z \bfkperp^{ }+ \bfpperp^{ }-\bfPhperp^{ })\;\omega
  	\; f^a(x,\bfkperp^2)\ D^a(z,\bfpperp^2)\;,
\ee
where $\omega$ is a weight function, which in general depends on
$\bfkperp$ and $\bfpperp$.
The 8 leading-twist structure functions are
\begin{subequations}
\label{Eqs:structure-functions-twist-2}
\ba
 F_{UU}	&=&{\cal C}\biggl[\;\omega^{\{0\}}\,f_1 D_1 \;\biggr] \, , \label{FUU}\\
  F_{UU}^{\cos 2\phi_h} 	
	&=& {\cal C}\biggl[\;\omega^{\{2\}}_{\rm AB}\,h_{1}^{\perp }\,H_{1}^{\perp }\;
	\biggr] \, , \label{F_UUcos2phi}\\
F_{UL}^{\sin 2\phi_h} 	
	&=& {\cal C}\biggl[\;\omega^{\{2\}}_{\rm AB}\,h_{1L}^{\perp } H_{1}^{\perp }\;
	\biggr] \, , \label{F_UUsin2phi}\\
 F_{LL}	&=&{\cal C}\biggl[\;\omega^{\{0\}}\,g_1 D_1 \;\biggr] \, , \label{FLL}\\
 F_{LT}^{\cos(\phi_h -\phi_S)}
	&=& {\cal C}\biggl[\,\omega^{\{1\}}_{\rm B} \,g_{1T}^\perp D_1\biggr] \, ,
	\label{Eq:FLT-twist-2}\\
 F_{UT}^{\sin\left(\phi_h +\phi_S\right)}
	&=& {\cal C}\biggl[\;\omega^{\{1\}}_{\rm A} \,h_{1} H_1^{\perp}\;\biggr] \, ,
	\label{Eq:FUTCol}\\
 F_{UT}^{\sin\left(\phi_h -\phi_S\right)}
	&=& {\cal C}\biggl[-\,\omega^{\{1\}}_{\rm B} \,f_{1T}^{\perp } D_1\,\biggr] \, ,
	\label{Eq:FUTSiv}\\
 F_{UT}^{\sin\left(3\phi_h -\phi_S\right)}
	&=& {\cal C}\biggl[\;\omega^{\{3\}}_{\rm { }}\,h_{1T}^{\perp } H_1^{\perp }\;
	\biggr] \, . \hspace{75mm} \label{Eq:FUTpretzel}
\ea\end{subequations}
At subleading-twist we have the structure functions
\begin{subequations}
\label{Eqs:structure-functions-twist-3}
\ba
	\ \hspace{-1cm}	F_{UU}^{\cos\phi_h}
	&=&
	\frac{2M_N}{Q}\,{\cal C}\biggl[\phantom{-}
   	\omega^{\{1\}}_{\rm A}
	\biggl( x h\,H_{1}^{\perp }
   	+ r_h^{ }\,\,f_1 \frac{\tilde{D}^{\perp }}{z}\biggr)
	- \omega^{\{1\}}_{\rm B} \biggl( x  f^{\perp } D_1
   	+ r_h^{ }\,\,h_{1}^{\perp } \frac{\tilde{H}}{z}\biggr)\biggr] ,
	\label{Eq:FUUcosphi}\\
	\ \hspace{-1cm}	F_{LU}^{\sin\phi_h}
	&=&
	\frac{2M_N}{Q}\,{\cal C}\biggl[ \phantom{-}
	\omega^{\{1\}}_{\rm A}
   	\biggl( x \, e \, H_1^{\perp }
   	+ r_h^{ }\,\,f_1\frac{\tilde{G}^{\perp }}{z}\,\biggr)
   	+\omega^{\{1\}}_{\rm B}
   	\biggl( x   g^{\perp }  D_1
   	+ r_h^{ }\,\, h_1^{\perp } \,\frac{\tilde{E}}{z} \biggr)\biggr] ,
	\label{FLUsinphi}\;\;\;\;\;\\
	\ \hspace{-1cm}	F_{UL}^{\sin\phi_h}
 	&=&
	\frac{2M_N}{Q}\,{\cal C}\biggl[\phantom{-}
   	\omega^{\{1\}}_{\rm A}
    	\biggl( x   h_L  H_1^{\perp } \!
   	+ r_h^{ }\,\,g_{1}\frac{\tilde{G}^{\perp } }{z}\biggr)
   	+\omega^{\{1\}}_{\rm B}
    	\biggl( x  f_{L}^{\perp }  D_1 \!
   	- r_h^{ }\,\, h_{1L}^{\perp }  \frac{\tilde{H}}{z}\biggr)\biggr] ,
	\label{FULsinphi}\\
	\ \hspace{-1cm}	F_{LL}^{\cos \phi_h}
 	&=&
	\frac{2M_N}{Q}\,{\cal C}\biggl[
	-\omega^{\{1\}}_{\rm A}
   	\biggl( x  e_L  H_1^{\perp }
   	- r_h^{ }\,\,g_{1}   \frac{\tilde{D}^{\perp }}{z}\biggr)
   	-\omega^{\{1\}}_{\rm B}
   	\biggl( x   g_L^{\perp }   D_1
   	+  r_h^{ }\,\,h_{1L}^{\perp } \frac{\tilde{E}}{z}\biggr)\biggr]  ,
	\label{FLLcosphi}\;\;\;\;\;\;\;\;\;\\
	\ \hspace{-1cm}	F_{UT}^{\sin \phi_S }
	&=&
	\frac{2M_N}{Q}\,{\cal C}\biggl[ \phantom{-}
	\omega^{\{0\}}\,\biggl(x   f_T   D_1 \;
   	- r_h^{ }\, \; h_{1} \, \frac{\tilde{H}}{z} \biggr)\nonumber\\
   	&&\hspace{1.2cm}
   	-\frac{\omega^{\{2\}}_{\rm B}}{2}
	\biggl( x   h_{T}  H_{1}^{\perp }
   	+ r_h^{ }\, g_{1T}^\perp \,\frac{\tilde{G}^{\perp }}{z}
   	- x   h_{T}^{\perp }  H_{1}^{\perp }
	+ r_h^{ }\, f_{1T}^{\perp } \,\frac{\tilde{D}^{\perp }}{z}
   	\biggr) \biggr]\, , \label{FUTsinphiS}
\ea
\ba
	\ \hspace{-1cm}	F_{LT}^{\cos \phi_S}
	&=&
	\frac{2M_N}{Q}\,{\cal C}\biggl[
   	- \omega^{\{0\}}\,\biggl(x   g_T   D_1
   	+ r_h^{ }\, \, h_{1}  \frac{\tilde{E}}{z} \biggr)\nonumber\\
   	&&\hspace{1.2cm}
	+\frac{\omega^{\{2\}}_{\rm B}}{2}
   	\biggl( x   e_{T}  H_{1}^{\perp }
   	- r_h^{ }\, g_{1T}^\perp \,\frac{\tilde{D}^{\perp }}{z}
   	+  x   e_{T}^{\perp }  H_{1}^{\perp }
   	+ r_h^{ }\,f_{1T}^{\perp }\,\frac{\tilde{G}^{\perp }}{z}\biggr)\biggr]
	\, , \label{FLTcosphiS} \\ \hspace{-5mm}
	\ \hspace{-1cm}	F_{UT}^{\sin(2\phi_h -\phi_S)}
	&=&
	\frac{2M_N}{Q}\,{\cal C}\biggl[\phantom{-}
   	\frac{\omega^{\{2\}}_{\rm AB}}{2}\,
   	\biggl( x   h_{T}  H_{1}^{\perp }
   	+ r_h^{ }\, g_{1T}^\perp \,\frac{\tilde{G}^{\perp }}{z}
        + x   h_{T}^{\perp }  H_{1}^{\perp }
   	- r_h^{ }\, f_{1T}^{\perp } \,\frac{\tilde{D}^{\perp }}{z}
	\biggr)\nonumber\\
	&&\hspace{1.3cm}
	+
   	\omega^{\{2\}}_{\rm C}
   	\biggl( x   f_T^{\perp } D_1
   	- r_h^{ }\, \, h_{1T}^{\perp }  \frac{\tilde{H}}{z}\biggr) \biggr] \, , \\
	\ \hspace{-1cm}	F_{LT}^{\cos(2\phi_h - \phi_S)}
	&=& \frac{2M_N}{Q}\,{\cal C}\biggl[
   	- \frac{\omega^{\{2\}}_{\rm AB}}{2}
   	\biggl( x   e_{T}  H_{1}^{\perp }
   	- r_h^{ }\, g_{1T}^\perp \,\frac{\tilde{D}^{\perp }}{z}
	- x   e_{T}^{\perp }  H_{1}^{\perp }
   	- r_h^{ }\, f_{1T}^{\perp } \,\frac{\tilde{G}^{\perp }}{z}\biggr)\nonumber\\
	&&\hspace{1.3cm}
   	- \omega^{\{2\}}_{\rm C}
   	\biggl( x   g_T^{\perp }   D_1
   	+ r_h^{ }\, \, h_{1T}^{\perp }  \frac{\tilde{E}}{z}\biggr)\biggr]\, ,
	\label{FLTsin(2phi-phiS)}
\ea\end{subequations}
where $r_h = \mh/M_N$ and
$F_{XY}^{\rm weight}\equiv F_{XY}^{\rm weight}(x,z,\Phperp)$. The
tilde-functions $\tilde{D}^{\perp },\,\tilde{G}^{\perp },\,\tilde{H},\,\tilde{E}$
are defined in terms of $\bar{q}gq$-correlators, see
Sec.~\ref{Sec-3.2:WW-type-TMD-FF}. The weight functions are defined as
\ba
&& \omega^{\{0\}}  	= 1 \, , \nonumber\\
&& \omega^{\{1\}}_{\rm A} 	= \frac{\bfhp\cdot\bfpperp^{ }}{z \mh}  \, , \;\;\;
   \omega^{\{1\}}_{\rm B} 	= \frac{\bfhp\cdot\bfkperp^{ }}{M_N}\,,\nonumber\\
&& \omega^{\{2\}}_{\rm A} 	=  \frac{2\, \bigl(\bfhp\cdot\bfpperp^{ }\bigr)\,
			\bigl(\bfhp\cdot\bfkperp^{ }\bigr)}{zM_N\mh}\,,\;\;\;
   \omega^{\{2\}}_{\rm B}	= -\frac{\bfpperp^{ }\cdot\bfkperp^{ }}{zM_N\mh}\,,\;\;\;
   \omega^{\{2\}}_{\rm C}	= \frac{2\,(\bfhp\cdot\bfkperp^{ })^2-\bfkperp^2}{2M_N^2}
			\, ,\nonumber\\
&& \omega^{\{3\}}_{\rm{ }}	= \frac{
			 4\,(\bfhp\cdot\bfpperp^{ })\,(\bfhp\cdot\bfkperp^{ })^2
			-2\,\bigl(\bfhp\cdot\bfkperp^{ }\bigr)\,
        		\bigl(\bfkperp^{ }\cdot\bfpperp^{ }\bigr)
   			-\bigl(\bfhp\cdot\bfpperp^{ }\bigr)\,\bfkperp^2\
   			}{2 z M_N^2 \mh}\,,
			\label{Eq:wi} \ea
and $\omega^{\{2\}}_{\rm AB} = \omega^{\{2\}}_{\rm A} + \omega^{\{2\}}_{\rm B}$.
In $\omega^{\{n\}}_{i}$ the index $n=0,\,1,\,2,\,3$ indicates the (maximal)
power $(\Phperp)^n$ with which the corresponding contribution scales,
and index $i$ (if any) distinguishes different types of contributions
at the given order $n$.
 Notice that twist-3 structure functions in
 Eqs.~(\ref{Eq:FUUcosphi}--\ref{FLTsin(2phi-phiS)}) contain an explicit
 factor $M_N/Q$. We also recall that we neglect two structure functions
 (denoted in  \cite{Bacchetta:2006tn}
 as $F_{UU,L}$ and $F_{UT,L}^{\sin( \phi_h-\phi_S)}$)
 due to longitudinal virtual-photon polarization, which are
 of order ${\cal O}(M^2/Q^2)$ in the TMD partonic description.

The structure functions that survive $\Phperp$--integration of the SIDIS cross section in
(\ref{Eq:SIDIS-leading-integrated},~\ref{Eq:SIDIS-subleading-integrated})
are associated with the trivial weights $\omega^{\{0\}}$ and expressed in
terms of collinear PDFs and FFs as follows (here the
sum rules (\ref{Eq:sum-rules-T-odd}) are used):
\begin{subequations}\ba
	F_{UU}(x,z) &=& x\sum\limits_ae_a^2\,f_1^a(x)\,D_1^a(z) \, ,
	\label{Eq:FUU-collinear}\\
	F_{LL}(x,z) &=& x\sum\limits_ae_a^2\,g_1^a(x)\,D_1^a(z) \, ,
	\label{Eq:FLL-collinear}\\
 	F_{LT}^{\cos\phi_S}(x,z) &=& -\,\frac{2M_N}{Q}\; x\sum_a e_a^2\,
	\biggl(x\,g_T^q(x)\,D_1^a(z)+r_h\,h_1^a(x)\frac{\tilde{E}^a(z)}{z}\biggr) \, ,
	\label{Eq:FLT-collinear}\\
	 F_{UT}^{\sin\phi_S}(x,z) &=& -\,\frac{2\,\mh}{Q}\; x\sum_a e_a^2\,
	h_1^a(x)\frac{\tilde{H}^a(z)}{z} \, .
	\label{Eq:FUT-collinear}
\ea\end{subequations}
Finally, integrating over $z$, summing over hadrons, and
using the sum rules for the T-odd FFs,
$\sum_h\int d z\;\tilde{E}^a(z)=0$ and
$\sum_h\int d z\;\tilde{H}^a(z)=0$, we recover
Eqs.~(\ref{Eq:DIS-F1}--\ref{Eq:DIS-zero}) and obtain for the DIS structure
functions
\begin{subequations}\ba
    F_1(x) & = & \frac{1}{2}\sum_a e_a^2\,f_1^a(x) \, , \label{Eq:DIS-F1-II}\\
    g_1(x) & = & \frac{1}{2}\sum_a e_a^2\,g_1^a(x) \, , \label{Eq:DIS-g1-II}\\
    g_2(x) & = & \frac{1}{2}\sum_a e_a^2\,g_T^a(x)\;-\;g_1(x)  \; .\label{Eq:DIS-g2}
	\hspace{25mm}
\ea\end{subequations}

Before introducing the WW-type approximations in the next section,
we would like to add a comment on TMD factorization: the
partonic description of the leading-twist structure functions in
(\ref{Eqs:structure-functions-twist-2}) is based on factorization
theorems \cite{Collins:1981uk,Ji:2004wu,Ji:2004xq,Collins:2011zzd,
Echevarria:2012js}. In contrast to this, the partonic description
of the subleading-twist structure functions in
(\ref{Eqs:structure-functions-twist-3}) is based on the
{\it assumption} that the SIDIS cross section factorizes.

A lot of progress has been achieved in recent years in the
theoretical understanding of leading-twist observables within the TMD
framework, including definition, renormalization and evolution of
leading-twist TMDs
\cite{Aybat:2011zv,Aybat:2011ge,Echevarria:2014xaa,Collins:2014jpa},
next-to-leading order corrections within the TMD framework
\cite{Ma:2013aca}, and phenomenological fits with evolution
\cite{Aybat:2011ta,Kang:2015msa}.
The matching of twist-2 collinear and TMD quantities was
studied to next-to-leading and next-to-next-to-leading order
\cite{Gutierrez-Reyes:2017glx,Gutierrez-Reyes:2018qez}.
The WW approximation has been used recently in Ref.~\cite{Scimemi:2018mmi}
to connect the twist-2 TMDs $f_{1T}^\perp$, $g_{1T}^\perp$, $h_{1}^\perp$,
$h_{1L}^\perp$ to certain higher-twist collinear matrix elements.

In contrast to this, the theory for subleading-twist TMD observables is
only poorly developed. Still to the present day, the state-of-the-art
approach to subleading-twist TMD observables is the one of
Refs.~\cite{Kotzinian:1994dv,Mulders:1995dh,Boer:1997nt,Goeke:2005hb,
Bacchetta:2006tn},
based on a TMD tree-level formalism, which we adopt here.
In fact, the results of Refs.~\cite{Metz:2004je,Gamberg:2006ru}
indicate doubts even in the tree-level formalism.
Recently, an attempt was made to remedy these doubts \cite{Chen:2016hgw}.
Keeping in mind these ``words of warning,'' still the formulas
(\ref{Eqs:structure-functions-twist-3})
are the best that theory has to offer currently. We may consider
(\ref{Eqs:structure-functions-twist-3}) as a model itself for
the twist-3 SIDIS observables. We hope that the phenomenological
approach based on WW-type approximations pursued in this work might
lead to more insight into these observables, and eventually might
trigger more theory efforts in the future.

%
\section{WW and WW-type approximations}
\label{Sec-3:WW}

In this section we will define the approximations and review what is
known about them.
The basic idea of the approximations is simple. One uses QCD
equations of motion to separate contributions from $\bar{q}q$--terms
and $\bar{q}gq$--terms and assumes that the latter can be neglected
with respect to the leading $\bar{q}q$--terms with a useful accuracy
(here the $\la\dots\ra$ denote symbolically the matrix elements
which enter the definitions of TMDs or FFs):
\be\label{Eq:WW-generic}
	\biggl|\frac{\la\bar{q}gq\ra}{\la\bar{q}q\ra}\biggr| \ll 1\,.
\ee

\subsection{WW approximation for PDF\lowercase{s}}
\label{Sec-3.1:WW-classic}

The WW approximation applies in principle to all twist-3 PDFs,
Eqs.~(\ref{Eq:sub-pdf-I},~\ref{Eq:sub-pdf-II},~\ref{Eq:sub-pdf-III}).
It was established first for $g_T^a(x)$ \cite{Wandzura:1977qf}, and
later for $h_L^a(x)$ \cite{Jaffe:1991ra}. The situation of $e^a(x)$
is somewhat special, see below and the review \cite{Efremov:2002qh}.

The origin of the approximations is as follows.
The operators defining $g_T^a(x)$ and~$h_L^a(x)$ can be decomposed by means
of QCD equations of motion in twist-2 parts, and pure twist-3
(interaction dependent) $\bar{q}gq$--terms and current-quark mass
terms. We denote $\bar{q}gq$--terms and mass terms collectively
and symbolically by functions with a tilde.\footnote{In the literature it is 
customary to reserve the term "tilde terms"
for matrix elements of $\bar{q}gq$ operators as done, e.g., in
Ref~\cite{Bacchetta:2006tn}. For convenience in this work "tilde terms"
refers to both $\bar{q}gq$ terms and current-quark mass terms as
done, e.g., in \cite{Metz:2008ib}.}  
Such decompositions are possible because $g_T^a(x)$ and $h_L^a(x)$ are
``twist-3'' not according to the ``strict QCD definition''
(twist $=$ mass dimension of associated local operator minus its spin).
Rather they are classified according to the ``working definition''
of twist \cite{Jaffe:1996zw}
(a function is ``twist $t$'' if, in addition to overall kinematic
prefactors, it contributes to cross sections in a partonic
description suppressed by $(M/Q)^{t-2}$ where $M$ is a generic
hadronic and $Q$ the hard scale).
The two definitions coincide for twist-2 quantities, but higher-twist
observables in general contain ``contaminations'' by leading twist.

In this way one obtains the decompositions and, if they apply, WW
approximations \cite{Wandzura:1977qf,Jaffe:1991ra} (keep in mind
here tilde terms contain pure twist-3 and current-quark mass terms)
\begin{subequations}\begin{alignat}{3}
   	g_T^a(x) &=&
        \phantom{2x}\;\int_x^1\frac{ d y}{y}\,g_1^a(y) + &\tilde{g}_T^a(x)
        \stackrel{\rm WW}{\approx}
        \phantom{2x}\;\int_x^1\frac{ d y}{y}\,g_1^a(y) \;,
	\label{Eq:WW-original1} \\
   	h_L^a(x) &=& 2x\int_x^1\frac{ d y}{y^2}\,h_1^a(y) + &\tilde{h}_L^a(x)
        \stackrel{\rm WW}{\approx} 2x\int_x^1\frac{ d y}{y^2}\,h_1^a(y)\;,
	\label{Eq:WW-original2}\\
   	x\,e^a(x) &=& x\,&\tilde{e}^a(x) \;\stackrel{\rm WW}{\approx}
	\;\;\;\; 0 \, , \label{Eq:WW-original-e}
\end{alignat}\end{subequations}
where we included $e^a(x)$ which is a special case in the sense that it
receives no twist-2 contribution.
A prefactor of $x$ is provided in (\ref{Eq:WW-original-e})
to cancel a $\delta(x)$--type singularity \cite{Efremov:2002qh}.

The relations (\ref{Eq:WW-original1}--\ref{Eq:WW-original-e})
have been derived basically using operator product expansion
techniques \cite{Wandzura:1977qf,Jaffe:1991ra}. Notice that the
operators defining $g_T^a$ and $h_L^a$ can also be decomposed
within the TMD framework by means of a combination of relations
derived from the QCD equations of motion and further
constraint relations, called Lorentz-invariance relations (LIRs),
into a twist-2 part, and dynamical twist-3 (interactions dependent)
$\bar{q}gq$-terms and current-quark mass terms (see recent review \cite{Kanazawa:2015ajw} and references therein).

We will come back to (\ref{Eq:WW-original1},~\ref{Eq:WW-original2}) and
review the theoretical predictions and supporting experiments, but before
we will introduce the WW-type approximations for TMDs and FFs.

\subsection{WW-type approximations for TMDs and FFs}
\label{Sec-3.2:WW-type-TMD-FF}

Analogous to WW approximations for PDFs discussed in
Sec.~\ref{Sec-3.1:WW-classic}, also certain TMDs and FFs
can be decomposed into twist-2 contributions and tilde terms.
The latter may be assumed, in the spirit of (\ref{Eq:WW-generic}),
to be small. Hereby it is important to keep in mind that for each TMD
or FF one deals
with different types of (``unintegrated'') $\bar{q}gq$--correlations,
and we prefer to refer to them as WW-type approximations.

In the T-even case one obtains the following approximations\footnote{%
	Notice that Ref.~\cite{Bacchetta:2006tn} uses four-vector notation
	for transverse vectors, while in this paper we always utilize
	two-vectors for transverse vectors, such that e.g.\
	${\bfkperp^2}_{\rm our} = -{{p}_T^2}_{\mbox{\tiny Ref.~\cite{Bacchetta:2006tn}}}$
	and analog for other scalar products of {\it transverse} vectors.
	In our notation transverse vectors
	are never understood as four-vectors such that
	${\bfkperp^2}_{\rm our} \equiv {\kperp^2}_{\rm our}$.},
where the terms on the left-hand-side are twist-3, those on the
right-hand-side (if any) are twist-2,
\begin{subequations}\ba
xe^q(x,\kperp^2)	&\WWtype&
			0,
			\label{Eq:WW-type-1}\\
xf^{\perp q}(x,\kperp^2)  &\WWtype&
			f_{1}^q(x,\kperp^2),
			\label{Eq:WW-type-Cahn}\phantom{\frac11}\\
xg_L^{\perp q}(x,\kperp^2)&\WWtype&
			g_{1}^q(x,\kperp^2),
			\label{Eq:WW-type-gLperp}\phantom{\frac11}\\
xg_T^{\perp q}(x,\kperp^2)&\WWtype&
			g_{1T}^{\perp q}(x,\kperp^2),\phantom{\frac11}
			\label{Eq:WW-type-gTperp}\\
xg_T^q(x,\kperp^2)   	&\WWtype&
             		g_{1T}^{\perp(1)q}(x,\kperp^2),
			\label{Eq:WW-type-gT}\\
xh_L^q(x,\kperp^2)	&\WWtype& -2 \,
                       	h_{1L}^{\perp(1)q}(x,\kperp^2),\phantom{\frac11}
                       	\label{Eq:WW-type-6}\\
xh_T^q(x,\kperp^2)      &\WWtype&
                       	- h_1^q(x,\kperp^2) - h_{1T}^{\perp(1)q}(x,\kperp^2),
                       	\label{Eq:WW-type-7}\\
xh_T^{\perp q}(x,\kperp^2)&\WWtype&
                       	\phantom{-}h_1^q(x,\kperp^2)-h_{1T}^{\perp(1)q}(x,\kperp^2).
                       	\phantom{\frac11} \label{Eq:WW-type-8}
\ea\end{subequations}
In the T-odd case one obtains the approximations
\begin{subequations}\ba
xe_L^q(x,\kperp^2)         	&\WWtype& 0,
			\label{Eq:WW-type-eLperp}\\
xe_T^q(x,\kperp^2)         	&\WWtype& 0, \\
xe_T^{\perp q}(x,\kperp^2) 	&\WWtype& 0, \\
xg^{\perp q}(x,\kperp^2)   	&\WWtype& 0,
                       	\label{Eq:WW-type-gperp}\\
xf_L^{\perp q}(x,\kperp^2) 	&\WWtype& 0,
			\label{Eq:WW-type-fLperp}\phantom{\frac11}\\
xf_T^{\perp q}(x,\kperp^2) 	&\WWtype&
                       	\phantom{-}\,f_{1T}^{\perp q}(x,\kperp^2),
			\phantom{\frac11}
                       	\label{Eq:WW-type-fTperp}\\
xf_T^{q}(x,\kperp^2)       	&\WWtype&
                       	-\,f_{1T}^{\perp(1)q}(x,\kperp^2), \label{Eq:WW-type-fT}\\
xh^q(x,\kperp^2)           	&\WWtype&
                       	- 2\,{h_1^{\perp(1)q}(x,\kperp^2)}.\phantom{\frac11 XXXXx}
                       	\label{Eq:WW-type-last}
\ea\end{subequations}
The superscript ``(1)'' denotes the first transverse
moment of TMDs defined generically as
\ba
f^{(1)}(x,\kperp^2) = \frac{\kperp^2}{2M^2}\,f(x, \kperp^2)\; , \;\;\;
f^{(1)}(x ) = \int d^2 \kperp f^{(1)}(x,\kperp^2) \; .
\ea

Two very useful WW-type approximations follow from combining the
WW approximations (\ref{Eq:WW-original1},~\ref{Eq:WW-original2}) with the
WW-type approximations (\ref{Eq:WW-type-gT},~\ref{Eq:WW-type-6}).
The resulting relations are the only WW-type relations applicable
to twist-2 TMDs and are given by 
\cite{Tangerman:1994bb,Mulders:1995dh,Avakian:2007mv}:
%
%
\begin{subequations}\ba
   	g_{1T}^{\perp(1)a}(x)&\WWtype&
        \phantom{-} x\,\int_x^1\frac{d y}{y\;}\,g_1^a(y) \;,
	\label{Eq:WW-approx-g1T}\\
    	h_{1L}^{\perp(1)a}(x)&\WWtype& -
	x^2\!\int_x^1\frac{d y}{y^2\;}\,h_1^a(y)\;.
	\label{Eq:WW-approx-h1L}
\ea\end{subequations}
Some of the above WW-type approximations were discussed in
\cite{Tangerman:1994bb,Kotzinian:1995cz,Mulders:1995dh,Kotzinian:1997wt,
Kotzinian:2006dw,Avakian:2007mv,Metz:2008ib,Teckentrup:2009tk}.
WW-relations for FFs are actually not needed: in Eqs.~%
(\ref{Eqs:structure-functions-twist-2},~\ref{Eqs:structure-functions-twist-3})
either twist-2 FFs $D_1^q$, $H_1^{\perp q}$ enter or tilde FFs, as a consequence
of how the azimuthal angles are defined \cite{Bacchetta:2006tn}.
For completeness we quote the WW-type approximations for FFs
\cite{Bacchetta:2006tn}
\begin{subequations}\ba
	E(z,\pperp^2)      &\WWtype& 0 ,
	\label{Eq:WW-type-FF-1}\\
	G^\perp(z,\pperp^2) &\WWtype& 0 ,
	\label{Eq:WW-type-FF-2}\\
	D^\perp(z,\pperp^2) &\WWtype
	& \hspace{11mm}z\,D_1(z,\pperp^2)\,, \label{Eq:WW-type-FF-3}\\
	H(z,\pperp^2) &\WWtype&
	-\frac{\pperp^2}{z\mh^2}\;H_1^\perp(z,\pperp^2)\,. \label{Eq:WW-type-FF-4}
\ea\end{subequations}
Having introduced the WW and WW-type approximations, we will review
in the following what is currently known from theory and experiment
about the WW(-type) approximations.

\subsection{Predictions from instanton vacuum model}
\label{Sec-3.3:WW-classic-instanton}

Insights into the relative size of hadronic matrix elements, such as
Eq.~(\ref{Eq:WW-generic}), require a non-perturbative approach. It is
by no means obvious which small parameter in the strong-interaction
regime would allow one to explain such results. An appealing
non-perturbative approach is provided by the instanton model of
the QCD vacuum \cite{Shuryak:1981ff,Diakonov:1983hh,Diakonov:1995qy}.
This semi-classical approach assumes that properties of the QCD vacuum
are dominated by instantons and anti-instantons, topological non-perturbative
gluon field configurations, which form a strongly interacting medium.
The approach provides a natural mechanism for dynamical chiral-symmetry
breaking, the dominant feature of strong interactions in the nonperturbative
regime. It was shown with variational and numerical methods that the
instantons form a dilute medium characterized by a non-trivial small parameter
$\rho/R\sim1/3$ \cite{Shuryak:1981ff,Diakonov:1983hh,Diakonov:1995qy},
where $\rho$ and $R$ denote respectively the average instanton size $\rho$
and separation $R$.

Applying the instanton vacuum model to studies of $g_T^a(x)$ and $h_L^a(x)$,
it was predicted that matrix elements of the $\bar{q}gq$ operators defining
$\tilde{g}_T^a(x)$ \cite{Balla:1997hf} and $\tilde{h}_L^a(x)$
\cite{Dressler:1999hc} are strongly suppressed by powers of the small
parameter $\rho/R$ with respect to contributions from the
respective twist-2 parts, which are of order $(\rho/R)^0$.
For the $n = 3$ Mellin moments (i.e.\ the lowest non-trivial ones for 
these tilde-functions) it was found \cite{Balla:1997hf,Dressler:1999hc}
\be\label{Eq:WW-instanoton}
	\frac{\tilde{g}_T^q}{g_T^q} \sim \frac{\tilde{h}_L^q}{h_L^q}
	\sim \frac{\la\bar{q}gq\ra}{\la\bar{q}q\ra} \sim
	\biggl(\frac{\rho}{R}\biggr)^{\!4} \log\biggl(\frac{\rho}{R}\biggr)
	\sim 10^{-2} \, ,
\ee
which strongly supports the generic approximation
in Eq.~(\ref{Eq:WW-generic}) with the instanton packing fraction providing
the non-trivial small parameter justifying the neglect of tilde terms.
The predictions for $\tilde{g}_T^a(x)$ \cite{Balla:1997hf}
were made before the advent of the first precise data on $g_2(x)$,
which we discuss next.
The instanton calculus has not yet been applied to $\tilde{e}^a(x)$.

\subsection{Tests of WW approximation in DIS experiments}
\label{Sec-3.4:WW-classic-experiment}

The presently available phenomenological information on $g_T^a(x)$ is due
to measurements of the structure function $g_2(x)$, Eq.~(\ref{Eq:DIS-g2}),
in DIS off various transversely polarized targets. In the WW-approximation
(\ref{Eq:WW-original1}) one can write $g_2(x)$ as a total derivative
expressed in terms of the experimentally well-known twist-2
structure function $g_1(x)$ as follows
\be
    	g_2(x) \stackrel{\rm WW}{\approx} g_2(x)_{\rm WW} \equiv
	\frac{ d\;}{ d x}\;\Biggl[x\int_x^1\frac{ d y}{y}
	\;g_1(y)\Biggr]\,.\label{Eq:g2-in-WW-approximation}
\ee
Data support (\ref{Eq:g2-in-WW-approximation}) to a good accuracy
\cite{Anthony:2002hy,Adams:1994id,Abe:1998wq,Airapetian:2011wu}, although especially
at smaller $x$ more stringent tests are not yet possible. Overall it has
been estimated that the WW approximation for $g_2(x)$ and $g_T^a(x)$ works
with an accuracy of about $40\,\%$ or better \cite{Accardi:2009au}.

\begin{figure}[b!]
\centering
\includegraphics[width=0.45\textwidth]{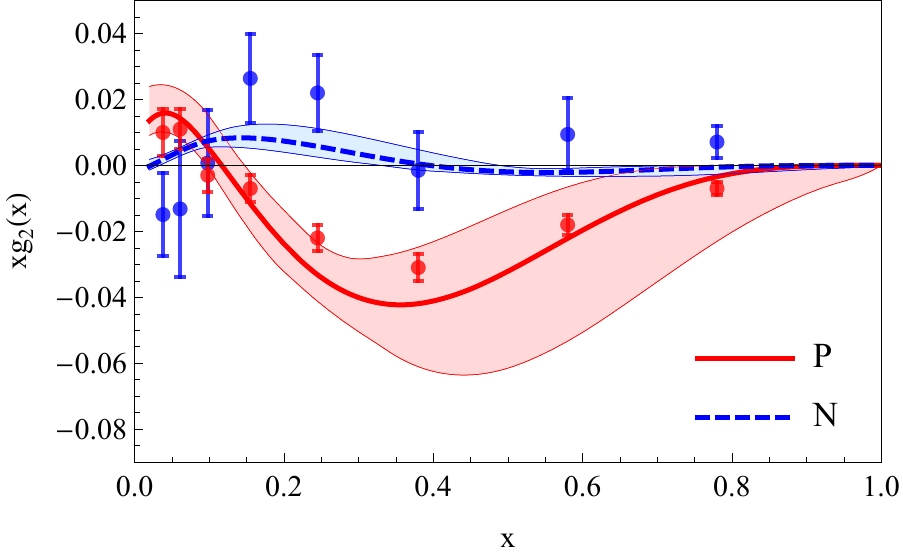}
\includegraphics[width=0.45\textwidth]{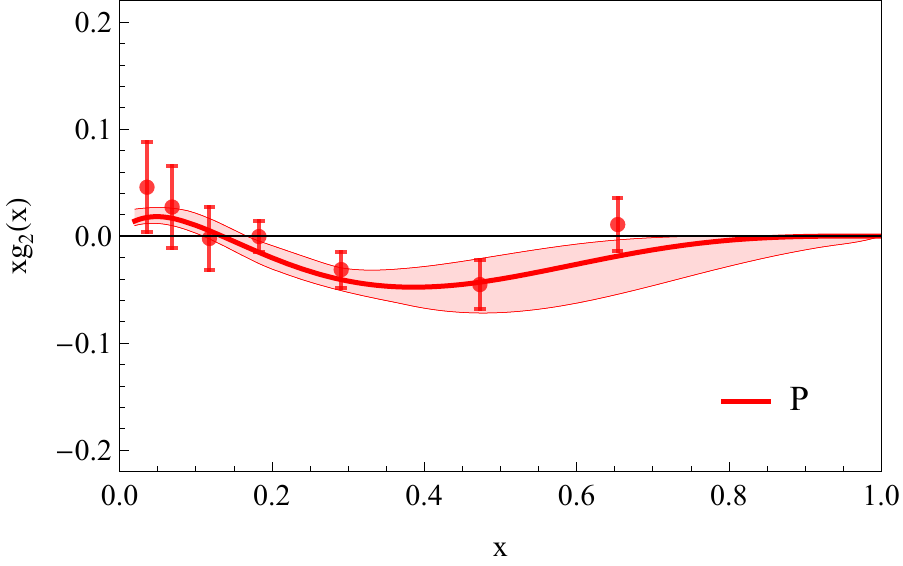}
\caption{\label{Fig:g2}
Left panel: The structure function $xg_2(x)$ in WW-approximation at $Q^2=7.1\,{\rm GeV}^2$,
Eq.~(\ref{Eq:g2-in-WW-approximation}), for proton (P) and neutron (N) targets,
and data from the SLAC E144 and E155 experiments at
$\la Q^2\ra=7.1\,{\rm GeV}^2$ \cite{Anthony:2002hy,Abe:1998wq}.
Right panel: HERMES data for $Q^2>1\,{\rm GeV}^2$ with
$\la Q^2\ra=2.4\,{\rm GeV}^2$ \cite{Airapetian:2011wu}.
The estimate of the theoretical uncertainty is described in the text.}
\end{figure}

We present calculations of $g_2(x)_{\rm WW}$ in Fig.~\ref{Fig:g2}.
This result is obtained using the LO $g_1^a(x)$-parametrization
\cite{Gluck:1998xa}. In order to display the theoretical ``uncertainty
band'' of this WW-approximation of about $40\,\%$ as deduced in
Ref.~\cite{Accardi:2009au} we proceed as follows: we split the
$40\,\%$ uncertainty into two parts: $\varepsilon_1=\pm 20\,\%$ and
$\varepsilon_2(x)=\pm 20\,\%(1-x)^\epsilon$ with a small $\epsilon>0$,
and estimate the impact of this uncertainty as
\be\label{Eq:g2-in-WW-model-violation}
    g_2(x)_{\rm WW} = (1\pm\varepsilon_1)\frac{ d\;}{ d x}\;\Biggl[
    x\int_x^1\frac{ d y}{y} \,\biggl(
    \frac12\sum_ae_a^2\,g_1^a(y(1\pm\varepsilon_2))\biggr)\Biggr]\,.
\ee
The effect of $\varepsilon_1$ is to change the magnitude
of $g_2(x)_{\rm WW}$, $\varepsilon_2$ varies the position of its zero.
The $x$--dependence of $\varepsilon_2$ preserves $\lim_{x\to1}g_2(x)= 0$;
we use $\epsilon=0.05$, which yields $\varepsilon_2\approx 20\,\%$ up to
the highest measured $x$--bin.
The good agreement of $g_2(x)_{\rm WW}$ with data is encouraging,
and in line with theory predictions \cite{Balla:1997hf}.
Our estimate with the split uncertainties
$\varepsilon_{1,2}$ may overestimate in certain $x$--bins the
$40\,\%$--``uncertainty band'' estimated in \cite{Accardi:2009au}.
This however helps us to display a conservative estimate of possible
uncertainties.
We conclude that the WW--approximation works reasonably well,
see Fig.~\ref{Fig:g2}.

Presently $h_L^a(x)$ is unknown.
With phenomenological information on $h_1^a(x)$
\cite{Efremov:2006qm,Anselmino:2007fs,Anselmino:2008jk},
the WW approximation (\ref{Eq:WW-original2}) for $h_L^a(x)$ could
be tested experimentally in Drell--Yan \cite{Koike:2008du}.

\subsection{Tests in lattice QCD}
\label{Sec-3.5:WW-lattice}

The lowest Mellin moments of the PDF $g_T^q(x)$ were studied in
lattice QCD in the quenched approximation \cite{Gockeler:2000ja}
and with $N_f = 2$ flavors of light dynamical quarks \cite{Gockeler:2005vw}.
The results obtained were compatible with a small $\tilde{g}_T^q(x)$.
We are not aware of lattice QCD studies related to the PDF $h_L^a(x)$,
and turn now our attention to TMD studies in lattice QCD.

After first exploratory investigations of TMDs on the lattice
\cite{Hagler:2009mb,Musch:2010ka}, recent years have witnessed considerable
progress and improvements with regard to rigor, realism and methodology.
For the latest developments we refer the
	interested reader to Refs.~\cite{Alexandrou:2017dzj,Ishikawa:2017faj,
	Engelhardt:2015xja,
	Ji:2014hxa,
	Musch:2011er,
   Ji:2017oey,Yoon:2017qzo,Green:2017xeu,Zhang:2017bzy}.
However, numerical results from recent calculations are only available
for a subset of observables, and the quantities calculated are not in a
form that lends itself to straightforward tests of the WW-type relations
as presented in this paper.

For the time being, we content ourselves with rather crude comparisons
based on the lattice data
published in Refs.~\cite{Hagler:2009mb,Musch:2010ka}.
These early works explored all nucleon and quark polarizations, but
they used a gauge link that does not incorporate the final or initial
state interactions present in SIDIS or Drell--Yan experiments. In other
words, the transverse momentum dependent quantities computed in
\cite{Hagler:2009mb,Musch:2010ka} are not precisely the TMDs measurable
in experiment. More caveats will be discussed along the way.

Let us now translate the approximations
(\ref{Eq:WW-approx-g1T},~\ref{Eq:WW-approx-h1L}) into expressions
for which we have a chance to compare them with available lattice data.
For that we multiply the
Eqs.~(\ref{Eq:WW-approx-g1T},~\ref{Eq:WW-approx-h1L}) by $x^N$
with $N=0,\,1,\,2,\,\dots$ and integrate over $x\in[-1,1]$ which yields
\ba
        \int_{-1}^1 d x\;x^N
       g_{1T}^{\perp(1)q}(x)&\WWtype&
        \phantom{-} \,\frac{1}{N+2} \int_{-1}^1 d x \;x^{N+1}g_1^q(x)
        \;,
    \label{Eq:WW-approx-g1T-d}\\
        \int_{-1}^1 d x\;x^N
        h_{1L}^{\perp(1)q}(x)&\WWtype&
        -\,\frac{1}{N+3} \int_{-1}^1 d x\: x^{N+1}\,h_1^q(x)
        \;.
    \label{Eq:WW-approx-h1L-d}
\ea
Here the 
negative $x$ refer to antiquark distributions
$g_1^{\bar q}(x) = +\,g_1^{q}(-x)$,
$h_1^{\bar q}(x) = -\,h_1^{q}(-x)$,
$g_{1T}^{\perp(1)\bar q}(x) =- g_{1T}^{\perp(1)q}(-x)$,
$h_{1L}^{\perp(1)\bar q}(x) = +\,h_{1L}^{\perp(1)q}(-x)$
depending on $C$--parity of the involved operators \cite{Mulders:1995dh}.
The right-hand sides of
Eqs.~(\ref{Eq:WW-approx-g1T-d},~\ref{Eq:WW-approx-h1L-d}) are $x$--moments
of parton distributions, and those can be obtained from lattice QCD using
well-established methods based on operator product expansion.
The left-hand sides are moments of TMDs in $x$ and $\bfkperp$. We have to
keep in mind that TMDs diverge for large $\bfkperp$. Therefore, without
regularizing these divergences in a scheme suitable for the comparison of
left and right hand side, a test of the above relations is meaningless,
even before we get to address the issues of lattice calculations. Let us
not give up at this point and take a look at the lattice observables of
Ref.~\cite{Musch:2010ka} where TMDs were obtained from amplitudes
$\tilde A_i(l^2,\ldots)$ in Fourier space, where $\bfkperp$ is encoded
in the Fourier conjugate variable $\bflperp$, which is the transverse
displacement of quark operators in the correlator evaluated on the lattice.
In Fourier space, the aforementioned divergent behavior for large $\bfkperp$
translates into strong lattice scale and scheme dependencies at short distances
$\bflperp$ between the quark operators. The $\bfkperp$ integrals needed for
the left-hand sides of Eqs.~(\ref{Eq:WW-approx-g1T-d},~\ref{Eq:WW-approx-h1L-d})
correspond to the amplitudes at $\bflperp = 0$, where scheme and
scale dependence is greatest.  In Ref.~\cite{Musch:2010ka} Gaussian fits
have been performed to the amplitudes \emph{excluding} data at short quark
separations $\bflperp$. The Gaussians describe the long-range data quite well
and bridge the gap at short distances $\bflperp$.
Taking the Gaussian fit at $\bflperp = 0$, we get a value that is
(presumably) largely lattice-scheme and scale independent. We have thus
swept the problem of divergences under the rug. The Gaussian fit acts as
a crude regularization of the divergences that appear in TMDs at large
$\bfkperp$ and manifest themselves as short range artifacts on the lattice.
Casting this line of thought into mathematics, we get
\begin{align}
    	\int_{-1}^1 \! d x\; g_{1T}^{\perp(1)q}(x)
	&=  \int_{-1}^1 \! d x \! \int \! d^2 \bfkperp
	\frac{\kperp^2}{2M^2} g_{1T}^{\perp q}(x,\kperp)
	= -2 \tilde{A}_{7,q}( \ell = 0 )
	\Gaussian
	 -c_{7,q}  \\
    	\int_{-1}^1 \! d x\; h_{1L}^{\perp(1)q}(x)
	&=  \int_{-1}^1 \! d x \! \int \!  d^2 \bfkperp
	\frac{\kperp^2}{2M^2} h_{1T}^{\perp q}(x,\kperp)
	= -2 \tilde{A}_{10,q}( \ell = 0 )
	\Gaussian
	-c_{10,q} 
\end{align}
where the amplitudes $\tilde{A}$ and constants $c$ are those of Ref.~\cite{Musch:2010ka}.
We have thus expressed the left-hand side of
Eqs.~(\ref{Eq:WW-approx-g1T-d},~\ref{Eq:WW-approx-h1L-d}) in terms of
amplitudes $c_{7,q}$ and $c_{10,q}$ of the Gaussian fits on the lattice.
Before quoting numbers, a few more comments are in order. The overall
multiplicative renormalization in Ref.~\cite{Musch:2010ka} was fixed by
setting the Gaussian integral $c_{2,u-d}$ of the unpolarized TMD $f_1$
in the isovector channel (u-d) to the nucleon quark content, namely to 1.
One then assumes that the normalization of the lattice results for the
unpolarized TMD $f_1$ also fixes the normalization for polarized quantities
correctly. 
This assumption holds if renormalization is multiplicative and
   flavor-independent for the non-local lattice operators. This is
   not true for all lattice actions \cite{Yoon:2017qzo}.    But presumably it is true
   if the lattice action preserves chiral symmetry, as it does in
   the present case.
The Gaussian fits along with the normalization prescription serve as
a crude form of renormalization, and this is needed to attempt
a comparison of left and right hand sides of equations
Eqs.~(\ref{Eq:WW-approx-g1T-d},~\ref{Eq:WW-approx-h1L-d}).

There is another issue to discuss.
The gauge link that goes into the evaluation of the quark-quark correlator
introduces a power divergence that has to be subtracted.
Ref.~\cite{Musch:2010ka} employs a subtraction scheme on the lattice
but establishes no connection with a subtraction scheme designed for
experimental TMDs and the corresponding gauge-link geometry.
The gauge-link renormalization mainly
influences the width of the Gaussian fits; the amplitudes are only slightly
affected, so it may not play a big role for our discussion. Altogether, the
significance of our numerical ``tests'' of WW relations should be taken
with a grain of salt.

For the test of (\ref{Eq:WW-approx-g1T-d}), we use the numbers
$\int d x\;g_{1T}^{\perp(1)u}(x)\Gaussian
-c_{7,u}= 0.1041(85)$ and
$\int d x\;g_{1T}^{\perp(1)d}(x)\Gaussian
-c_{7,d}=-0.0232(42)$
from \cite{Musch:2010ka}. 
Lattice data for
$\int d x \,x^{N}g_1^q(x)$
\cite{Hagler:2003is,Hagler:2007xi} and
$\int d x \,x^{N}h_1^q(x)$
\cite{Gockeler:2005cj} are available for $N=0,\,1,\,2,\,3$ .
These values have been computed using (quasi-)local operators that
have been renormalized to the $\overline{MS}$ scheme at the scale
$\mu^2 = 4\,\text{GeV}^2$.
According to \cite{Hagler:2007xi} (data set 4:
with $a\,m_{u,d} = 0.020$ with $m_\pi\approx 500\,{\rm MeV}$)
one has $\int d x \;x\,g_1^{u-d}(x)= 0.257(10)$ and
$\int d x \;x\,g_1^{u+d}(x)= 0.159(14)$.
Decomposing the results from  \cite{Hagler:2007xi} into
individual flavors and inserting them into (\ref{Eq:WW-approx-g1T-d}), we obtain
\ba
        \underbrace{\int \! d x\;g_{1T}^{\perp(1)u}(x)}
        _{= 0.1041(85) \;\mbox{\footnotesize Ref.~\cite{Musch:2010ka}}}
        &\stackrel{!}{\approx}
        \underbrace{\frac{1}{2}\int \! d x\;x\,g_1^u(x)}
        _{= 0.104(9) \;\mbox{\footnotesize Ref.~\cite{Hagler:2007xi}}}
        \hspace{2mm} , \nonumber \\
        \underbrace{\int \! d x\;g_{1T}^{\perp(1)d}(x)}
        _{= -0.0232(42) \;\mbox{\footnotesize Ref.~\cite{Musch:2010ka}}}
        &\stackrel{!}{\approx}
        \underbrace{\frac{1}{2}\int \!d x\;x\,g_1^d(x)}
        _{= -0.025(9) \;\mbox{\footnotesize Ref.~\cite{Hagler:2007xi}}}
        \hspace{2mm},
        \label{Eq:test-WW-type-lattice-g1T}
\ea
which confirms the approximation (\ref{Eq:WW-approx-g1T-d}) for $N=0$
within the statistical uncertainties of the lattice calculations.
In order to test (\ref{Eq:WW-approx-h1L-d}) we use
$\int dx\;h_{1L}^{\perp(1)u}(x)\Gaussian -c_{10,u}=-0.0881(72)$
and
$\int dx\;h_{1L}^{\perp(1)d}(x)\Gaussian -c_{10,d}=0.0137(34)$
from \cite{Musch:2010ka} and the lattice data
$\int d x \;x\,h_1^u(x)= 0.28(1)$ and
$\int d x \;x\,h_1^d(x)= -0.054(4)$
from QCDSF \cite{Gockeler:2005cj}.\footnote{
  These numbers are read off from a figure in \cite{Gockeler:2005cj},
  and were computed on a different lattice. We interpolate them to a
  common value of the pion mass $m_\pi\approx500\,{\rm MeV}$, and
  estimate the uncertainty conservatively in order to take systematic effects
  into account due to the use of a different lattice.}
Inserting these numbers into  (\ref{Eq:WW-approx-h1L-d}) for the case
$N=0$ we obtain
\ba
        \underbrace{\int \! d x\;h_{1L}^{\perp(1)u}(x)}
        _{= -0.0881(72)\;\mbox{\footnotesize Ref.~\cite{Musch:2010ka}}}
        &\stackrel{!}{\approx}&
        \underbrace{-\,\frac{1}{3}\int \! d x\;x\,h_1^u(x)}
        _{= -0.093(3) \;\mbox{\footnotesize Ref.~\cite{Gockeler:2005cj}}}
        \hspace{2mm} , \hspace{7mm} \nonumber \\
        \underbrace{\int \! d x\;h_{1L}^{\perp(1)d}(x)}
        _{= 0.0137(34) \;\mbox{\footnotesize Ref.~\cite{Musch:2010ka}}}
        &\stackrel{!}{\approx}&
        \underbrace{-\,\frac{1}{3}\int \! d x\;x\,h_1^d(x)}
        _{= 0.018(1) \;\mbox{\footnotesize Ref.~\cite{Gockeler:2005cj}}}
        \hspace{2mm},
        \label{Eq:test-WW-type-lattice-h1L}
\ea
which again confirms the WW-type approximation within the statistical
uncertainties of the lattice calculations.

Several more comments are in order concerning the, at first glance, remarkably
good confirmation of the  WW-type approximations by lattice data in
Eqs.~(\ref{Eq:test-WW-type-lattice-g1T},~\ref{Eq:test-WW-type-lattice-h1L}).

First, the relations refer to lattice parameters corresponding
to pion masses of $500\,{\rm MeV}$. We do not
need to worry about that too much. The lattice results do provide
a valid test of the approximations in a ``hadronic world'' with
somewhat heavier pions and nucleons. All that matters in our
context is that the relative size of $\bar{q}gq$--matrix elements
is small with respect to $\bar{q}q$--matrix elements.

Second, we have to revisit carefully which approximations the above
lattice calculations actually test. As mentioned above, in
the lattice study \cite{Hagler:2009mb,Musch:2010ka}, a specific choice for
the path of the gauge link was chosen, which is actually different
from the paths required in SIDIS or Drell--Yan. With the path choice of
\cite{Hagler:2009mb,Musch:2010ka} there are effectively only (T-even)
$A_i$ amplitudes, the $B_i$ amplitudes are absent.
Therefore the test (\ref{Eq:test-WW-type-lattice-g1T}) of the WW-type
approximation (\ref{Eq:WW-approx-g1T-d}) actually constitutes a test
of the WW-approximation (\ref{Eq:WW-original1}) and confirms
earlier lattice work \cite{Gockeler:2000ja,Gockeler:2005vw},
(cf.~Refs.~\cite{Metz:2008ib,Teckentrup:2009tk} and
Sec.~\ref{Sec-3.6:models}).
Similarly, the test (\ref{Eq:test-WW-type-lattice-h1L}) of the
WW-type approximation (\ref{Eq:WW-approx-h1L-d}) actually constitutes
a test of the WW-approximation (\ref{Eq:WW-original2}). The latter,
however, has not been reported previously in literature, and constitutes a
new result.

Third, to be precise,
(\ref{Eq:test-WW-type-lattice-g1T},~\ref{Eq:test-WW-type-lattice-h1L})
test the first Mellin moments of the WW approximations
(\ref{Eq:WW-original1},~\ref{Eq:WW-original2}), which corresponds to the
Burkhardt-Cottingham sum rule for $g_T^a(x)$ and an analogous sum rule for
$h_L^a(x)$ (see \cite{Jaffe:1996zw} and references therein).
In view of the long debate on the validity of those sum rules
\cite{Burkardt:2001iy,Bass:2003vp,Efremov:2002qh}, this is
an interesting result in itself.

It is important to stress that in view of the pioneering and
exploratory status of the TMD lattice calculations
\cite{Hagler:2009mb,Musch:2010ka}, this is already a remarkable and very
interesting result. Thus, apart from the instanton calculation
\cite{Dressler:1999hc}, also lattice data provide support for
the validity of the WW approximation (\ref{Eq:WW-original2}).
At the same time, however, we also have to admit that we do
not really reach our goal of testing the WW-type approximations
on the lattice. We have to wait for better lattice data.
Meanwhile we may try to gain insights into the quality of
WW-type approximations from models.

\subsection{Tests in models}
\label{Sec-3.6:models}

Effective approaches and models such as bag
\cite{Jaffe:1991ra,Stratmann:1993aw,Signal:1996ct,Avakian:2010br},
spectator \cite{Jakob:1997wg}, chiral quark-soliton
\cite{Wakamatsu:2000ex}, or light-cone
constituent \cite{Pasquini:2008ax,Lorce:2011dv} models
support the approximations (\ref{Eq:WW-original1},~\ref{Eq:WW-original2})
for PDFs within an accuracy of $(10-30)\,\%$ at low hadronic scale
below $1\,{\rm GeV}$.

Turning to TMDs, we recall that in models without gluon
degrees of freedom certain relations among TMDs hold, the
so-called quark-model Lorentz-invariance relations (qLIRs)
\cite{Tangerman:1994bb,Mulders:1995dh}.\footnote{Notice that
	the qLIRs of \cite{Tangerman:1994bb,Mulders:1995dh} are
	valid only in quark models with no gluons and should not
	be confused with the LIRs of \cite{Kanazawa:2015ajw}, which
	are exact relations in QCD, see Sec.~\ref{Sec-3.1:WW-classic}.
	In the literature, both are often simply referred to as LIRs.
	This ambiguity is unfortunate.}
Initially thought to be exact \cite{Tangerman:1994bb,Mulders:1995dh},
qLIRs were shown  to be invalid in models with gluons
\cite{Kundu:2001pk,Schlegel:2004rg} and in QCD \cite{Goeke:2003az}.
They originate from decomposing the (completely unintegrated)
quark correlator in terms of Lorentz-invariant amplitudes, and
TMDs are certain integrals over those amplitudes.
When gluons are absent, the correlator consists
of twelve amplitudes \cite{Tangerman:1994bb,Mulders:1995dh}, i.e., fewer
amplitudes than TMDs, which implies relations: the qLIRs.
In QCD, the correct Lorentz decomposition requires the consideration of
gauge links, which introduces further amplitudes. As a result one has
as many amplitudes as TMDs and no relations exist \cite{Goeke:2003az}.
However, qLIRs ``hold'' in QCD in the WW-type approximation
\cite{Metz:2008ib}. In models without gluon degrees of freedom
they are exact
\cite{Metz:2008ib,Teckentrup:2009tk,Avakian:2010br,Jakob:1997wg}.

The bag, spectator, and light-cone constituent-quark models support
the approximations (\ref{Eq:WW-approx-g1T},~\ref{Eq:WW-approx-h1L})
within an accuracy of $(10-30)\,\%$
\cite{Jakob:1997wg,Pasquini:2008ax,Avakian:2010br,Lorce:2011dv}.
The spectator and bag model support WW-type approximations
within $(10-30)\,\%$ \cite{Avakian:2010br}.
As they are defined in terms of quark bilinear expressions
(\ref{Eq:correlator}), it is possible to evaluate twist-3 functions
in quark models \cite{Jaffe:1991ra}. The tilde-terms arise due to
the different model interactions, and it is important to discuss
critically how realistically they describe the $\bar{q}gq$--terms
of QCD \cite{Lorce:2014hxa,Lorce:2016ugb}.

In the covariant parton model with intrinsic 3D-symmetric parton
orbital motion \cite{Zavada:1996kp}, quarks are free, $\bar{q}gq$
correlations absent, and all WW and WW-type relations exact
\cite{Efremov:2010mt,Efremov:2009ze}.
The phenomenological success of this approach \cite{Zavada:1996kp} may
hint at a general smallness of $\bar{q}gq$ terms, although some of the
predictions from this model have yet to be tested \cite{Efremov:2010mt}.

Noteworthy is the result from the chiral quark-soliton
model where the WW-type approximation (\ref{Eq:WW-type-Cahn})
happens to be exact: $xf^{\perp q}(x,\kperp^2)=f_{1}^q(x,\kperp^2)$
for quarks and antiquarks \cite{Lorce:2014hxa}. The degrees of freedom
in this model are quarks, antiquarks, and Goldstone bosons, which are
strongly coupled (the coupling constant is $\sim 4$) and has to be
solved using nonperturbative techniques (expansion in $1/N_c$, where
$N_c$ is the number of colors) with the nucleon described as a
chiral soliton. In general, the model predicts non-zero tilde-terms, for
instance $\tilde{e}^a(x)\neq 0$
\cite{Schweitzer:2003uy,Ohnishi:2003mf,Cebulla:2007ej}.
However, despite strong interactions in this effective theory, the tilde
term $\tilde{f}^{\perp q}(x,\kperp^2)$ vanishes exactly in this model
\cite{Lorce:2014hxa} and the WW-type approximation (\ref{Eq:WW-type-Cahn})
becomes exact at the low initial scale of this model of
$\mu_0\sim 0.6\,{\rm GeV}$.

Let us finally discuss quark-target models,
where gluon degrees of freedom are included and WW(-type)
approximations badly violated
\cite{Kundu:2001pk,Schlegel:2004rg,Meissner:2007rx,Mukherjee:2009uy}.
This is natural in this class of models for two
reasons. First, quark-mass terms are of ${\cal O}(m_q/M_N)$
and negligible in the nucleon case, but of ${\cal O}(100\,\%)$
in a quark target where $m_q$ plays also the role of $M_N$.
Second, even if one refrains from mass terms the approximations are
spoiled by gluon radiation, see for instance \cite{Harindranath:1997qn}
in the context of (\ref{Eq:WW-original1}).
This means that perturbative QCD does not support the WW-approximations:
they certainly are not preserved by evolution. However, scaling violations
{\it per se} do not need to be large. What is crucial in this context are
dynamical reasons for the smallness of the {\sl matrix elements} of
$\bar{q}gq$--operators. This requires the consideration of chiral symmetry
breaking effects reflected in the hadronic spectrum, as considered in the
instanton vacuum model \cite{Balla:1997hf,Dressler:1999hc} but
out of scope in quark-target models.

We are not aware of systematic tests of WW-type approximations for FFs. One
information worth mentioning in this context is that in spectator models
\cite{Jakob:1997wg} tilde-contributions to FFs are proportional to the
offshellness of partons 
\cite{Lorce:2014hxa,Lorce:2016ugb}. This
natural feature may indicate that in the region dominated by effects of
small $P_\perp$ tilde-terms might be small. On the other hand, quarks have
sizable constituent masses of the order of few hundred MeV in spectator models
and the mass-terms are not small.
The applicability of WW-type approximations to FFs
remains the least tested point in our approach.

\subsection{Basis functions for the WW-type approximations}
\label{Sec-3.7:basis}

The 6 leading--twist TMDs
$f_1^a, \; f_{1T}^{\perp a}, \; g_1^a, \; h_1^a, \;h_1^{\perp a},\; h_{1T}^{\perp a}$
and 2 leading--twist FFs $D_1^a, \; H_1^{\perp a}$ provide a basis
in the sense that in WW-type approximation all other TMDs and FFs
can either be expressed in terms of these basis functions or vanish.
Below we shall see that, under the assumption of the validity of WW-type
approximations, it is possible to express all SIDIS structure functions
in terms of the basis functions.\footnote{%
   Notice that SIDIS alone is not sufficient to uniquely determine
   the eight basis functions that appear in six SIDIS
   leading-twist structure functions. It is thus crucial to take
   advantage of other processes (like Drell-Yan and hadron production in
   $e^+e^-$ annihilation, which are indispensable for the determination
   of $f_1^a$, $D_1^a$, $H_1^{\perp a}$).}
These basis functions allow us to describe, in WW-type approximation,
all other TMDs. The experiment will tell us how well the approximations work.
In some cases, however, we know in advance that the WW-type approximations
have limitations, see next Section~\ref{Sec-3.8:limitations}.

\subsection{Limitation of WW-type approximations}
\label{Sec-3.8:limitations}

The approximation may work in the case when a TMD or FF
$=\la\bar{q}q\ra + \la\bar{q}gq\ra \approx \la\bar{q}q\ra \neq 0$ with
a ``controlled approximation'' in the spirit of Eq.~(\ref{Eq:WW-generic}).
We know cases where this works, see
Secs.~\ref{Sec-3.3:WW-classic-instanton}, \ref{Sec-3.4:WW-classic-experiment},
but it has to be checked case by case whether
$| \la\bar{q}gq\ra  | \ll  |\la\bar{q}q\ra|$ for a given operator.
At least in such cases the approximation has a chance to work.

However, it may happen that after applying the QCD equations
of motion one ends up in the situation that a given function
$=\la\bar{q}q\ra + \la\bar{q}gq\ra$ with $\la\bar{q}q\ra = 0$.
This happens for the T-even TMD
	$e^a$ in Eqs.~(\ref{Eq:WW-original-e},~\ref{Eq:WW-type-1}),
for the T-odd TMDs
	$e_L^q$,
	$e_T^q$,
	$e_T^{\perp q}$,
	$f_L^{\perp q}$,
	$g^{\perp q}$ in Eqs.~(\ref{Eq:WW-type-eLperp}--\ref{Eq:WW-type-fLperp}),
and for the FFs
	$E^q$,
	$G^{\perp q}$ in Eqs.~(\ref{Eq:WW-type-FF-1},~\ref{Eq:WW-type-FF-2})
(actually, all twist-3 FFs are affected, we will discuss this in detail below).
In this situation the ``leading term'' is absent, so neglecting the
``subleading (pure twist-3) term'' actually constitutes an error of $100\,\%$
even if the neglected matrix element  $\la\bar{q}gq\ra$ is very small.
Notice that this occurs for all subleading-twist FFs that enter
SIDIS structure functions only in the shape of tilde-FFs, see
Sec.~\ref{Sec-2:SIDIS+TMDs+FF} and
Eqs.~(\ref{Eqs:structure-functions-twist-3}).
We shall see that some structure functions are potentially more and
others potentially less affected by this generic limitation. In any
case, phenomenological work has to be carried out to find out whether
or not the approximation works.

For both FFs and TMDs there are also limitations
which go beyond this generic issue. To illustrate this for FFs
we recall that both $H_1^{\perp(1)q}$ and $\tilde H_1^{q}$ are
related to integrals of an underlying function $H_{FU}^{q,\Im}(z,z_1)$
as pointed out in Ref.~\cite{Kanazawa:2015ajw}. Therefore, if one
literally assumed $\tilde H^q(z)$ to be zero, this would imply that
also $H_1^{\perp(1)q}$ would vanish, indicating that the WW-type
approximation has to be used with care for chiral-odd FFs.

Similar limitations exist also for TMDs. This is manifest in particular
for those twist-3 T-odd TMDs that appear in the decomposition of the
correlator (\ref{Eq:correlator}) with no prefactor of $\kperp$.
There are three cases: $f_T^a(x,k_\perp^{2})$, $h^a(x,\kperp^{2})$, and $e_L^a(x,\kperp^{2})$.
Such TMDs in principle survive integration of the correlator over $\kperp$
and would have PDF counterparts if there were not the sum rules in
Eq.~(\ref{Eq:sum-rules-T-odd}). These sum rules arise because hypothetical
PDF versions of T-odd TMDs vanish: they have a simple straight gauge link
along the lightcone, and such objects vanish due to parity and time-reversal
symmetry of strong interactions. This argument does not apply to other T-odd
TMDs because they drop out from the $\kperp$--integrated correlator due to
explicit factors of, e.g., $\kperp^j$ in the case of the Sivers function.

Let us first discuss the case of $f_T^a(x,k_\perp^{2})$. Taking the
WW-type approximation (\ref{Eq:WW-type-fT}) literally means
$x\int d^2 k_\perp\,f_T^a(x,k_\perp^{2})\,\stackrel{!?}{=}
-f_{1T}^{\perp(1)a}(x)\neq0$,
at variance with the sum rule (\ref{Eq:sum-rules-T-odd}). We
have $xf_T^a(x,k_\perp)=x\tilde{f}_T^a(x,k_\perp^{2})-f_{1T}^{\perp(1)a}(x,k_\perp^{2})$
from QCD equations of motion \cite{Bacchetta:2006tn}, which yields
(\ref{Eq:WW-type-fT}). The point is that in this case it is
essential to keep the tilde-function.
The situation for the chirally and T-odd twist-3
TMD $h^a(x,k_\perp^{2})$ is analogous. The third
function in (\ref{Eq:sum-rules-T-odd}) causes no issues since
$e_L^a(x,k_\perp^{2})=\tilde{e}_L^a(x,k_\perp^{2})\approx0$ in WW-type approximation.

Does it mean WW-type approximations fail for $f_T^a(x,k_\perp^{2})$
and $h^a(x,k_\perp^{2})$? Not necessarily! The approximations may
work in some but not all regions of $\kperp$, but the sum rules
(\ref{Eq:sum-rules-T-odd}) include integration over all $k_\perp$.
Notice also that, e.g., $f_{1T}^{\perp (1),q}(x)$ is related to the
soft-gluon-pole matrix element $T_{F}(x,x)$ \cite{Boer:2003cm,Ji:2006ub},
which is a $\bar{q}gq$-term that one would naturally neglect
in WW-type approximation.
In this sense (\ref{Eq:WW-type-fT}) could be consistent.
Thus, issues with the sum rules
(\ref{Eq:sum-rules-T-odd}) do not need to exclude the
possibility that the WW-type approximations for $f_T^a(x,k_\perp^{2})$ and
$h^a(x,k_\perp^{2})$ in (\ref{Eq:WW-type-fT},~\ref{Eq:WW-type-last})
may work at small $k_\perp$ where we use them in our TMD approach.
This would mean that the UV region is essential to realize the sum rules
(\ref{Eq:sum-rules-T-odd}). Alternatively, one could also envision
the sum rules (\ref{Eq:sum-rules-T-odd}) to be sensitive to the
IR region through gluonic or fermionic pole contributions manifest
in tilde-terms.
Presently too little is known in the theory of subleading-twist TMDs.
In Secs.~\ref{Sec-7.6:FUTsinphiS} and \ref{Sec-7.7:FUUcosphi} we
will present pragmatic solutions for how to deal with the TMDs
$f_T^a(x,k_\perp^{2})$ and $h^a(x,k_\perp^2)$ phenomenologically.
For now let us keep in mind that
one has to keep a vigilant eye on all WW-type approximations, and
especially on those for $f_T^a(x,k_\perp^{2})$ and $h^a(x,k_\perp^{2})$.

As it was mentioned in the Introduction
one important limitation concerns the fact that the WW-type approximations
are not preserved under $Q^2$ evolution. Still some intuition can be 
obtained from the collinear case: the evolution equations for $g_T^a(x)$ 
and $h_L^a(x)$ exhibit complicated mixing patterns typical for higher 
twist functions, which simplify to DGLAP-type evolutions in the limit 
of a large number of colors $N_c$ and in the limit of large-$x$ 
\cite{Ali:1991em,Koike:1994st,Balitsky:1996uh,Belitsky:1997zw}. 
These evolution equations differ from those of the leading-twist
functions $g_1^a(x)$ and $h_1^a(x)$. 
However, since $Q^2$ varies moderately in the considered experiments 
(e.g.\ for common values of $x$ the $Q^2$ at COMPASS is only about a factor 2-3
larger than at HERMES), this point is not a major uncertainty in our study.
More theoretical work will be required to understand $k_\perp$-evolution 
effects of subleading twist TMDs in future experiments (EIC) 
covering kinematic regions that vary by orders of magnitude in $Q^2$.

%
\section{SIDIS in the WW-type approximation and Gaussian model}
\label{Sec-4:SIDIS-in-WW-approximation}

In this section, we consequently apply the WW and WW-type approximation
to SIDIS, and describe our procedure to evaluate the structure
functions in this approximation and the Gaussian Ansatz which we use
to model the $k_\perp$ dependence of TMDs.

\subsection{Leading structure functions amenable to WW-type approximations}
\label{Sec-4.1:WW-twist-2}

The WW and WW-type approximations are useful for the following
two leading-twist structure functions:
\begin{subequations}\ba
 F_{LT}^{\cos(\phi_h -\phi_S)}
	&\stackrel{\rm WW}{=}&
	{\cal C}\biggl[\omega^{\{1\}}_{\rm B}\, {g_{1T}^\perp}D_1 \biggr]
        \with{ }{g_{1T}^{\perp a}\to g_1^a}{Eq.~(\ref{Eq:WW-approx-g1T})} \; ,
        \label{F_LTcos(phi-phiS)-WW} \\
 F_{UL}^{\sin 2\phi_h} 	
        &\stackrel{\rm WW}{=}&
	{\cal C}\biggl[\omega^{\{2\}}_{\rm AB}\,
    	{h_{1L}^{\perp }} H_{1}^{\perp }\biggr]
        \with{ }{h_{1L}^{\perp a}\to h_1^a}{Eq.~(\ref{Eq:WW-approx-h1L})}  \; .
        \label{F_UUsin2phi-WW}
\ea\end{subequations}

\subsection{Subleading structure functions in WW-type approximations}
\label{Sec-4.2:WW-twist-3}

In the case of the subleading-twist structure functions the WW-type
approximations in (\ref{Eq:WW-type-1}--\ref{Eq:WW-type-last})
lead to considerable simplifications. We obtain the approximations
\begin{subequations}\ba
F_{UU}^{\cos\phi_h} &\stackrel{\rm WW}{=}&\frac{2M_N}{Q}\,{\cal C}\biggl[
	 \omega^{\{1\}}_{\rm A}\,x\,h\,H_{1}^{\perp }
   	-\omega^{\{1\}}_{\rm B}\,x\,f^\perp D_1\biggr]
        \with{ }
	{f^{\perp a}\to f_1^a, \; h^a\to h_1^{\perp a}}
	{with Eqs.~(\ref{Eq:WW-type-Cahn},~\ref{Eq:WW-type-last})} 
	\label{Eq:WW-type-FUUcosphi}\\
F_{UL}^{\sin\phi_h} &\stackrel{\rm WW}{=}& \frac{2M_N}{Q}\,{\cal C}\biggl[
   	\omega^{\{1\}}_{\rm A}\,
    	x\,h_L  H_1^{\perp } \biggr]
        \with{ }
	{h_L^a\to h_{1L}^{\perp a}}
	{(\ref{Eq:WW-type-6})} 
	\label{Eq:WW-type-FULsinphi}\\
F_{UT}^{\sin \phi_S } &\stackrel{\rm WW}{=}& \frac{2M_N}{Q}\,
	{\cal C}\biggl[ \omega^{\{0\}} \, x\,f_TD_1
	-\frac{\omega^{\{2\}}_{\rm B}}{2}\,(xh_T-xh_T^\perp)\,H_{1}^{\perp } \biggr]
        \with
	{f_T^a \to f_{1T}^{\perp a},}
	{h_T^a-h_T^{\perp a}\to h_1^a}
	{(\ref{Eq:WW-type-fT},~\ref{Eq:WW-type-7},~\ref{Eq:WW-type-8})} 
	\label{Eq:WW-type-FUTsinphiS}\\
F_{UT}^{\sin(2\phi_h -\phi_S)} &\stackrel{\rm WW}{=}& \frac{2M_N}{Q}\,{\cal C}\biggl[
   	\omega^{\{2\}}_{\rm C}\,
   	{  x\,f_T^\perp}D_1
        + \frac{\omega^{\{2\}}_{\rm AB}}{2}
	{x(h_T+h_T^\perp)}H_1^\perp \biggr]
        \with
	{f_T^{\perp a}\to f_{1T}^{\perp a},}
	{(h_T^a+h_T^{\perp a})\to h_{1T}^{\perp a}}{
	(\ref{Eq:WW-type-fTperp},~\ref{Eq:WW-type-7},~\ref{Eq:WW-type-8})} 
	\\
F_{LU}^{\sin\phi_h} &\stackrel{\rm WW}{=}& 0 \phantom{\frac11}
	\label{Eq:WW-type-FLUsinphi}\\
F_{LT}^{\cos \phi_S}&\stackrel{\rm WW}{=}& \frac{2M_N}{Q}\,
	{\cal C}\biggl[-  \omega^{\{0\}}\, x\,g_T D_1 \biggr]
        \with{ }
	{g_T^a\to g_1^a}
	{(\ref{Eq:WW-original1})}  \\
F_{LL}^{\cos \phi_h} &\stackrel{\rm WW}{=}& \frac{2M_N}{Q}\,{\cal C}\biggl[
   	-\omega^{\{1\}}_{\rm B}\,
   	xg_L^{\perp} D_1 \biggr]
        \with{ }
	{g_L^{\perp a}\to g_1^a}
	{(\ref{Eq:WW-type-gLperp})}  \\
F_{LT}^{\cos(2\phi_h - \phi_S)} &\stackrel{\rm WW}{=}& \frac{2M_N}{Q}\,{\cal C}\biggl[
   	- \omega^{\{2\}}_{\rm C}\,
   	x g_T^{\perp } D_1 \biggr]
        \with{ }
	{g_T^{\perp a}\to g_1^a}
	{(\ref{Eq:WW-type-gTperp},~\ref{Eq:WW-approx-g1T})} 
\ea\end{subequations}

\subsection{Gaussian Ansatz for TMDs and FFs}
\label{Sec-4.3:evaluation}

In this work we will use the so-called Gaussian Ansatz for the TMDs and FFs.
This Ansatz, which for a generic TMD or FF is given by
\be\label{Eq:Gauss-generic}
    f(x,\kperp^2) = f(x)\;
    \frac{e^{-\kperp^2/\avkperp}}{\pi\avkperp} \;,\;\;\;
    D(z,\pperp^2) = D(z)\,
    \frac{e^{-\pperp^2/\avpperp}}{\pi\avpperp}\;,
\ee
is popular not only because it considerably simplifies the
calculations. In fact, all convolution integrals of the type
(\ref{Eq:def-convolution-integral}) can be solved analytically
with this Ansatz. Far more important is the fact that it works
phenomenologically with a good accuracy in many practical applications
\cite{Anselmino:2005nn,Collins:2005ie,D'Alesio:2007jt,Schweitzer:2010tt,
Signori:2013mda,Anselmino:2013lza}.
Of course this Ansatz is only a rough approximation. For instance,
it is not consistent with general matching expectations for large
$\kperp$ \cite{Bacchetta:2008xw}.

Nevertheless, if one limits oneself to work in a regime where the
transverse momenta (of hadrons produced in SIDIS, dileptons produced
in the Drell--Yan process, etc.) are small compared to the hard
scale in the process, then the Ansatz works quantitatively
very well. The most recent and detailed tests were reported in
\cite{Schweitzer:2010tt}, where the Gaussian Ansatz was shown to
describe the most recent SIDIS data: no deviations were observed
within the error bars of the data provided one takes into account
the broadening of the Gaussian widths with increasing energy
\cite{Schweitzer:2010tt} according
with expectations from QCD \cite{Aybat:2011zv}.
The Gaussian Ansatz is approximately compatible with
the $\kperp$--shapes obtained from evolution \cite{Aybat:2011zv}
or fits to high-energy Tevatron data on weak-boson production
\cite{Landry:2002ix}. Effective models at
low \cite{Pasquini:2008ax,Avakian:2010br,Lorce:2011dv} and
high \cite{Efremov:2009ze} renormalization scales support this
Ansatz as a good approximation.

\subsection{Evaluation of structure functions in WW-type \&
 Gaussian approximation}
\label{Sec-4.4:evaluation}

The Gaussian Ansatz is compatible with many WW-type approximations,
but not all. The trivial approximations (\ref{Eq:WW-type-1}) and
(\ref{Eq:WW-type-eLperp}--\ref{Eq:WW-type-fLperp}) cause no issue.
The Gaussian Ansatz can also be applied to the nontrivial approximations
in Eqs.~(\ref{Eq:WW-type-Cahn}--\ref{Eq:WW-type-gTperp})
and (\ref{Eq:WW-type-fTperp}), provided the corresponding Gaussian
widths are defined to be equal to each other: for example, in the
WW-type approximation (\ref{Eq:WW-type-Cahn}),
$xf^{\perp q}(x,\kperp^2)\approx f_1^q(x,\kperp^2)$, one may
assume Gaussian $\kperp$--dependence for $f^{\perp q}(x,\kperp^2)$
and for $f_1^q(x,\kperp^2)$ as long as the Gaussian widths
of these two TMDs are assumed to be equal.

In the case of the approximations
(\ref{Eq:WW-type-gT}--\ref{Eq:WW-type-8})
the situation is different because here twist-3 TMDs
are related to transverse moments of twist-2 TMDs. In such cases the
Gaussian Ansatz is not compatible with the WW-type approximations:
for instance, the approximation (\ref{Eq:WW-type-gT}) relates
$xg_T^q(x,\kperp^2)\approx\frac{k_\perp^2}{2M_N^2}\,g_{1T}^{q}(x,\kperp^2)$,
e.g., if $g_{1T}^q(x,\kperp^2)$ was exactly Gaussian then
$g_T^q(x,\kperp^2)$ certainly could not be Gaussian. If one wanted to take
the Gaussian Ansatz and WW-type approximations literally, one clearly
would deal with an incompatibility. However, we of course must keep
in mind that both are approximations.

Some comments are in order to understand how the usage of the Gaussian
Ansatz and the WW-type approximations can be reconciled.
First, let us remark that the individual TMDs, say
$g_T^q(x,\kperp^2)$ and $g_{1T}^{q}(x,\kperp^2)$ in our example,
may each by itself be assumed to be approximately Gaussian in $k_\perp$,
which is supported by quark model calculations \cite{Avakian:2010br}.
Second, we actually do not need the unintegrated WW-type approximations.
For phenomenological applications we can use the WW-type approximations
in ``integrated form.''

Let us stress that if one took an unintegrated WW-type approximation of the
type $xg_T^q(x,\kperp^2)\approx\frac{k_\perp^2}{2M_N^2}\,g_{1T}^{q}(x,\kperp^2)$
literally and assumed both TMDs to be exactly Gaussian, one would find
``incompatibilities'', perhaps most strikingly in the limit $k_\perp\to 0$
where the left-hand side is finite while the right-hand side vanishes.
%
%
Notice that the failure of the WW-type approximations
(\ref{Eq:WW-type-gT}--\ref{Eq:WW-type-8}) in the limit $k_\perp\to 0$ is
not specific to the Gaussian model,
but a general feature caused by neglecting tilde-terms. This indicates
a practical scheme how to use responsibly the WW-type approximations in
Eqs.~(\ref{Eq:WW-type-gT}--\ref{Eq:WW-type-8}).

Our procedure is as follows. In a first step we assume that all TMDs and
FFs are (approximately) Gaussian and solve the convolution integrals.
In the second step we use the integrated WW-type approximations to
simplify the results for the structure functions.

Notice that in some cases (when T-even TMDs are involved)
one could choose a different order of the steps: first apply
WW-type approximations and then solve convolution integrals
with Gaussian Ansatz.
In general, this would yield different (and bulkier) analytical
expressions, but we convinced ourselves that the differences
are numerically within the accuracy expected for this approach.
However, for the structure functions discussed in
Secs.~\ref{Sec-7.6:FUTsinphiS} and \ref{Sec-7.7:FUUcosphi},
such an ``alternative scheme'' would give results at
variance with the sum rules for the twist-3 T-odd TMDs
in Eq.~(\ref{Eq:sum-rules-T-odd}), as discussed in
Sec.~\ref{Sec-3.8:limitations}.
The scheme presented here will allow us to implement those sum
rules in a convenient and consistent way. We will follow up on
this in more detail in
Secs.~\ref{Sec-7.6:FUTsinphiS} and \ref{Sec-7.7:FUUcosphi}.

To summarize, our procedure is to solve first the convolution
integrals with a Gaussian Ansatz, and use then WW-type approximations.
When implementing this procedure we will see that the results
for the structure functions can be conveniently expressed in
terms of the basis TMDs or their adequate transverse moments.

\subsection{Phenomenological information on basis functions}
\label{Sec-4.3:plot-basis-functions}

\begin{figure}[b!]
\centering
\includegraphics[width=0.32\textwidth]{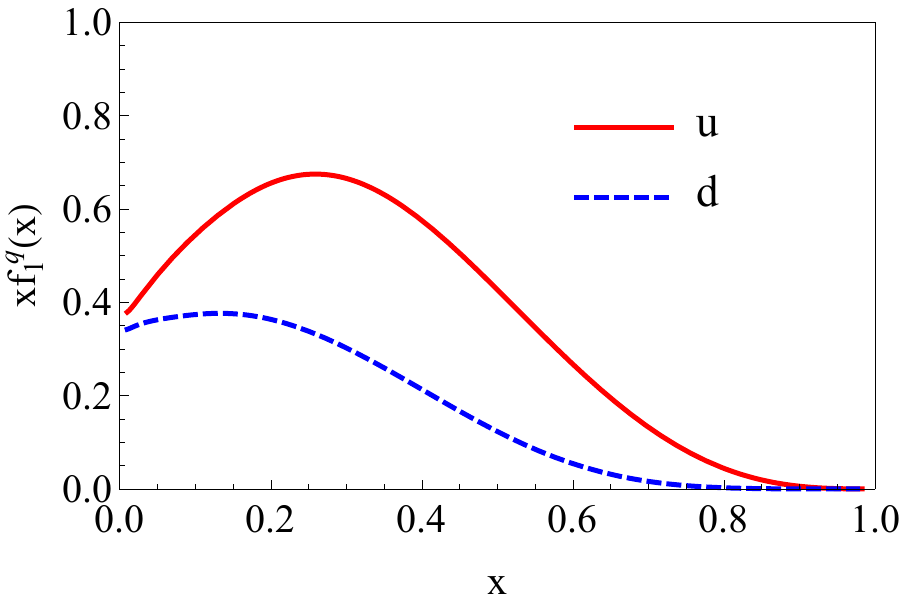}
\includegraphics[width=0.32\textwidth]{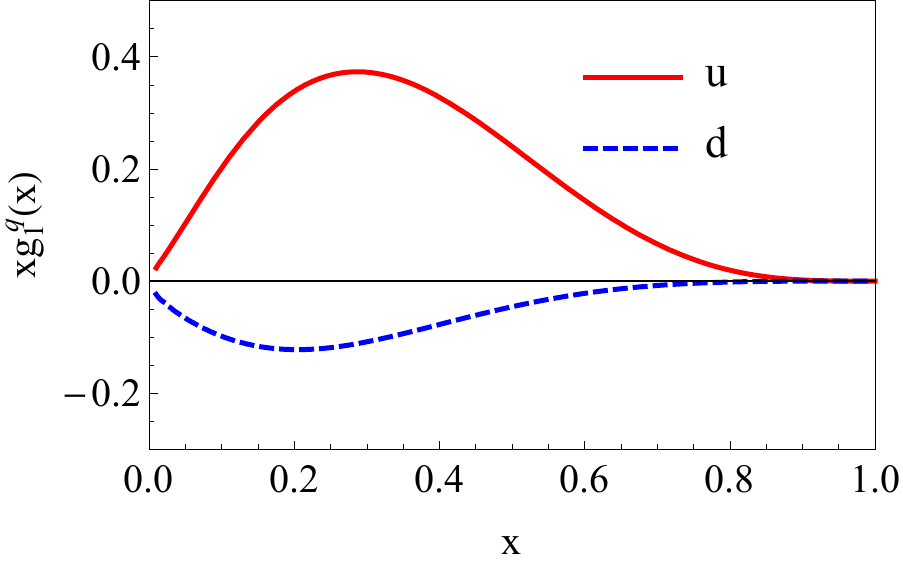}
\includegraphics[width=0.32\textwidth]{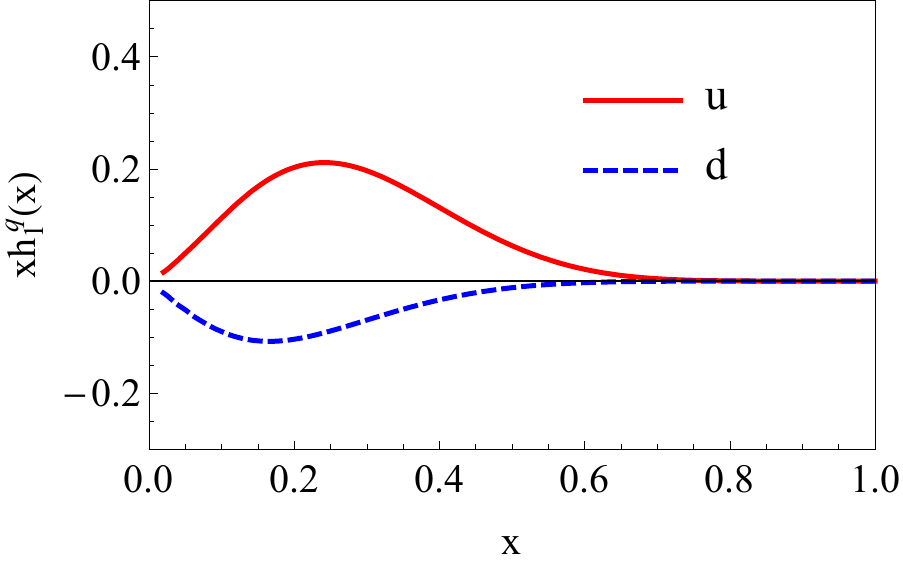}

\includegraphics[width=0.32\textwidth]{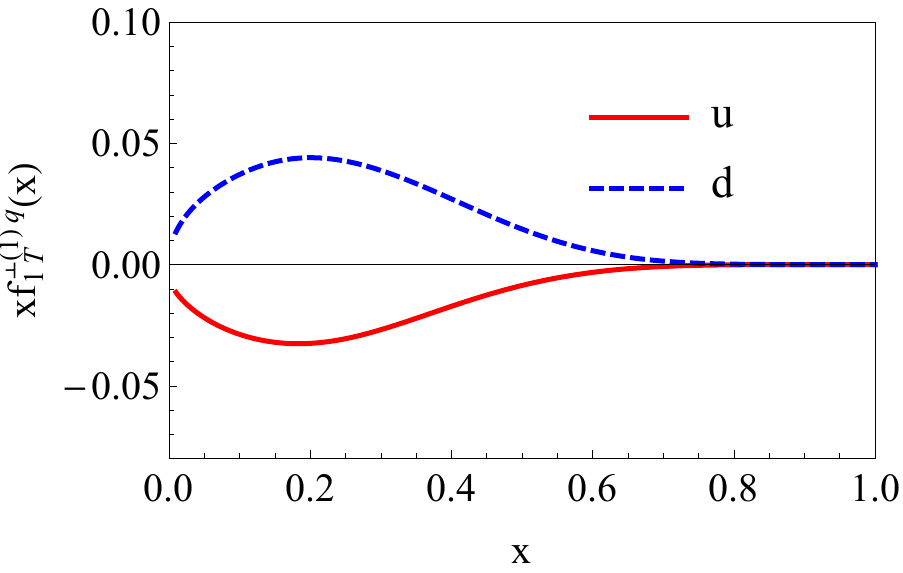}
\includegraphics[width=0.32\textwidth]{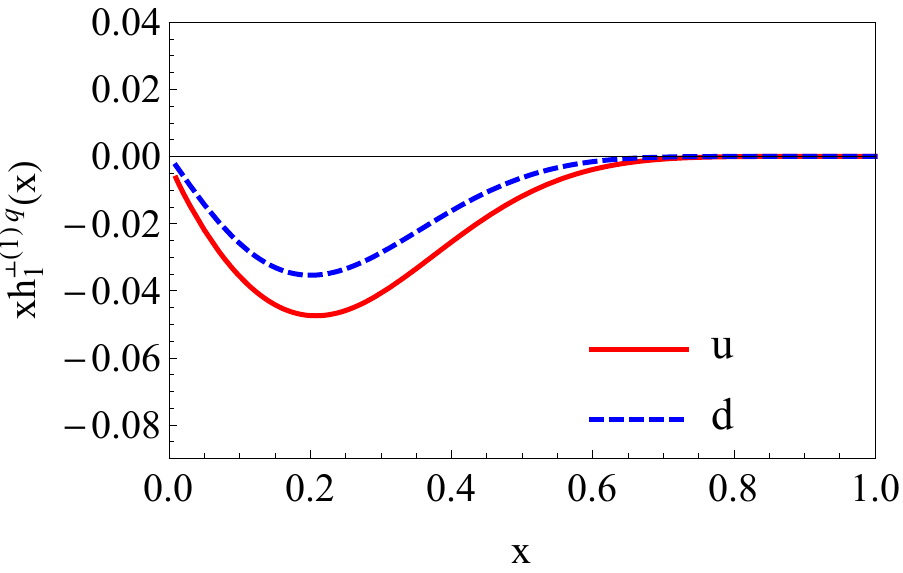}
\includegraphics[width=0.32\textwidth]{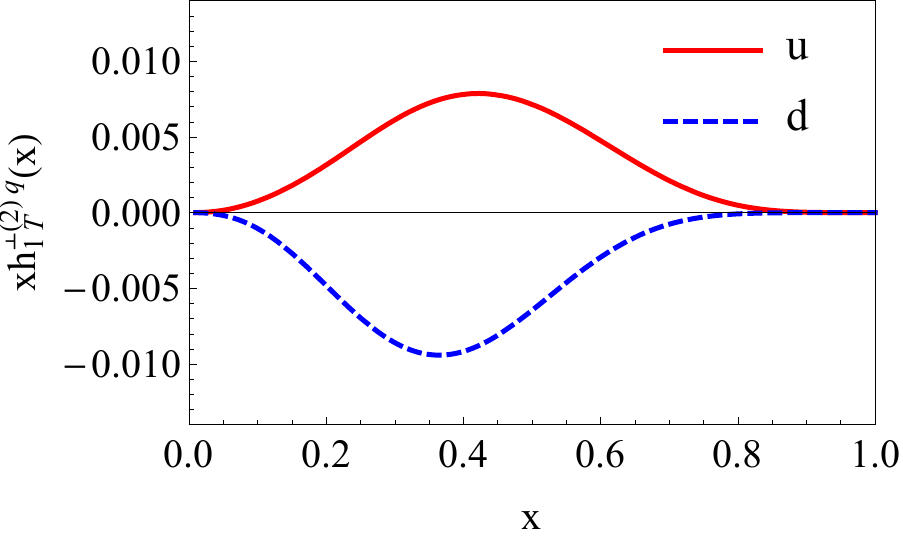}

\includegraphics[width=0.33\textwidth]{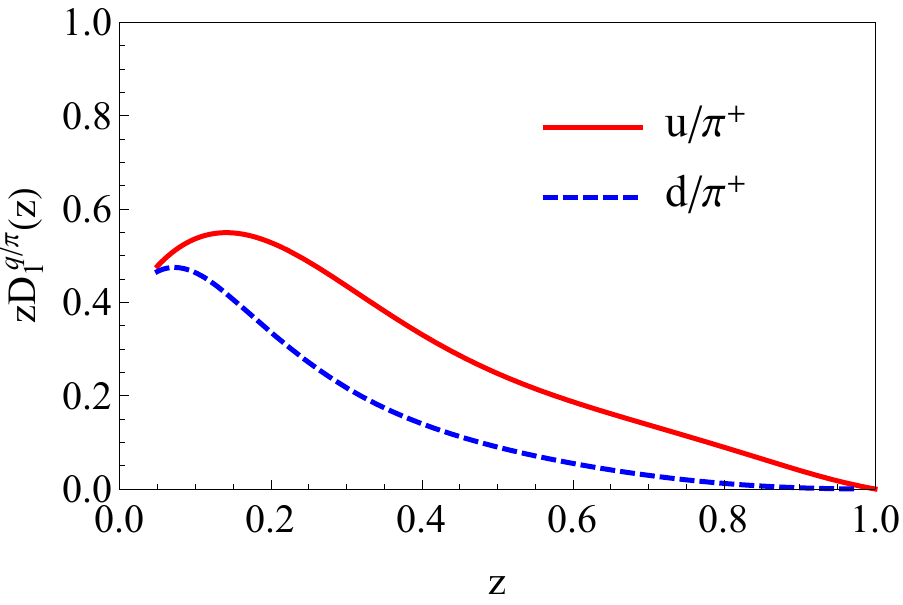}
\includegraphics[width=0.33\textwidth]{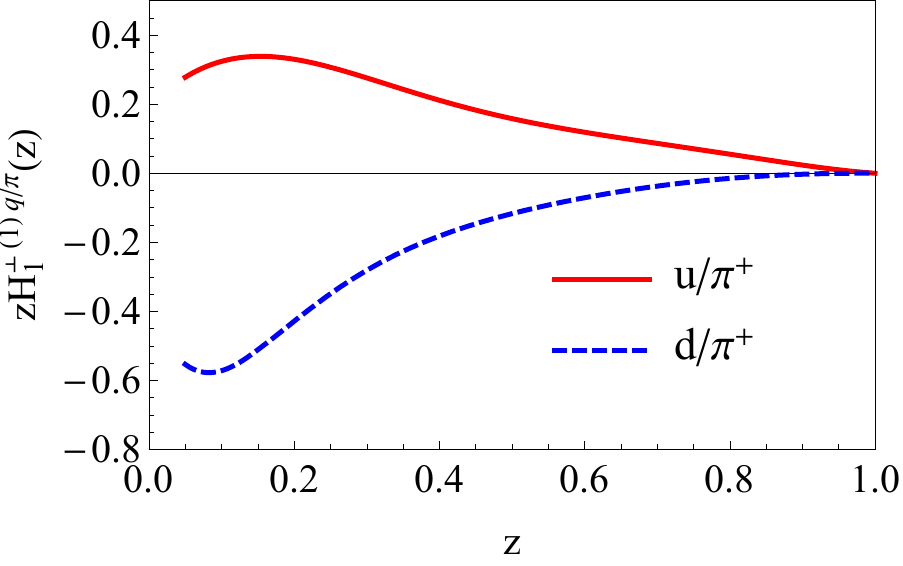}
\caption{\label{basis}
	The basis functions $f_1^a, \; g_1^a, \; h_1^a,
	f_{1T}^{\perp a}, \;h_1^{\perp a},\; h_{1T}^{\perp a}; \;
	D_1^a, \; H_1^{\perp a} \,$.
	The parametrizations of the basis functions and the Gaussian
	model parameters are described in detail in App.~\ref{App:basis}.}
\end{figure}

We have seen that the following 6 TMDs and 2 FFs provide a basis
(Sec.~\ref{Sec-3:WW}) and allow us to express all SIDIS structure
functions (Sec.~\ref{Sec-4:SIDIS-in-WW-approximation})
in WW-type approximation:
\be\label{Eq:basis}
   \mbox{basis: \ \ }
   f_1^a, \; f_{1T}^{\perp a}, \; g_1^a, \; h_1^a, \;h_1^{\perp a},\; h_{1T}^{\perp a};
   \; D_1^a, \; H_1^{\perp a} \, .
\ee
Phenomenological information is available for all basis functions at
least to some extent.
In Fig.~\ref{basis} we present plots of the basis functions, and refer
to App.~\ref{App:basis} for details.
The four functions $f_1^a, \; g_1^a, \; h_1^a,\; D_1^a$  are related to
twist-2 collinear functions. All collinear functions are calculated at
$Q^2 = 2.4$ GeV$^2$ with $f_1^a(x)$ from \cite{Martin:2009iq},
$g_1^a(x)$ from \cite{Gluck:1998xa}, and $D_1^a(z)$ from
\cite{deFlorian:2007aj}. The other four TMDs
have no collinear counterparts.
For $f_{1T}^{\perp a}$, $h_1^{\perp a}$, and $H_1^{\perp a}$ it is convenient to
consider their (1)-moments, for $ h_{1T}^{\perp a}$  the (2)-moment;
see (\ref{eq:moments}) for definitions.
This has two important advantages. First, this step simplifies
the Gaussian model expressions, and the Gaussian width parameters are
largely absorbed in the definitions of the transverse moments. Second,
the $k_\perp$--moments of these TMDs have in principle simple definitions
in QCD (whereas, e.g., the function $f_{1T}^{\perp a}(x)$ can be computed in
models but is very cumbersome to define in QCD).
The parametrizations for the basis functions read
\begin{subequations}\ba
	f^a_1(x,\kperp^2) &=& f^a_1(x)\;
    	\frac{1}{\,\pi\avkperp_{f_1}}\;e^{-\kperp^2/\avkperp_{f_1}} \, ,
	\label{Eq:Gauss-f1}\\
    	D^a_1(z,\pperp^2) &=& D_1^a(z)\,
    	\frac{1}{\,\pi\avpperp_{D_1}}\;e^{-\pperp^2/\avpperp_{D_1}} \, ,
	\label{Eq:Gauss-D1}\\
	g^a_1(x,\kperp^2) &=& g^a_1(x)\;
    	\frac{1}{\,\pi\avkperp_{g_1}}\;e^{-\kperp^2/\avkperp_{g_1}} \, ,
	\label{Eq:Gauss-g1}\\
	h_{1}^{a} (x, \kperp^2) &=& h_{1}^{a} (x)\;
  	\frac{1}{\,\pi \avkperp_{h_1}}\;e^{-{\kperp^2}/{\avkperp_{h_1} }} \, ,
	\label{Eq:Gauss-h1}\\
	H_{1}^{\perp a}(z,\pperp^2) &=&  H_{1}^{\perp (1) a}(z) \;
	\frac{2 z^2 \mh^2}{\pi \avpperp_{H_{1}^\perp}^2} \;
	e^{-\pperp^2/{\avpperp_{H_{1}^\perp}}}\, ,\\
	f_{1T}^{\perp a}(x,\kperp^2) &=&  f_{1T}^{\perp (1) a}(x)   \;
	\frac{2 M^2}{\pi \avkperp_{f_{1T}^\perp}^2} \;
	e^{-\kperp^2/{\avkperp_{f_{1T}^\perp}}}
	\label{Eq:Gauss-f1Tperp}\, ,\\
	h_{1}^{\perp a}(x,\kperp^2) &=&  h_{1}^{\perp (1) a}(x)\;
   	\frac{2 M^2}{\pi \avkperp_{h_{1}^\perp}^2}\;
 	e^{-\kperp^2/{\avkperp_{h_{1}^\perp}}}\,
	\label{Eq:Gauss-h1perp}\, ,\\
	h_{1T}^{\perp a}(x,\kperp^2) &=&  h_{1T}^{\perp (2) a}(x)\;
   	\frac{2 M^4}{\pi \avkperp_{h_{1T}^\perp}^3} \;
	e^{-\kperp^2/{\avkperp_{h_{1T}^\perp}}}
	\label{Eq:Gauss-h1Tperp}\, .
\ea\end{subequations}


%
\section{Leading-twist asymmetries and basis functions}
\label{Sec-5:twist-2+basis}
In this section we review how the basis functions describe available
SIDIS data. This is of importance to assess the reliability of the
predictions presented in the next sections.

\subsection{\boldmath Leading-twist $F_{UU}$ and Gaussian Ansatz}
\label{Sec-5.1:FUU-basis}

As explained in Sec.~\ref{Sec-4.3:evaluation} the Gaussian Ansatz is chosen
not only because it considerably simplifies the calculations, but more
importantly because it works phenomenologically with a good accuracy
in many processes including SIDIS
\cite{Anselmino:2005nn,Collins:2005ie,D'Alesio:2007jt,Schweitzer:2010tt,
Signori:2013mda,Anselmino:2013lza}.

The Gaussian Ansatz for the unpolarized TMD and FF
is given by Eqs.~(\ref{Eq:Gauss-f1},~\ref{Eq:Gauss-D1}).
The parameters $\avkperp_{f_1}$ and $\avpperp_{D_1}$ can be
assumed to be flavor- and $x$-- or $z$--independent, as present
data hardly allow us to constrain too many parameters, see
App.~\ref{App:basis-f1-D1} for a review. This assumption can be
relaxed, e.g., theoretical studies in chiral effective theories
predict a strong flavor-dependence in the $\kperp$--behavior
of sea and valence quark TMDs \cite{Schweitzer:2012hh}.

\begin{figure}[t!]
\centering
\includegraphics[height=3.2cm]{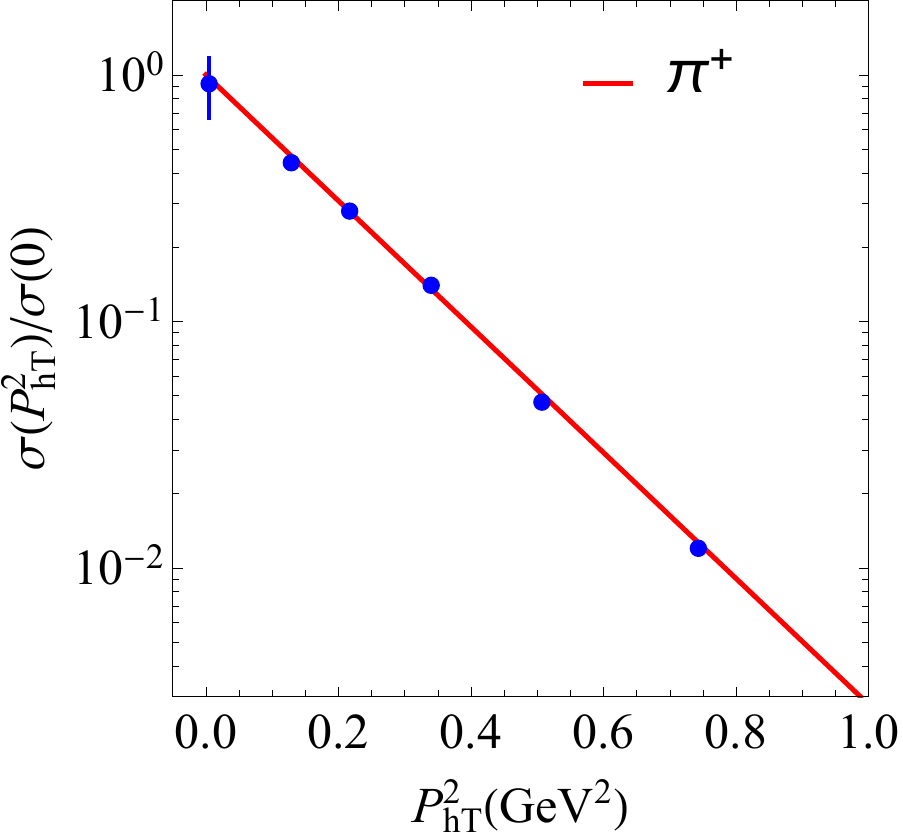}  \quad
\includegraphics[height=3.2cm]{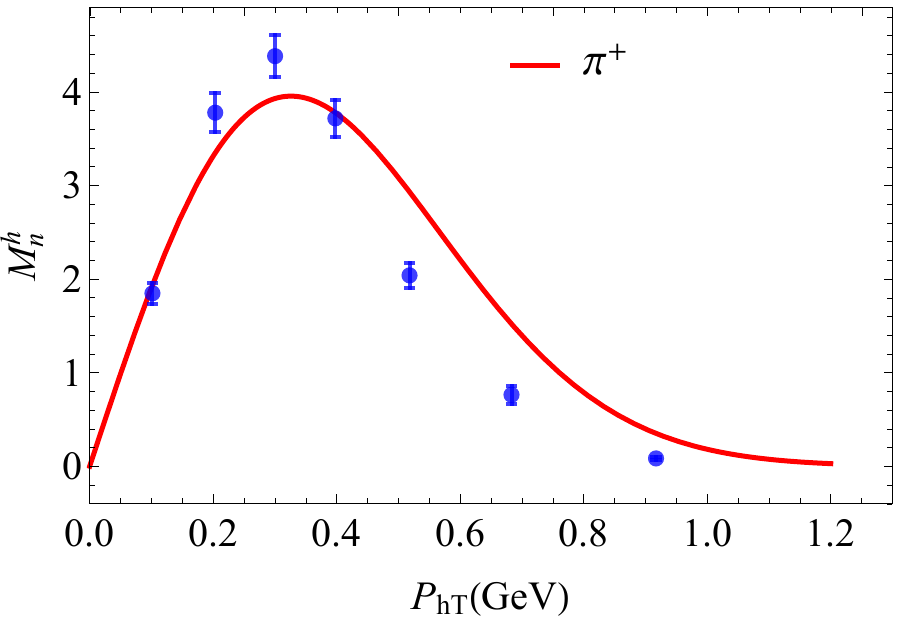} \quad
\includegraphics[height=3.2cm]{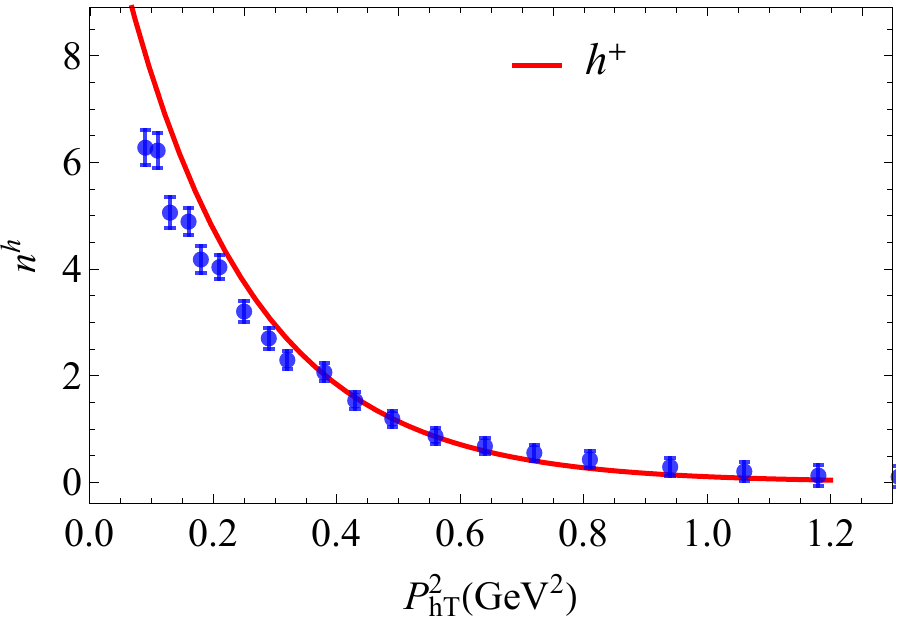}
\caption{\label{FUU-show-pT-dependence}
Left panel:
	$F_{UU}(P_{hT}^2)/F_{UU}(0)$ for $\pi^+$
	production
	at JLab with a 5.75 GeV $e^{-}$ beam \cite{Osipenko:2008aa}.
Middle panel:
	HERMES multiplicity (\ref{Eq:multiplicity-HERMES}) at
	$\la Q^2\ra=2.87\,{\rm GeV}^2$, $\la x\ra  =0.15$, $\la z\ra  =0.22$
	 \cite{Airapetian:2012ki}.
Right panel:
	COMPASS multiplicity (\ref{Eq:multiplicity-COMPASS}) at
	$\la Q^2\ra=20\,{\rm GeV}^2$, $\la x\ra  =0.15$, $\la z\ra  =0.2$
	 \cite{Aghasyan:2017ctw}.}
\end{figure}

The structure function $F_{UU}$ needed for our analysis reads
\begin{subequations}\ba
	F_{UU}(x,z,\Phperp)
	&=& x \sum_q e_q^2\,f^q_1(x)\,D_1^q(z)\,{\cal G}(\Phperp)\,,
	\label{Eq:FUU-Phperp}\\
	F_{UU}(x,z) 
	&=& x \sum_q e_q^2\,f^q_1(x)\,D_1^q(z)  \, ,
	\label{Eq:FUU}
\ea\end{subequations}
where we introduce the notation ${\cal G}(\Phperp)$, which is defined as
\be\label{Eq:def-Gaussian-lambda}
	{\cal G}(\Phperp) = \frac{\exp(-\Phperp^2/\lambda)}{\pi\,\lambda}
	\, , \;\;\;
	\lambda = z^2\,\la\kperp^2\ra_{f_1} + \la\pperp^2\ra_{D_1} \,,
\ee
with the understanding that the convenient abbreviation $\lambda$ is expressed
in terms of the Gaussian widths of the {\it preceding} TMD and FF. Notice
that ${\cal G}(\Phperp)\equiv {\cal G}(x,z,\Phperp)$ and that in general
${\cal G}(\Phperp)$ appears under the flavor sum due to a possible
flavor-dependence of the involved Gaussian widths.
The normalization $\int d^2\Phperp \,{\cal G}(\Phperp)=1$
correctly connects the structure function $F_{UU}(x,z,\Phperp)$
in (\ref{Eq:FUU-Phperp}) with its $\Phperp$--integrated counterpart
(\ref{Eq:FUU}). In our effective description this step is trivial. In
QCD the connection of TMDs to PDFs is subtle \cite{Collins:2016hqq}.
Figure~\ref{FUU-show-pT-dependence} illustrates how the Gaussian Ansatz
describes selected SIDIS data.

Let us begin with JLab where, in the pre-12$\,$GeV era, electron beams
from CEBAF with energies in the range $4.3$ to $5.7$ GeV were scattered
off proton or deuterium targets in the typical kinematics
$1 \,{\rm GeV}^2 < Q^2 < 4.5 \,{\rm GeV}^2$, $W > 2\,{\rm GeV}$,
$0.1 < x < 0.6$, $y < 0.85$, $0.5<z<0.8$.
The left panel of Fig.~\ref{FUU-show-pT-dependence} shows basically
the SIDIS structure function $F_{UU}(P_{hT}^2)$ normalized with respect
to its value at zero transverse hadron momentum\footnote{Strictly
	speaking in \cite{Osipenko:2008aa} data for the normalized
	SIDIS cross section was presented. But these data correspond
	to $F_{UU}(P_{hT}^2)/F_{UU}(0) \equiv
	F_{UU}(\la x\ra,\la z\ra,P_{hT}^2)/F_{UU}(\la x\ra,\la z\ra,0)$
	up to $1/Q^2$-suppressed terms.}
for $\pi^+$ production from a proton target measured in the CLAS experiment
with a $5.75$ GeV beam for the kinematics $\la Q^2\ra=2.37\,{\rm GeV}^2$,
$\la x\ra=0.24$, $\la z\ra=0.30$ \cite{Osipenko:2008aa}. Clearly, the
Gaussian model works for the entire region of $P_{hT}$ covered in this
experiment, in which the structure function $F_{UU}$ falls down by 2 orders of
magnitude \cite{Schweitzer:2010tt}.

Next we discuss a representative plot from the HERMES experiment
where pions or kaons were measured in the scattering of 27.6 GeV
electrons or positrons of HERA's polarized lepton storage ring
off proton and deuteron targets in the SIDIS kinematics
$Q^2 > 1 \,{\rm GeV}^2$, $W^{2} > 10\,{\rm GeV}^{2}$,
$0.023 < x < 0.4$, $y < 0.85$, $0.2<z<0.7$.
The middle panel of Fig.~\ref{FUU-show-pT-dependence} displays the
HERMES multiplicity~\cite{Airapetian:2012ki}
\be\label{Eq:multiplicity-HERMES}
	M_n^h(x,z,\Phperp) \equiv
	\frac{d\sigma_{\rm SIDIS}(x,z,\Phperp)/dx\,dz\,d\Phperp}
	{d\sigma_{\rm DIS}(x)/dx} =
	2 \pi \Phperp \frac{F_{UU}(x,z,\Phperp) }{ x \sum_q e_q^2\,f^q_1(x)}
\ee
at $\la Q^2\ra=2.87\,{\rm GeV}^2$, $\la x\ra=0.15$, $\la z\ra=0.22$
for $\pi^+$ production on the proton target~\cite{Airapetian:2012ki}.

Finally we show also a representative plot from the COMPASS experiment
where charged pions, kaons, or hadrons were measured with 160 GeV
longitudinally polarized muons scattered off proton and deuteron
targets in the typical SIDIS kinematics
$Q^2 > 1 \,{\rm GeV}^2$, $W > 5\,{\rm GeV}$,
$0.003 < x < 0.7$, $0.1<y < 0.9$, $0.2<z<1$.
The right panel of
Fig.~\ref{FUU-show-pT-dependence} shows the COMPASS multiplicity
\cite{Aghasyan:2017ctw}
\be\label{Eq:multiplicity-COMPASS}
	n^h(x,z,\Phperp^2)  \equiv
	\frac{d\sigma_{\rm SIDIS}(x,z,\Phperp^2)/dx\,dz\,d\Phperp^2}
	{d\sigma_{\rm DIS}(x)/dx} =
	\pi \frac{F_{UU}(x,z,\Phperp^2) }{ x \sum_q e_q^2\,f^q_1(x)}\;
\ee
at $\la Q^2\ra=20\,{\rm GeV}^2$, $\la x\ra  =0.15$, $\la z\ra  =0.2$
for $h^+$ production on the deuterium target \cite{Aghasyan:2017ctw}.

To streamline the presentation we refer to the comprehensive App.~\ref{App:basis}
on the parametrizations used, 
and for technical details on the Gaussian Ansatz to App.~\ref{App:factor}.

The description of the HERMES and COMPASS multiplicities in
Fig.~\ref{FUU-show-pT-dependence} is good and sufficient for our
purposes, but it is not perfect. The descriptions of the COMPASS
data in the region of small $\Phperp^2$ and that of the HERMES data
for $\Phperp \gtrsim 0.3\,{\rm GeV}$ are not ideal.
However, notice that in our description we use the Gaussian widths
as fitted and employed in the original extractions of the TMDs. These
values were not optimized to fit the HERMES or COMPASS multiplicities.
Keeping this in mind, the description in Fig.~\ref{FUU-show-pT-dependence}
can be considered as satisfactory. We also remark that we do not take
into account $\kperp$-broadening effects between HERMES and
COMPASS energies and that the HERMES data actually represent
multiplicities integrated (separately for numerator and denominator)
over the kinematic ranges of each bin while the curve is plotted for
a fixed set of kinematics. Through dedicated fits to the HERMES, COMPASS
(and other) data and consideration of $\kperp$-evolution effects
it is possible to obtain a better description than in
Fig.~\ref{FUU-show-pT-dependence}, see \cite{Bacchetta:2017gcc}.

\subsection{\boldmath Leading-twist $A_{LL}$ and first test of Gaussian Ansatz
	in polarized scattering}
\label{Sec-5.2:FLL-basis}

The Gaussian Ansatz is useful in unpolarized case
\cite{Anselmino:2005nn,Collins:2005ie,D'Alesio:2007jt,Schweitzer:2010tt,
Signori:2013mda,Anselmino:2013lza}, but nothing is known about its
applicability to spin asymmetries. The JLab data \cite{Avakian:2010ae}
on $A_{LL}(\Phperp)$ put us in the position to conduct a first ``test''
for polarized partons. We assume Gaussian form for $g_{1}^{a}(x,\kperp^2)$,
Eq.~(\ref{Eq:Gauss-g1}), and use lattice QCD results \cite{Hagler:2009mb}
to estimate the width $\avkperp_{g_{1}}$, see App.~\ref{App:basis-g1}.
With $\lambda=z^2\la\kperp^2\ra_{g_1}+\la\pperp^2\ra_{D_1}$ implicit
in ${\cal G}(\Phperp)$, the structure function $F_{LL}$  reads
\begin{subequations}\ba
	F_{LL}(x,z,\Phperp)
	&=& x \sum_q e_q^2\,g^q_1(x)\,D_1^q(z)\,{\cal G}(\Phperp)\,,
	\label{Eq:FLL-Phperp}\\
	F_{LL}(x,z)
	&=& x \sum_q e_q^2\,g^q_1(x)\,D_1^q(z)  \, .
	\label{Eq:FLL}
\ea\end{subequations}

\begin{figure}[b]
\centering
\includegraphics[width=0.333\textwidth]{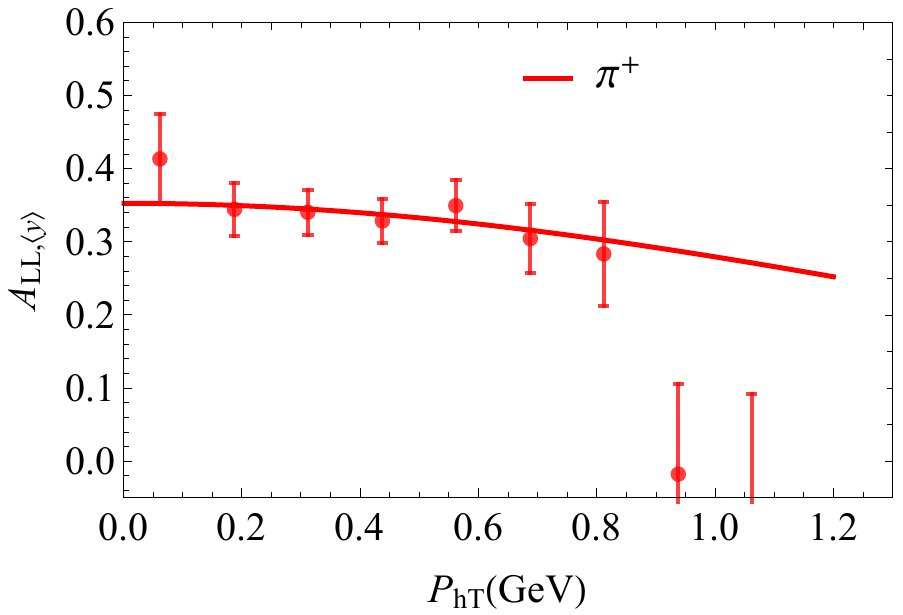}%
\includegraphics[width=0.333\textwidth]{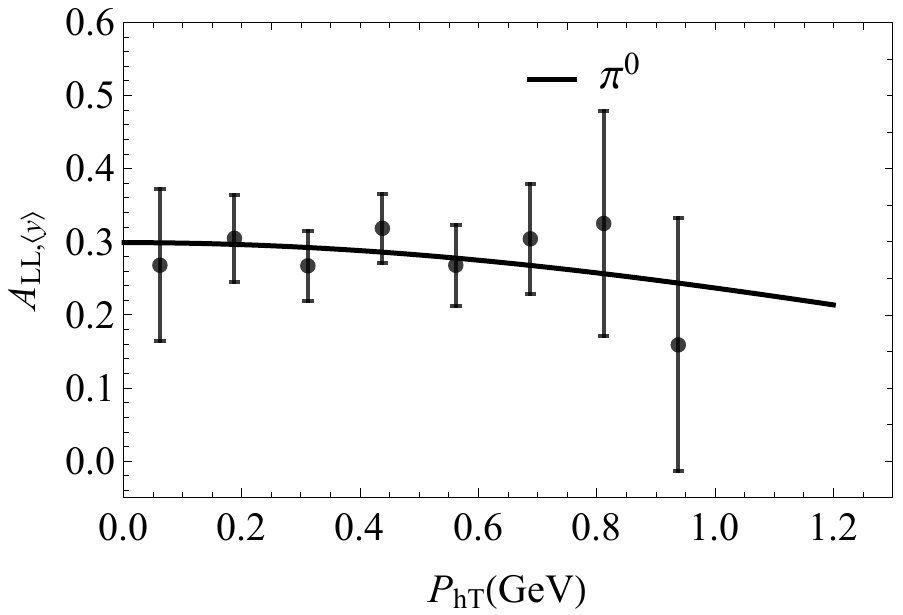}%
\includegraphics[width=0.333\textwidth]{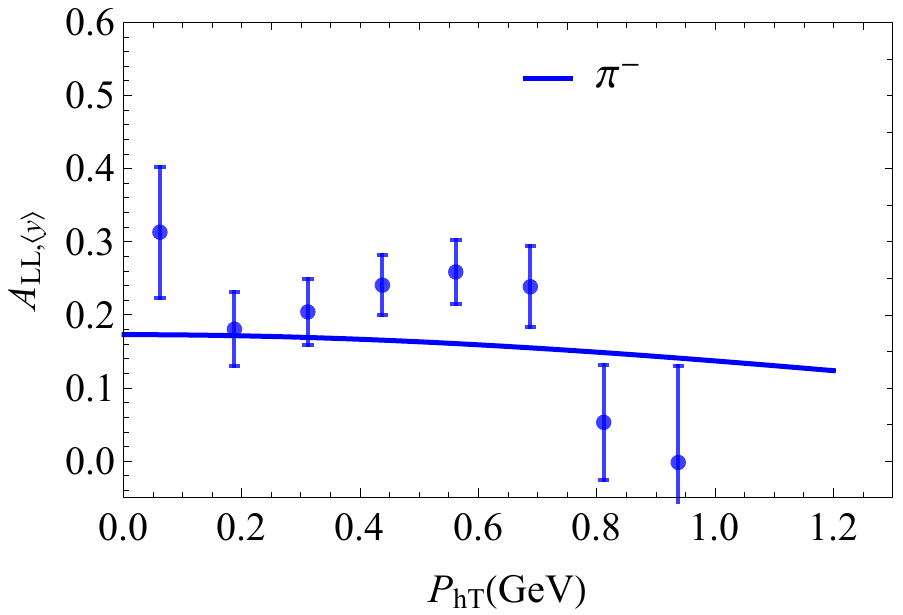}
\caption{\label{jlab_ALL}
	$A_{LL,\la y\ra}$ as function of $P_{hT}$ vs.~JLab
	data \cite{Avakian:2010ae}
 	for $\pi^+$, $\pi^0$, $\pi^-$. The solid lines are
	our results for the mean values of kinematical variables
	$\langle x \rangle = 0.25$,
	$\langle z \rangle = 0.5$, and $\langle Q^2 \rangle = 1.67$ GeV$^2$.
}
\end{figure}

The definition of the asymmetry used by the JLab experiment \cite{Avakian:2010ae} is
\be\label{Eq:ALLy}
	A_{LL,\la y\ra}(x,z,\Phperp)
	= \la p_2 \,A_{LL}(x,z,\Phperp) \ra \,
	= \frac{\la y (2-y) \; F_{LL}(x,z,\Phperp)\ra}
	{\la(1+(1-y)^2) \; F_{UU}(x,z,\Phperp)\ra} \, ,
\ee
where $p_2 = y (2-y)/(1+(1-y)^2)$ and averaging (separately in numerator
and denominator) over the kinematics of \cite{Avakian:2010ae} is implied.
We use the lattice data \cite{Hagler:2009mb} to
constrain the Gaussian width $\la\kperp^2\ra_{g_1}$ as described in
App.~\ref{App:basis-g1}. All other ingredients in (\ref{Eq:ALLy}) are known
and tested through other observables in Sec.~\ref{Sec-5.1:FUU-basis}.
Therefore the comparison of our results to the JLab data \cite{Avakian:2010ae}
shown in Fig.~\ref{jlab_ALL} provides several important tests.
First, the JLab data \cite{Avakian:2010ae} are compatible
with the Gaussian Ansatz within uncertainties. Second, the lattice
results---in the way we use them in App.~\ref{App:basis-g1}---give an 
appropriate description of the data.
	(Another important test was already presented in
	\cite{Avakian:2010ae}: the $\Phperp$--integrated (``collinear'')
	asymmetry (\ref{Eq:FLL}) is compatible with data
	from other experiments and theoretical results obtained from
	parametrizations of $f_1^a(x)$, $g_1^a(x)$, $D_1^a(z)$. This
	shows that in the pre-12~GeV era one was, to a good
	approximation, indeed already probing DIS \cite{Avakian:2010ae}.)
We remark that HERMES and COMPASS data also show flat 
$P_{hT}$-distributions \cite{Airapetian:2018rlq,Adolph:2016vou}.

Encouraged by these findings we will use lattice predictions from
Ref.~\cite{Hagler:2009mb} below also for the Gaussian widths of
$g_{1T}^{\perp(1)a}$ and $h_{1L}^{\perp(1)a}$.
Of course, at this point one could argue that the WW and WW-type
approximations (\ref{Eq:WW-approx-g1T},~\ref{Eq:WW-approx-h1L}) also
dictate that $g_{1T}^\perp$ and $h_{1L}^\perp$ have the same Gaussian
widths as $g_1$ and $h_1$. In fact, the lattice results for the
respective widths are numerically similar, which can be interpreted as
yet another argument in favor of the usefulness of the approximations.
The practical predictions depend only weakly on the choice of parameters.

\subsection{\boldmath Leading-twist $A_{UT}^{\sin(\phi_h-\phi_S)}$ Sivers asymmetry}
\label{Sec-5.3:Sivers-basis}

The $F_{UT}^{\sin(\phi_h-\phi_S)}$ structure function is related to the
Sivers function~\cite{Sivers:1989cc}, which describes the distribution
of unpolarized quarks inside a transversely polarized proton. It has so far
received the widest attention, from both phenomenological and experimental
points of view.

The Sivers function $f_{1T}^\perp$ is related to initial and final-state
interactions of the struck quark and the rest of the nucleon and could
not exist without contributions of orbital angular momentum of
partons to the spin of the nucleon. As such it encodes the correlation
between the partonic intrinsic motion and the transverse spin of the
nucleon, and it generates a dipole deformation in momentum space.
The Sivers function has been extracted from SIDIS data
by several groups, with consistent results
\cite{Anselmino:2010bs,Anselmino:2005ea,Anselmino:2005an,Collins:2005ie,Vogelsang:2005cs,Anselmino:2008sga,Bacchetta:2011gx,Echevarria:2014xaa}.

The structure function $F_{UT}^{\sin(\phi_h-\phi_S)}$ reads
\begin{subequations}\ba
	F_{UT}^{\sin(\phi_h-\phi_S)}(x,z,\Phperp)
	&=& - x\sum_q e_q^2\,f_{1T}^{\perp (1) q}(x)\,D_1^{q}(z)\;
	b^{(1)}_{\rm B}\,\biggl(\frac{z \Phperp} {\lambda}\biggr)\,
	{ \cal G}(\Phperp ) \, , \label{FUTsiv-Gauss-Phperp}\\
	F_{UT}^{\sin(\phi_h-\phi_S)}(x,z,\la\Phperp\ra)
	&=& - x\sum_q e_q^2\,f_{1T}^{\perp (1) q}(x)\,D_1^{q}(z)\;
	c^{(1)}_{\rm B}\,\biggl(\frac{z} {\lambda^{1/2}}\biggr)\,,
	\label{FUTsiv-Gauss}
\ea\end{subequations}
where $\lambda=z^2 \avkperp_{f_{1T}^\perp} + \avpperp_{D_1}$ and
$b^{(1)}_{\rm B}=2M_N$ and $c^{(1)}_{\rm B} = \sqrt{\pi}\,M_N$,
see App.~\ref{App:convol-details} for details.

Notice that integrating structure functions over $\Phperp$
is different from integrating the cross section over $\Phperp$
where azimuthal hadron modulations drop out.
Only if the relevant weight is $\omega^{\{0\}}$ we obtain 
``collinear structure functions'':  $F_{UU}(x,z)$, $F_{LL}(x,z)$
in Secs.~\ref{Sec-5.1:FUU-basis}, \ref{Sec-5.2:FLL-basis},
and below in Secs.~\ref{Sec-7.2:FLTcosphiS}, \ref{Sec-7.6:FUTsinphiS}.
In all other cases, despite integration over $\Phperp$, we end up
always with true convoluted TMDs (here within Gaussian model).
We stress this important point by displaying the dependence of
the structure functions on the mean transverse momentum, e.g.,
$F_{UT}^{\sin(\phi_h-\phi_S)}(x,z,\la\Phperp\ra) =
\int d^2\Phperp F_{UT}^{\sin(\phi_h-\phi_S)}(x,z,\Phperp)$
in (\ref{FUTsiv-Gauss}).

\begin{figure}[t!]
\centering
\includegraphics[width=0.35\textwidth]{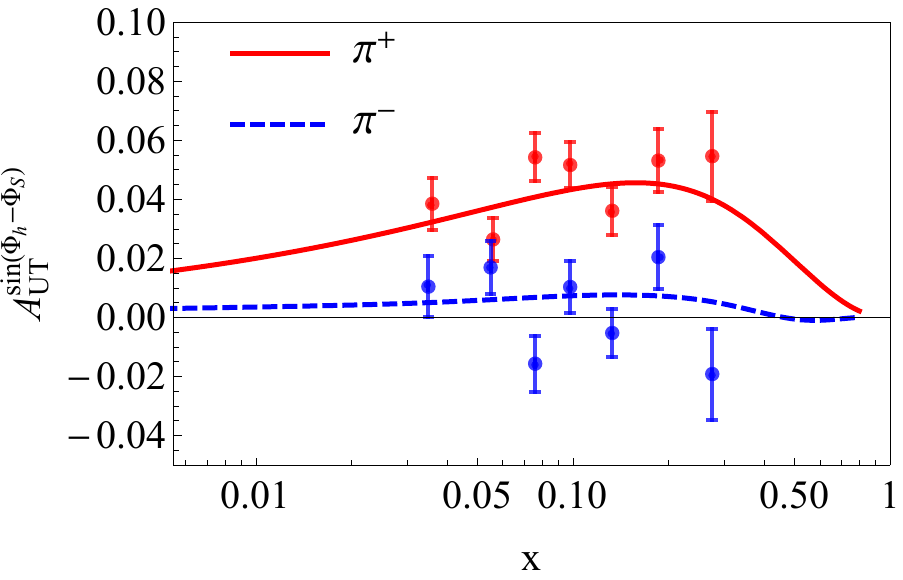}  \hspace{5mm}
\includegraphics[width=0.35\textwidth]{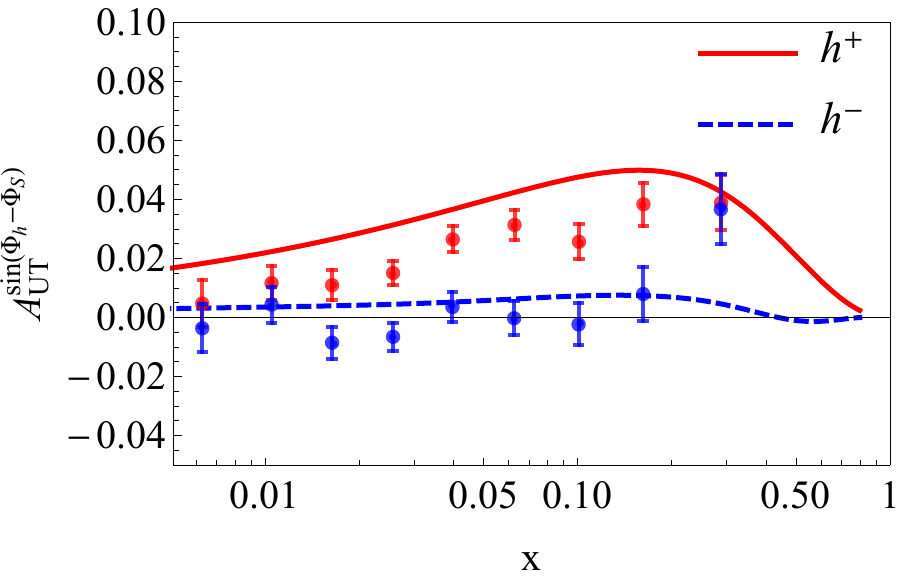}
\caption{\label{aut_f1t_jlab} Sivers asymmetry
	$A_{UT}^{\sin(\phi_h-\phi_S)}$ for a proton target as function of $ x $
	based on the fit \cite{Anselmino:2011gs} in comparison to
	(left panel) HERMES \cite{Airapetian:2009ae}
	and (right panel) COMPASS  \cite{Adolph:2012sp} data.}
\end{figure}

The asymmetries $A_{UT}^{\sin(\phi_h-\phi_S)}= F_{UT}^{\sin(\phi_h-\phi_S)}/F_{UU}$
obtained from the fit \cite{Anselmino:2011gs}
are plotted in Fig.~\ref{aut_f1t_jlab} as functions of $x$ in comparison
to HERMES \cite{Airapetian:2009ae} and COMPASS \cite{Adolph:2012sp} data
on respectively charged pion and hadron production from a proton target.
	Notice that the COMPASS data points seem to be below the
	theoretical curves which may indicate evolution
	effects~\cite{Aybat:2011ta,Anselmino:2012aa}.
We do not show here the description of the $P_{hT}$--dependence of the
data but it is well described by the fit of \cite{Anselmino:2011gs}
which confirms that the Gaussian model works also in this case.
The Sivers function is predicted to enter the description of hadron-hadron 
collisions (with transversely polarized protons) with an opposite sign 
compared to SIDIS \cite{Collins:2002kn,Brodsky:2002cx,Brodsky:2002rv}.
Recent results on single-spin asymmetries in weak-boson production 
from RHIC \cite{Adamczyk:2015gyk} and Drell--Yan from COMPASS 
\cite{Aghasyan:2017jop,Parsamyan:2018zju} are consistent 
with this prediction.

\subsection{\boldmath Leading-twist $A_{UT}^{\sin(\phi_h+\phi_S)}$ Collins asymmetry}
\label{Sec-5.4:Collins-basis}

The $F_{UT}^{\sin(\phi_h+\phi_S)}$ structure function of the SIDIS cross section is
due to the convolution of the transversity distribution $h_1$ and the Collins
FF $H_1^\perp$. 
It describes the distribution of transversely polarized quarks 
in a transversely polarized nucleon, and is the only source of information
on the tensor charge of the nucleon. 
Transversity can also be accessed as a PDF in Drell--Yan or dihadron 
production \cite{Bacchetta:2002ux,Bacchetta:2003vn,Bacchetta:2011ip,
Bacchetta:2012ty,Radici:2015mwa,Radici:2018iag}. 
The Collins FF $H_1^\perp$ decodes the
fundamental correlation between the transverse spin of a fragmenting quark
and the transverse momentum of the produced final hadron~\cite{Collins:1992kk}.
There are many extractions of $h_1$ and $H_1^\perp$ from 
combined fits of SIDIS and $e^+e^-$ data, for instance those of
Refs.~\cite{Anselmino:2013vqa,Kang:2014zza,Anselmino:2015sxa}.
In this work we will use the extractions of $h_1$ and $H_1^\perp$
from Ref.~\cite{Anselmino:2013vqa}.

The structure function $F_{UT}^{\sin(\phi_h+\phi_S)}$ reads
\begin{subequations}\ba
	F_{UT}^{\sin(\phi_h+\phi_S)}(x,z,\Phperp)
	&=& x\sum_q e_q^2\,h_{1}^{q}(x)\,H_1^{\perp(1)q}(z)\;
	b^{(1)}_{\rm A}\,\biggl(\frac{z \Phperp} {\lambda}\biggr)\,
	{ \cal G}(\Phperp ) \, , \\
	F_{UT}^{\sin(\phi_h+\phi_S)}(x,z,\la\Phperp\ra)
	&=& x\sum_q e_q^2\,h_{1}^{q}(x)\,H_1^{\perp(1)q}(z)\;
	c^{(1)}_{\rm A}\,\biggl(\frac{z} {\lambda^{1/2}}\biggr)\,,
	\label{eq:asymmetry_aut_h1}
\ea\end{subequations}
where $\lambda=z^2 \avkperp_{h_1} + \avpperp_{H_1^\perp}$ and
$b^{(1)}_{\rm A}=2\mh$ and $c^{(1)}_{\rm A} = \sqrt{\pi}\,\mh$,
see App.~\ref{App:convol-details} for details.

\begin{figure}[t!]
\centering
\includegraphics[width=0.45\textwidth]{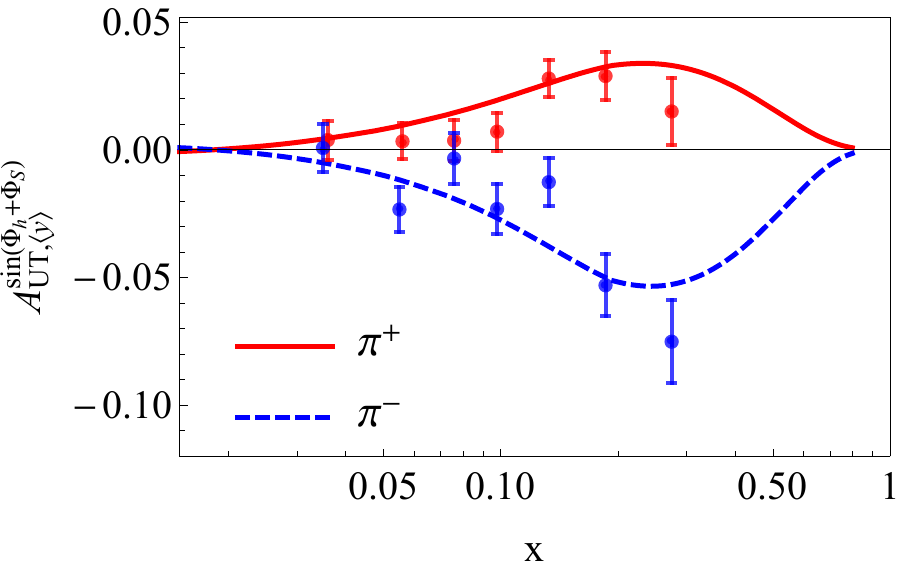}
{\tiny(a)}
\includegraphics[width=0.45\textwidth]{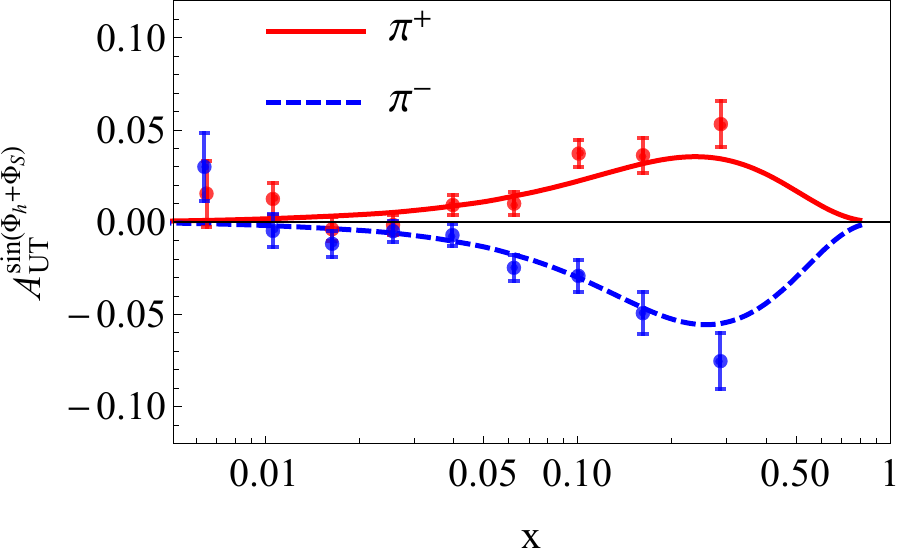}
{\tiny(b)}
\caption{\label{aut_h1_jlab}  Collins asymmetry for a proton target 
	vs $x$ based on the fit \cite{Anselmino:2013vqa}.
	(a) $A_{UT,  \langle y\rangle}^{\sin(\phi_h+\phi_S)}$ in comparison
	to HERMES \cite{Airapetian:2010ds} data.
	(b) $A_{UT}^{\sin(\phi_h+\phi_S)}$ in comparison 
	to COMPASS \cite{Adolph:2014zba} data.}
\end{figure}

The asymmetries $A_{UT, \langle y \rangle}^{\sin(\phi_h+\phi_S)}= \langle(1-y)F_{UT}^{\sin(\phi_h+\phi_S)}\rangle/\langle(1-y + y^2/2)F_{UU}\rangle$
are plotted in Fig.~\ref{aut_h1_jlab} as functions of $x$ in comparison
to HERMES \cite{Airapetian:2010ds} and $A_{UT}^{\sin(\phi_h+\phi_S)}= F_{UT}^{\sin(\phi_h+\phi_S)}/F_{UU}$  for COMPASS \cite{Adolph:2014zba}
data on charged-pion production from proton targets.
We remark that the description of the $P_{hT}$--dependences of
this azimuthal spin asymmetry is equally satisfactory by the
fit of Ref.~\cite{Anselmino:2013vqa}, which implies that the
data are compatible with the Gaussian Ansatz also in this case.

\subsection{\boldmath Leading-twist $A_{UU}^{\cos(2\phi_h)}$ Boer--Mulders asymmetry}
\label{Sec-5.5:BM-basis}

The structure function $F_{UU}^{\cos(2\phi_h)}$ arises from a convolution of
the Collins fragmention function and the Boer--Mulders TMD $h_{1}^{\perp }$,
which describes
the distribution of transversely polarized partons inside an unpolarized
target. The expression of this structure function is given by
\begin{subequations}\ba
	F_{UU}^{\cos(2\phi_h)}(x,z,\Phperp)
	&=& x \sum_q e_q^2\,h_{1}^{\perp (1) q}(x)\,H_1^{\perp(1) q}(z)\;
	b^{(2)}_{\rm AB}\,\biggl(\frac{z \Phperp} {\lambda}\biggr)^{\!2}\,
	{ \cal G}(\Phperp)\, , \\
	F_{UU}^{\cos2\phi_h}(x,z,\la\Phperp\ra)
	&=& x\sum_q e_q^2\,h_{1}^{\perp(1) q}(x)\,H_1^{\perp(1)q}(z)\;
	c^{(2)}_{\rm AB}\,\biggl(\frac{z} {\lambda^{1/2}}\biggr)^{\!2}\,,
	\label{eq:asymmetry_auu_cos2phi}
\ea\end{subequations}
where $\lambda=z^2 \avkperp_{h_1^\perp} + \avpperp_{H_1^\perp}$ and
$b^{(2)}_{\rm AB}=4M_N\mh$ and $c^{(2)}_{\rm AB} = 4M_N\mh$,
see App.~\ref{App:convol-details}.

\begin{figure}[t!]
\centering
\includegraphics[width=0.45\textwidth]{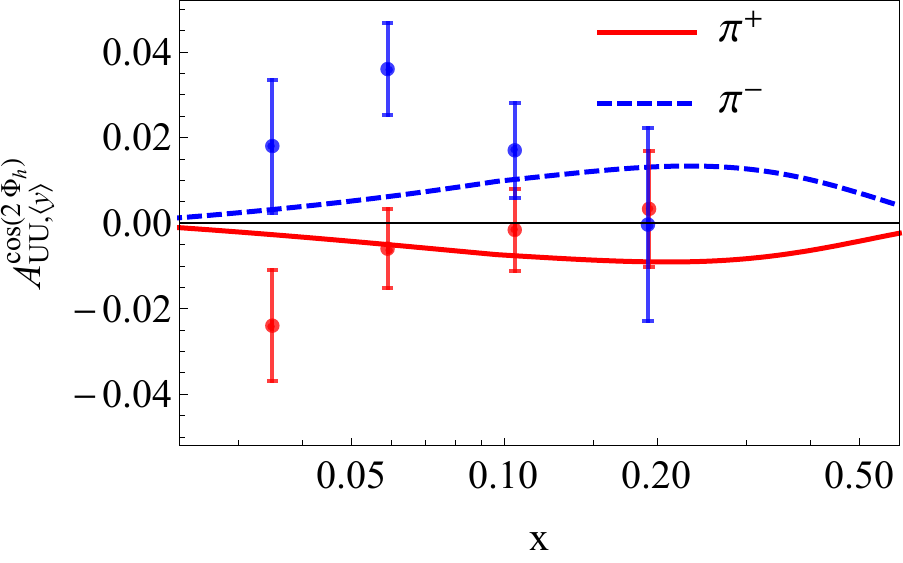}
\includegraphics[width=0.45\textwidth]{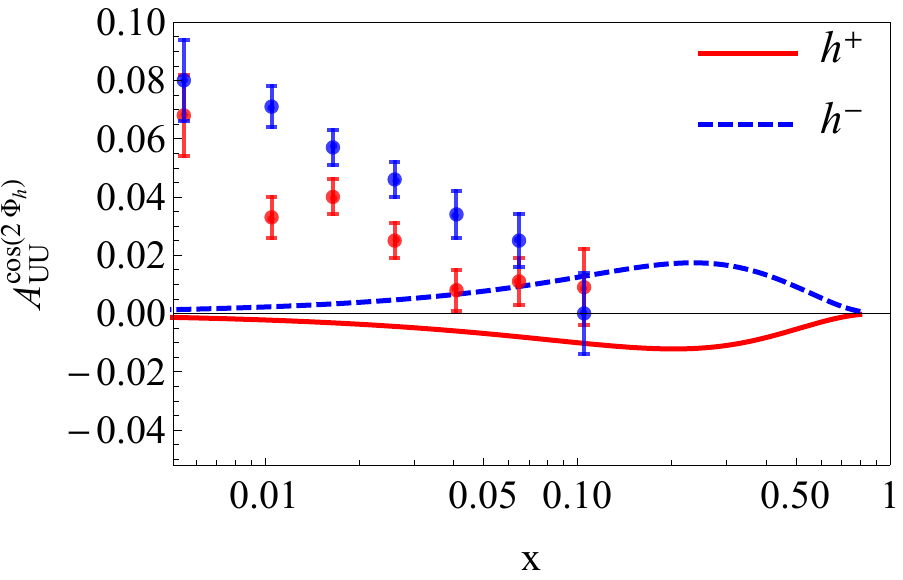}
\caption{\label{auu_cos2phi_jlab} The asymmetry
	$A_{UU, \langle y\rangle}^{\cos(2\phi_h)}$ for a proton target as function 
	of $ x $ based on the fit \cite{Barone:2009hw} in comparison to
	(left panel) HERMES \cite{Airapetian:2012yg} and $A_{UU}^{\cos(2\phi_h)}$
	(right panel) COMPASS data \cite{Adolph:2014pwc}.}
\end{figure}

The asymmetries  $A_{UU, \langle y\rangle}^{\cos(2\phi_h)}=\langle(1-y)F_{UU}^{\cos(2\phi_h)}\rangle/\langle(1-y + y^2/2)F_{UU}\rangle$
for HERMES \cite{Airapetian:2012yg} and $A_{UU}^{\cos(2\phi_h)}=F_{UU}^{\cos(2\phi_h)}/F_{UU}$ for COMPASS \cite{Adolph:2014pwc}
are plotted in Fig.~\ref{auu_cos2phi_jlab}, where we only considered the
Boer--Mulders contribution to $A_{UU}^{\cos(2\phi_h)}$, which does not describe
the data well. Especially for COMPASS one can see that calculation and
data are of opposite signs. In fact, it is suspected that this observable
receives a significant contribution from the Cahn effect \cite{Cahn:1978se},
a term of higher-twist character of the type $\la\Phperp^2\ra/Q^2$, which
is not negligible in fixed-target experiments as shown in
phenomenological \cite{Schweitzer:2010tt} and model \cite{Cao:2018qqf} studies.
In the phenomenological works
\cite{Barone:2009hw,Barone:2010gk,Barone:2015ksa}, an attempt was
made to estimate and correct for this contribution in order
to obtain a picture of the Boer--Mulders function undistorted
from Cahn effect. The point is that this substantial twist-4 contamination
can be estimated phenomenologically, even though there is no rigorous
theoretical basis for the description of such power-suppressed terms.
In this work we consistently neglect power-suppressed contributions of
order $1/Q^2$, and do so also in Fig.~\ref{auu_cos2phi_jlab}.
Nevertheless, we of course use the parametrizations of
\cite{Barone:2009hw,Barone:2010gk,Barone:2015ksa} offering
the best currently available parametrizations for $h_1^{\perp}$,
which were corrected for the Cahn effect as good as it is possible at
the current state of art. It is unknown whether other asymmetries
could be similarly effected by such type of power corrections.
This is an important point to be kept in mind as the lesson
from Fig.~\ref{auu_cos2phi_jlab} shows.

\subsection{\boldmath Leading-twist $A_{UT}^{\sin(3\phi_h-\phi_S)}$  asymmetry}
\label{Sec-5.6:pretzel-basis}

The pretzelosity TMD $h_{1T}^{\perp q}$ is the least known basis
function. It is of interest as it allows one to measure the deviation
of the nucleon spin density from spherical shape \cite{Miller:2007ae},
is related to the only leading-twist SIDIS structure function where the
small-$P_{hT}$ description in terms of TMDs and the large-$P_{hT}$ expansion
in perturbative QCD mismatch \cite{Bacchetta:2008xw}, and is the only TMD
where a clear relation to quark orbital angular momentum
could be established (albeit only within quark models)
\cite{Avakian:2008dz,She:2009jq,Avakian:2010br,Lorce:2011kn}.

The structure function $F_{UT}^{\sin(3\phi_h-\phi_S)}$ reads
\begin{subequations}\ba
	F_{UT}^{\sin(3\phi_h-\phi_S)}(x,z,\Phperp)
	&=& x \sum_q e_q^2\,h_{1T}^{\perp (2) q}(x)\,H_1^{\perp(1) q}(z)\;
	b^{(3)}\,\biggl(\frac{z \Phperp} {\lambda}\biggr)^{\!3}\,
	{ \cal G}(\Phperp) \, , \;\;\;\;
	\label{eq:asymmetry_aut_h1tp} \\
	F_{UT}^{\sin(3\phi_h-\phi_S)}(x,z,\la\Phperp\ra)
	&=& x\sum_q e_q^2\,h_{1T}^{\perp (2) q}(x)\,H_1^{\perp(1)q}(z)\;
	c^{(3)}\,\biggl(\frac{z} {\lambda^{1/2}}\biggr)^{3}\,,
\ea\end{subequations}
where $\lambda=z^2 \avkperp_{h_{1T}^\perp} + \avpperp_{H_1^\perp}$ and
$b^{(3)}=2M_N^2\mh$ and $c^{(3)}_{\rm  } = 3/2\sqrt{\pi} \,M_N^2\mh$,
see App.~\ref{App:convol-details}.
In Eq.~(\ref{eq:asymmetry_aut_h1tp}) we see that this structure
function suffers a cubic suppression for small transverse hadron
momenta.

\begin{figure}[t!]
\centering
\begin{tabular}{cc}
\includegraphics[height=3.5cm]{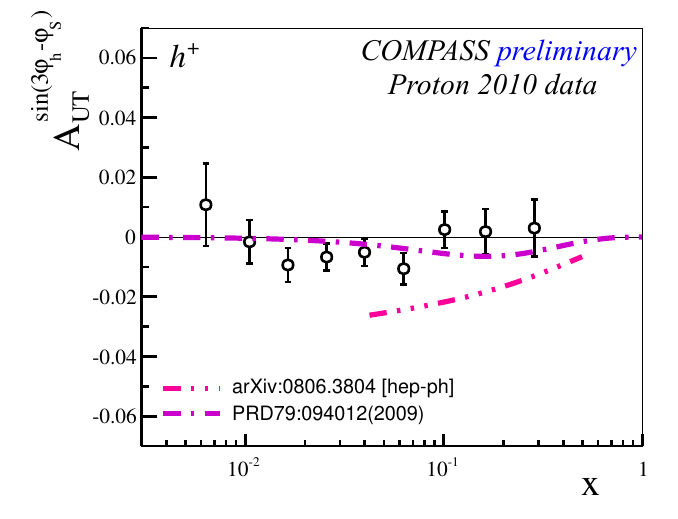}{\tiny(a)}&
\includegraphics[height=3.4cm]{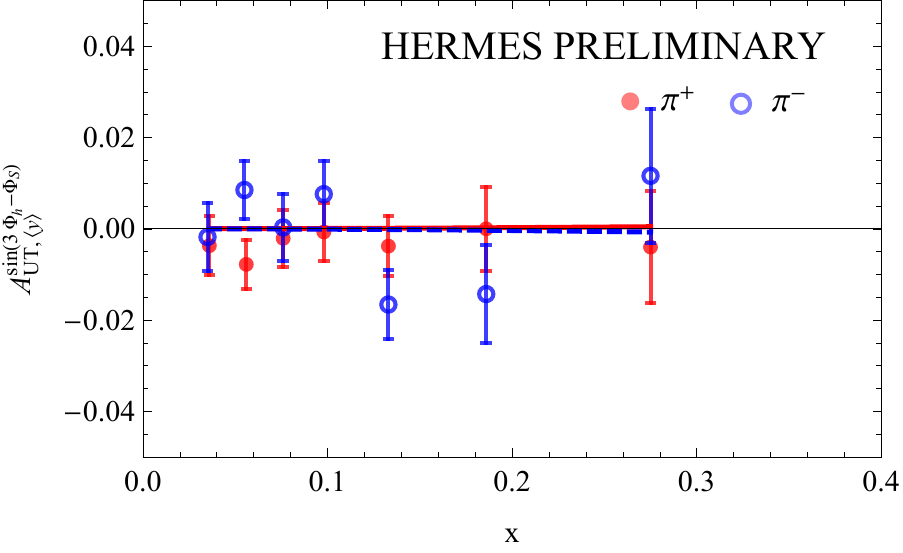}{\tiny(c)}
\\
\includegraphics[height=3.5cm]{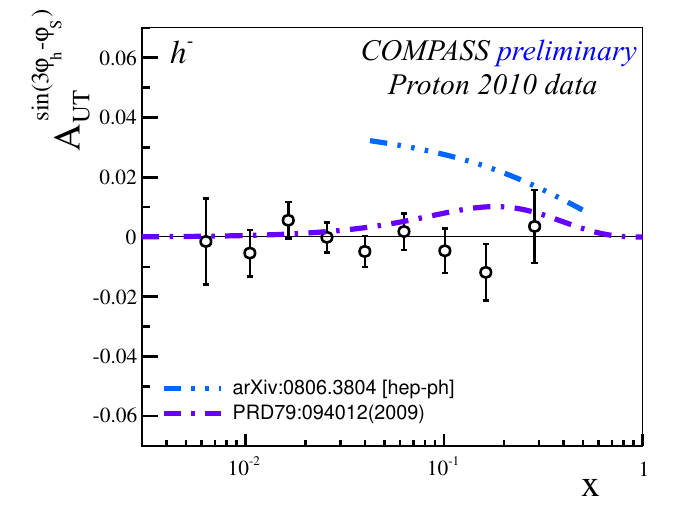}{\tiny(b)}&
\includegraphics[height=3.4cm]{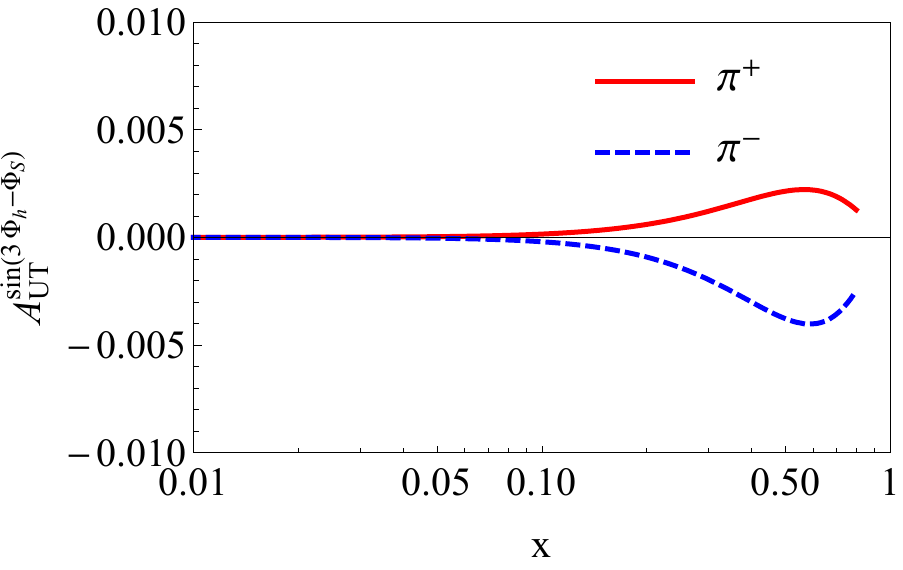}{\tiny(d)}
\end{tabular}
\caption{\label{aut_h1tp_jlab} $A_{UT}^{\sin(3 \phi_h - \phi_S)}$
	as a function of $x$ from preliminary COMPASS \cite{Parsamyan:2013fia} 
	(a,b) and $A_{UT,\langle y\rangle}^{\sin(3 \phi_h - \phi_S)}$ for HERMES
	\cite{Schnell:2010zza} (c) in comparison to the best fit from 
	\cite{Lefky:2014eia}
	(whose 1-$\sigma$ uncertainty band is compatible with zero), and 
	our calculation for COMPASS kinematics (d). For comparison the 
	COMPASS plots show the model results 
	\cite{Kotzinian:2008fe,Boffi:2009sh}.
        [We remark that in this and several subsequent figures we have 
        the permission to show the preliminary data \cite{Parsamyan:2013fia} 
	only in the official figures provided by the COMPASS collaboration
 	in (a,b), and have to display our results separately in (d).
	Notice also the different scale on the y-axis in panel (d) 
	as compared to (a,b).]}
\end{figure}

Preliminary COMPASS data \cite{Parsamyan:2013fia} for 
$A_{UT}^{\sin(3 \phi_h - \phi_S)}=F_{UT}^{\sin(3 \phi_h - \phi_S)}/F_{UU}$ and 
preliminary HERMES data \cite{Schnell:2010zza} for$
A_{UT, \langle y \rangle}^{\sin(3 \phi_h - \phi_S)}=\la
(1-y)F_{UT}^{\sin(3 \phi_h - \phi_S)}\ra/\la(1-y + y^2/2)F_{UU}\ra$
are plotted in Fig.~\ref{aut_h1tp_jlab}. 
Clearly, the pretzelosity TMD is the least known of the basis TMDs.
Nevertheless it is of fundamental importance, as it provides one of the
basis functions in our approach. It is so difficult to access it
experimentally, because its contribution to the SIDIS cross section
is proportional to $\Phperp^3$, the TMD approach requires us to
necessarily operate at $\Phperp\ll Q$, and so far only moderate
values of $Q$ could be achieved in the fixed target experiments.
Future high-luminosity data from JLab 12 are expected
to significantly improve our knowledge of this TMD.

For comparison Fig.~\ref{altcos2phihphis} shows also the results
from the quark-diquark model calculation of Ref.~\cite{Kotzinian:2008fe},
and from the light-cone constituent quark model of Ref.~\cite{Boffi:2009sh}.


\subsection{Statistical and systematic uncertainties of basis functions}

Even the well-known collinear functions $f_1^a(x)$ and $g_1^a(x)$ have
statistical uncertainties and systematic uncertainties (the latter
introduced by choosing a certain fit Ansatz, which however can be
avoided through neural-network techniques \cite{Ball:2014uwa}).
These uncertainties as well as those of $D_1^a(z)$ can safely
be neglected for our purposes.
For TMDs the situation is different. Already the transverse
momentum descriptions of $f_1^a(x,\kperp^{2})$ and $D_1^a(z,\pperp^{2})$
are associated with non-negligible statistical uncertainties,
which are reviewed in App.~\ref{App:basis}, and with systematic
uncertainties that are very difficult to assess as they are
related to model bias (because of Gaussian model and its limitations).
The statistical and systematic uncertainties are significant
when we deal with the basis functions
$f_{1T}^{\perp q}$, $h_{1}^{q}$, $H_{1}^{\perp q}$.
The least well-controlled uncertainties are associated with the
Boer--Mulders function $h_1^{\perp q}$ and pretzelosity $h_{1T}^{\perp q}$.

In this work we are not interested in these uncertainties, which
future data will allow us to diminish, even though in practice
they may be sizable. Rather in this work we are interested in
making projections based on the WW-type approximation. To avoid
cumbersome and difficult-to-interpret figures, we will therefore
refrain from indicating the uncertainties associated with our
current knowledge of the basis functions. In the following
we will only display the estimated systematic uncertainty
associated with the WW-type approximations {\it assuming}
it works within a relative accuracy of $\pm40\%$.
We stress that this is only to simplify the presentation. The
actual uncertainty of the predictions may be larger.

\section{Leading-twist asymmetries in WW-type approximation}
\label{Sec-6:twist-2-and-WW}

\begin{figure}[b!]
\centering
\includegraphics[width=0.45\textwidth]{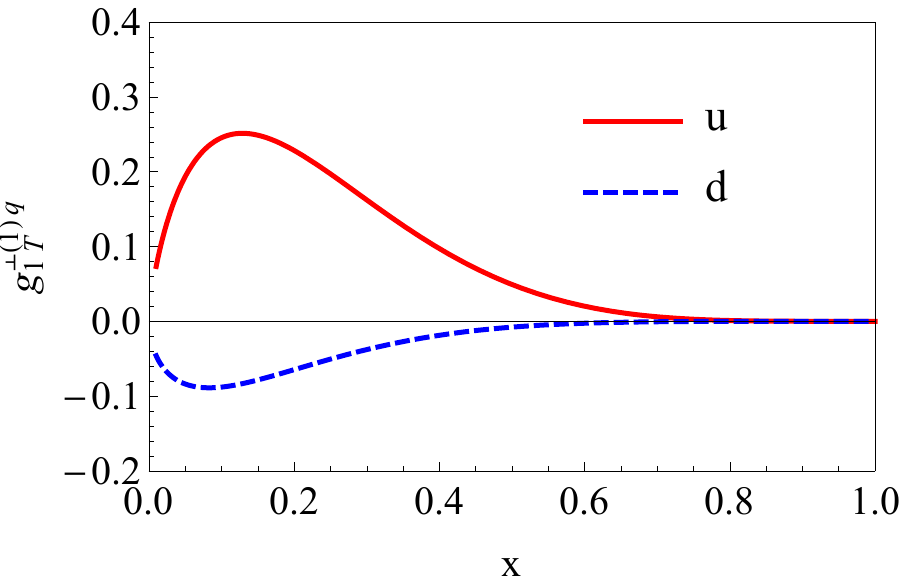} \quad
\includegraphics[width=0.45\textwidth]{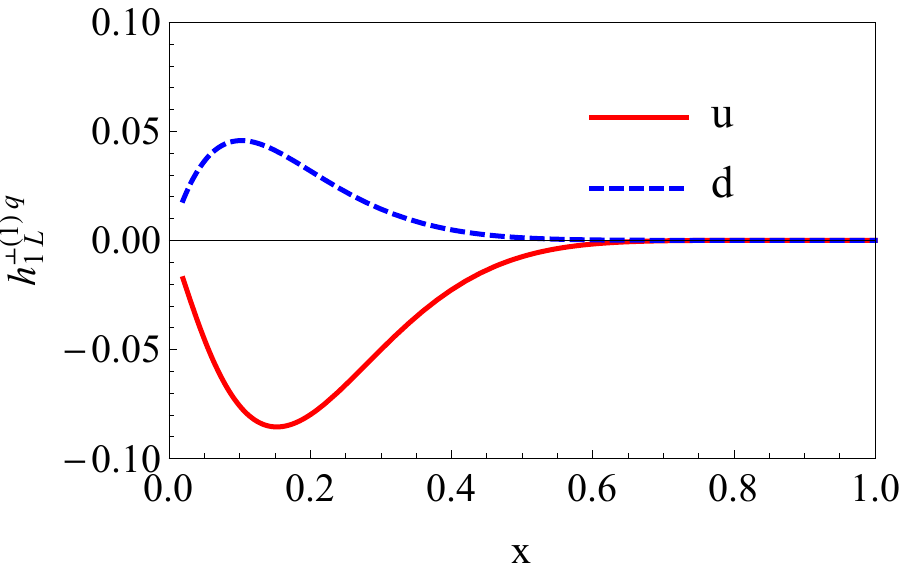}
	\caption{\label{g1t_h1l_functions}
	$g^{\perp (1) q}_{1T}(x)$ (left panel) and
	$h^{\perp (1) q}_{1L}(x)$ distributions (right panel)
	for up and down flavor
	as predicted by the WW-type approximations in
	Eqs.~(\ref{Eq:WW-approx-g1T},~\ref{Eq:WW-approx-h1L}).}
\end{figure}

Two out of the eight leading-twist structure functions
can be described in the WW-type approximation thanks to
Eqs.~(\ref{Eq:WW-approx-g1T},~\ref{Eq:WW-approx-h1L}), which
relate $g_{1T}^{\perp (1) q}(x)$ and $h_{1L}^{\perp (1) q}(x)$ to the
basis functions $g_1^q(x)$ and $h_1^a(x)$, see Fig.~\ref{g1t_h1l_functions}.
These TMDs are sometimes referred to as ``worm gear'' functions, where
$g_{1T}^{\perp q}$ describes the distributions of longitudinally
polarized quarks inside a transversely polarized nucleon,
and $h_{1L}^{\perp q}$ the opposite configuration, namely transversely
polarized quarks inside a longitudinally polarized nucleon.
It is interesting that both cases can be expressed in the
WW-type approximation in terms of the basis functions.
In this section we discuss the associated asymmetries.

\subsection{\boldmath Leading-twist $A_{LT}^{\cos(\phi_h-\phi_S)}$}
\label{Sec-6.1:FLTcosphi-phiS}

We assume for $g^{\perp}_{1T}$ the Gaussian Ansatz as shown
in (\ref{eq:g1t}) of App.~\ref{App-B:Gauss-Ansatz-non-basis-TMDs},
see also \cite{Kotzinian:2006dw}, and evaluate $g^{\perp (1) q}_{1T}(x)$
using (\ref{Eq:WW-approx-g1T}), which yields the result
shown in Fig.~\ref{g1t_h1l_functions}.
For our numerical estimates we use $\avkperp_{g_{1T}^\perp} = \avkperp_{g_{1}}$,
which is supported by lattice results \cite{Hagler:2009mb}.

In the Gaussian Ansatz the structure function $F_{LT}^{\cos(\phi_h -\phi_S)}$
has the form
\begin{subequations}\ba
	F_{LT}^{\cos(\phi_h -\phi_S)}(x,z,\Phperp)
	&=& x\sum_q e_q^2\,g_{1T}^{\perp (1) q}(x)\,D_1^{q}(z)\;
	b^{(1)}_{\rm B}\,\biggl(\frac{z \Phperp} {\lambda}\biggr)\,
	{ \cal G}(\Phperp )  \\
	F_{LT}^{\cos(\phi_h-\phi_S)}(x,z,\la\Phperp\ra)
	&=&  x\sum_q e_q^2\,g_{1T}^{\perp (1) q}(x)\,D_1^{q}(z)\;
	c^{(1)}_{\rm B}\,\biggl(\frac{z} {\lambda^{1/2}}\biggr)\,
	\label{eq:asymmetry_g1t_pt-II}
\ea\end{subequations}
where
$\lambda  = z^2 \avkperp_{g_{1T}^\perp} + \avpperp_{D_1}$,
$b^{(1)}_{\rm B} = 2M_N$,
$c^{(1)}_{\rm B} = \sqrt{\pi}\,M_N$,
see App.~\ref{App:convol-details} for details.

\begin{figure}
\centering
\begin{tabular}{ccc}\ \hspace{-8mm}
\includegraphics[width=0.31\textwidth]{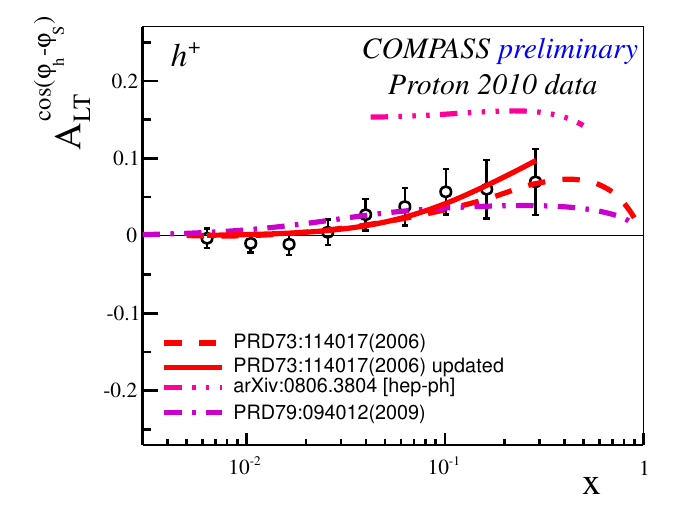}&
\ \hspace{-3mm}
\includegraphics[width=0.31\textwidth]{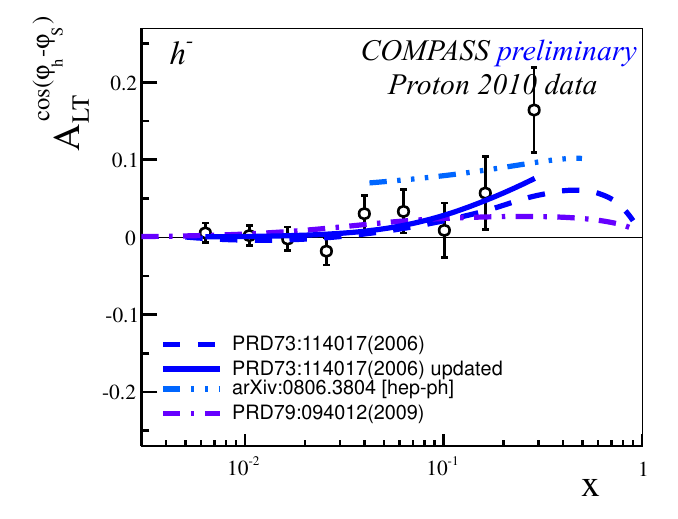}&
\ \hspace{-3mm}
\includegraphics[width=0.36\textwidth]{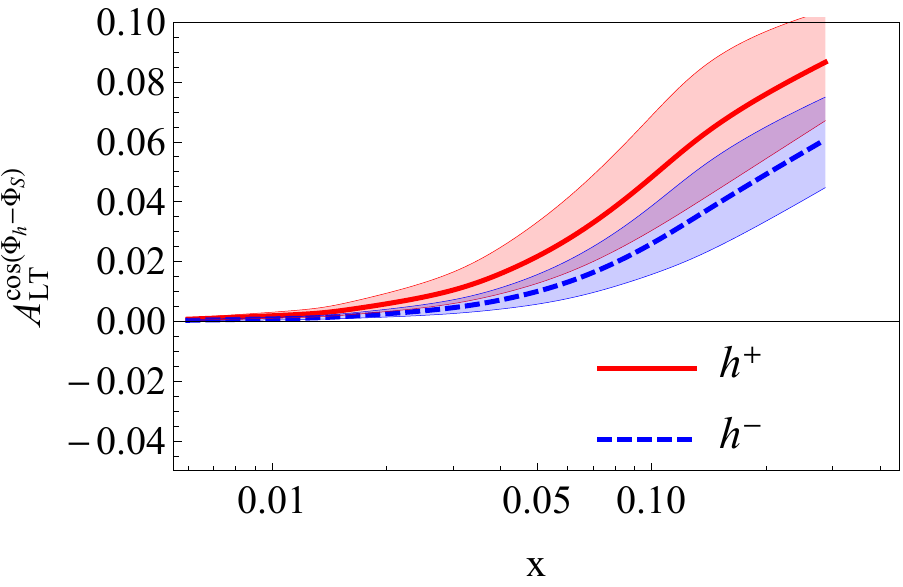} \\[-3mm]
 {\tiny (a)}&~~~~{\tiny (b)}&~~~~~~{\tiny (c)}
\end{tabular}
	\caption{\label{cosphihphis}
	Leading-twist $A_{LT}^{\cos(\phi_h-\phi_S)}$:
	preliminary COMPASS data \cite{Parsamyan:2013fia} (a,b); 
	and our calculation for COMPASS kinematics (c)
	shown separately for reasons explained in the
	caption of Fig.~\ref{aut_h1tp_jlab}.}
\end{figure}

This asymmetry was measured at JLab~\cite{Huang:2011bc}, COMPASS~\cite{
Parsamyan:2007ju,Parsamyan:2015dfa,Parsamyan:2018evv} and HERMES~\cite{Pappalardo:2011cu,Pappalardo:2012zz} (for the latter two experiments only preliminary results are available so far).
Figure~\ref{cosphihphis} shows the preliminary results from the 2010 COMPASS data \cite{Parsamyan:2013fia},
in addition to our calculation, where 
we approximate the charged hadrons (70--80~\% of which are $\pi^\pm$ at COMPASS) by
charged pions, see App.~\ref{App:basis-f1-D1}.
We observe that the WW-type approximation describes the data within
their experimental uncertainties.
For comparison also results from the theoretical works
\cite{Kotzinian:2006dw,Kotzinian:2008fe,Boffi:2009sh} are shown.
Our results are also compatible with the JLab data, which were
taken with a neutron ($^3$He) target \cite{Huang:2011bc} and have
larger statistical uncertainties than the preliminary COMPASS and HERMES data.


\subsection{\boldmath
	Leading-twist $A_{UL}^{\sin2\phi_h}$ Kotzinian--Mulders  asymmetry}
	\label{Sec-6.2:FULsin2phi}

We use the Gaussian form for the Kotzinian--Mulders function
$h_{1L}^{\perp a}$, (\ref{eq:h1l_final}) in
App.~\ref{App-B:Gauss-Ansatz-non-basis-TMDs}, with
$\avkperp_{h_{1L}^\perp} = \avkperp_{h_{1}}$ as
supported by lattice data \cite{Hagler:2009mb}.
From (\ref{Eq:WW-approx-h1L}) we obtain the WW-type estimate
for $h_{1L}^{\perp(1) a}(x)$ shown in Fig.~\ref{g1t_h1l_functions}.
The structure function $F_{UL}^{\sin(2\phi_h)}$ reads
\begin{subequations}\ba
	F_{UL}^{\sin(2\phi_h)}(x,z,\Phperp)
	&=&
	x \sum_q e_q^2\,h_{1L}^{\perp (1) q}(x)\,H_1^{\perp(1) q/h}(z)
	\biggl(\frac{z\Phperp}{\lambda}\biggr)^{\!\!2} \;
	b^{(2)}_{\rm AB}\;{ \cal G}(\Phperp )\, , \\
	F_{UL}^{\sin(2\phi_h)}(x,z,\la\Phperp\ra)
	&=&
	x \sum_q e_q^2\,h_{1L}^{\perp (1) q}(x)\,H_1^{\perp(1) q/h}(z)
	\biggl(\frac{z}{\lambda^{1/2}}\biggr)^{\!\!2} \;
	c^{(2)}_{\rm AB}\,,
\ea\end{subequations}
where $\lambda= z^2 \avkperp_{h_{1L}^\perp} + \avpperp_{H_1^\perp}$ and
$b^{(2)}_{\rm AB}=c^{(2)}_{\rm AB}=4M_N\mh$,
see App.~\ref{App:convol-details} for details.

\begin{figure}
\centering \ \hspace{-6mm}
\includegraphics[height=5cm]{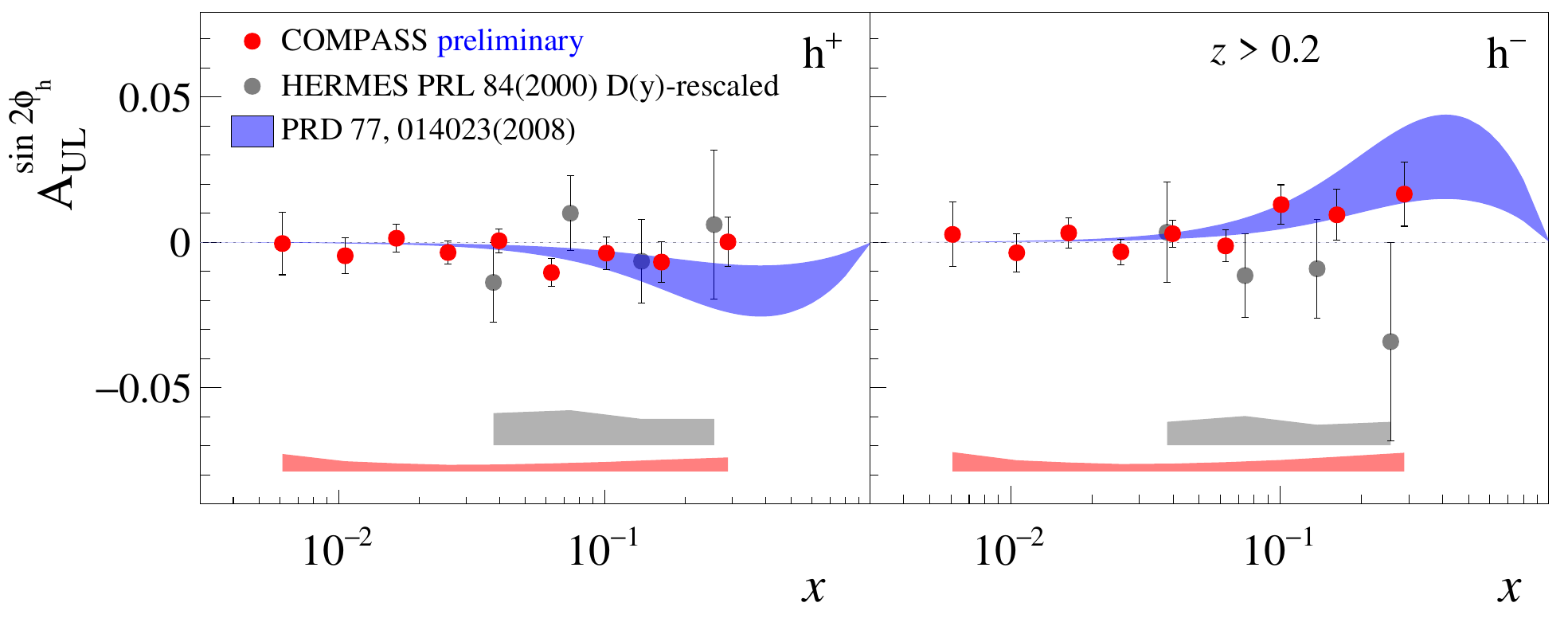} \\

\vspace{-0.5cm}

\ \hspace{1cm} {\tiny (a)} \hspace{6cm} {\tiny (b)} 

\vspace{1cm}

\includegraphics[height=3.6cm]{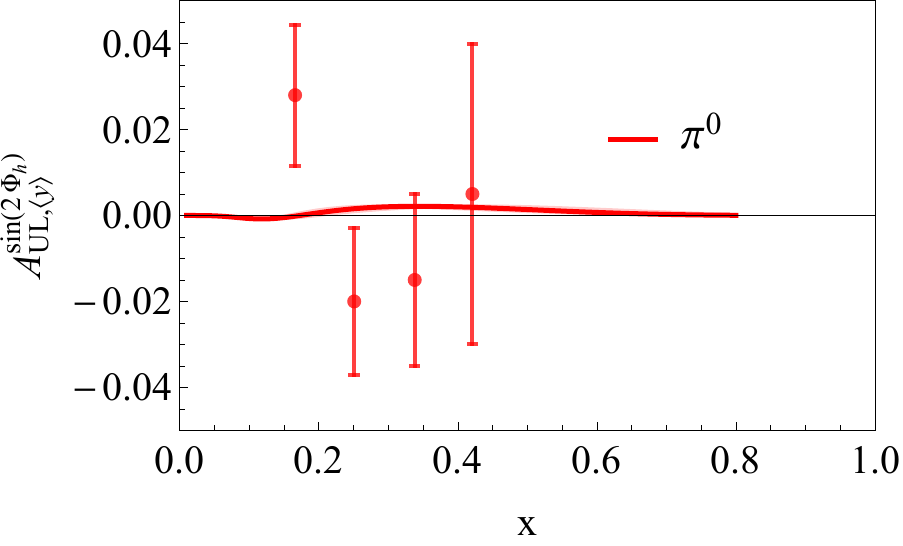} \
\includegraphics[height=3.6cm]{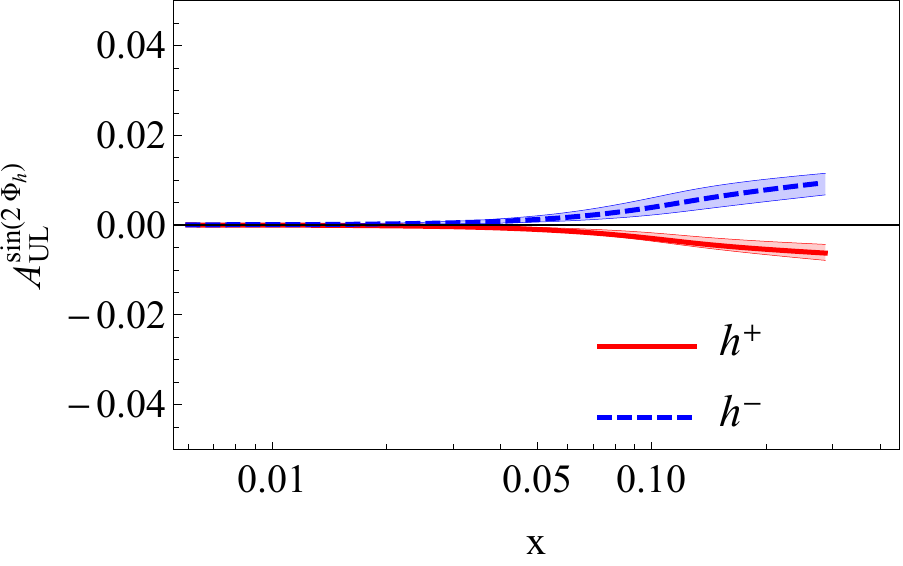} \\
\ \hspace{1cm} {\tiny (c)} \hspace{6cm} {\tiny (d)}
	\caption{\label{aul_jlab}
	Leading twist $A_{UL}^{\sin(2\phi_h)}$ vs.~$x$ from
	HERMES \cite{Airapetian:1999tv} and 
	COMPASS \cite{Parsamyan:2018ovx,Parsamyan:2018evv} (a, b),
	$A_{UL,  \langle y\rangle}^{\sin(2\phi_h)}$ from
	JLab \cite{Jawalkar:2017ube} (c),  and from our calculation 
	for COMPASS kinematics in the WW-type approximation (d)
	shown separately for reasons explained in the
	caption of Fig.~\ref{aut_h1tp_jlab}. Results
	from \cite{Avakian:2007mv} are also shown in comparison to the 
	COMPASS data.}
\end{figure}

The asymmetry $A_{UL}^{\sin(2\phi_h)}=F_{UL}^{\sin(2\phi_h)}/F_{UU}$  was
studied at HERMES \cite{Airapetian:1999tv,Airapetian:2002mf}, COMPASS
\cite{Parsamyan:2018ovx,Parsamyan:2018evv}, and JLab \cite{Avakian:2010ae,Jawalkar:2017ube}.
In Fig.~\ref{aul_jlab} proton data are shown for $\pi^\pm$ in the
HERMES experiment measured with the 27.6 GeV positron
beam of HERA for
$1\,{\rm GeV}^2 < Q^2 < 15 \,{\rm GeV}^2$, $W > 2\,{\rm GeV}$,
$0.023 < x < 0.4$, and $y < 0.85$.
The COMPASS data were taken in 2007 (160 GeV) and 2011 (200 GeV) and show
the asymmetry for charged hadrons (in practice mainly pions).
Since $y$--dependent prefactors were included in the analyses
(see Sec.~\ref{Sec-2.1:SIDIS+structure-functions}),
the HERMES data are adequately (``$D(y)$~--'')rescaled.
The CLAS $\pi^0$ data in the right panel were measured using 6$\,$GeV
longitudinally polarized electrons scattering off
longitudinally polarized protons in a cryogenic $^{14}$NH$_3$
target in the kinematic region of $1.0\,{\rm GeV}^2 < Q^2 < 3.2\,{\rm GeV}^2$,
$0.12 < x < 0.48$ and $0.4 < z < 0.7$ \cite{Jawalkar:2017ube}.

$A_{UL}^{\sin(2\phi_h)}$  can be expected to be smaller than
$A_{LT}^{\cos(\phi_h -\phi_S)}$ discussed in Sec.~\ref{Sec-6.1:FLTcosphi-phiS},
even though both are leading twist. This is because
$F_{UL}^{\sin(2\phi_h)}$ is quadratic in the hadron transverse
momenta $P_{hT}$, while $F_{LT}^{\cos(\phi_h -\phi_S)}$ is linear.
In addition, the former is proportional to the Collins function,
which is smaller than $D_1^q(z)$, and the WW-type approximation
predicts the magnitude of $h_{1L}^{\perp(1)q}(x)$ to be about half
of the size of $g_{1T}^{\perp(1)q}(x)$.
The data support these expectations. HERMES and JLab data are compatible
with zero for this asymmetry. So are the preliminary COMPASS data except
for the region
$x>0.1$ for negative hadrons, where the trend of the data provides a first
encouraging confirmation for our results. Thus current data are compatible
with the WW-type approximation for $h_{1L}^{\perp(1)q}(x)$.

\subsection{Inequalities and a cross check}

We discussed WW-type approximations for the twist-2 TMDs
$g^{\perp q}_{1T}$ and $h^{\perp q}_{1L}$
in Secs.~\ref{Sec-6.1:FLTcosphi-phiS}, \ref{Sec-6.2:FULsin2phi}.
Before proceeding with twist-3 let us pause and revisit positivity
bounds \cite{Bacchetta:1999kz}. 

The Kotzinian--Mulders function $h^{\perp q}_{1L}$ in conjunction with
the Boer--Mulders function, and the TMD $g^{\perp q}_{1T}$ in conjunction
with the Sivers function obey the positivity bounds
\cite{Bacchetta:1999kz}
\begin{subequations}\ba
	\frac{\kperp^2}{4M_N^2}\;
	\left((f_{1}^{q}(x,\kperp^2))^2 -(g_{1}^{q}(x,\kperp^2))^2\right)
	- (h^{\perp(1)q}_{1L}(x,\kperp^2))^2
	- (h_{1}^{\perp(1)q}(x,\kperp^2))^2
	& \ge & 0\,, \quad \label{eq:positivity}\\
	\frac{\kperp^2}{4M_N^2}\;
	\left((f_{1}^{q}(x,\kperp^2))^2 -(g_{1}^{q}(x,\kperp^2))^2\right)
	- (f_{1T}^{\perp(1)q}(x,\kperp^2))^2
	- (g^{\perp(1)q}_{1T}(x,\kperp^2))^2
	& \ge & 0\,, \quad \label{eq:positivity1}
\ea\end{subequations}
where $f^{(1)}(x,\kperp^2) \equiv \frac{\kperp^2}{2M_N^2} f(x,\kperp^2)$.
The inequalities provide a non-trivial test of our approach.
The inequalities (\ref{eq:positivity},~\ref{eq:positivity1})
must be strictly satisfied by the {\it exact} leading-order QCD
expressions for the TMDs.
(For PDFs it is known that positivity can be preserved in some
schemes and violated in others. For TMDs not much is
known about positivity conditions beyond leading order.)
However, here we do not deal with exact TMDs but (i) we invoked
strong model assumptions (WW-type approximations for $g^{\perp q}_{1T}$
and $h^{\perp q}_{1L}$ and Gaussian Ansatz for all TMDs), and (ii) we deal
with first extractions, which have sizable uncertainties including
poorly controlled systematic uncertainties.
Therefore, considering that we deal with approximations
(WW-type, Gauss) and considering the current state of TMD extractions,
the inequalities (\ref{eq:positivity},~\ref{eq:positivity1}) constitute
a challenging test for the approach.

\begin{figure}[b!]
\centering
\includegraphics[width=0.45\textwidth]{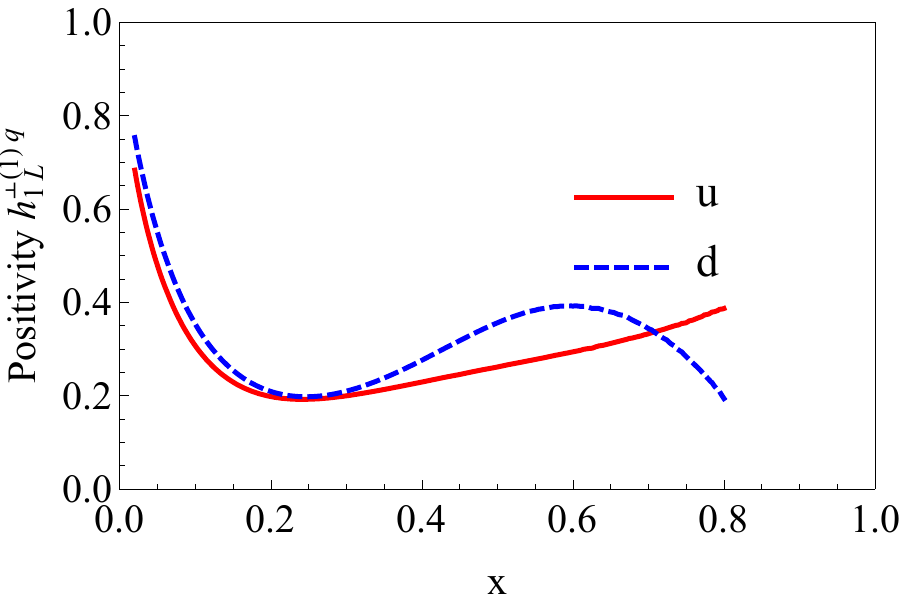} \quad
\includegraphics[width=0.45\textwidth]{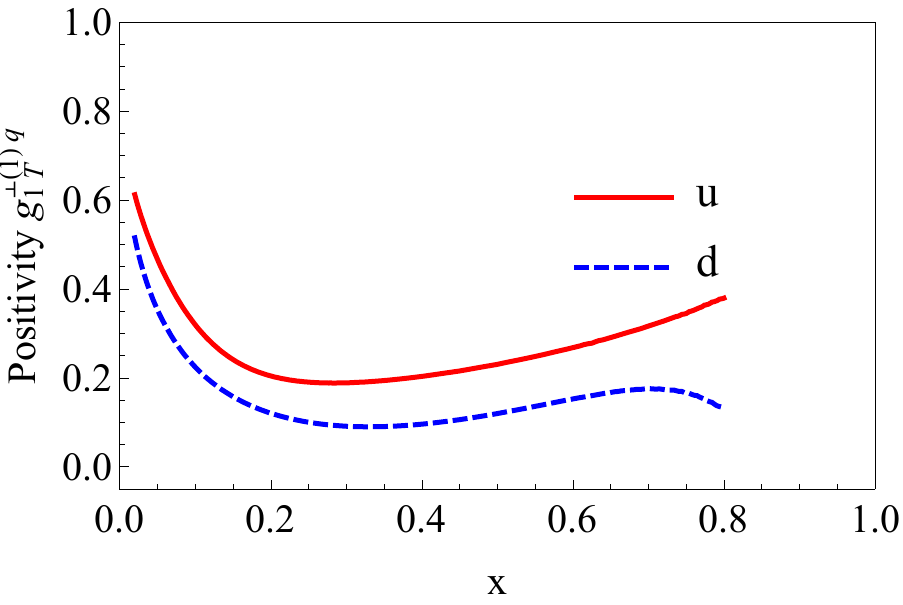}
	\caption{\label{h1l_pos}
	The normalized inequalities for $g^{\perp(1)q}_{1T}(x)$ and
	$h^{\perp(1)q}_{1L}(x)$ vs.~$x$ which are obtained by integrating
	(\ref{eq:positivity}) and (\ref{eq:positivity1}) over $\kperp$
	and normalizing with respect to the sum of the absolute
	values of the individual terms. The result must be positive and
	smaller than unity {\it if} the WW-type approximations
	and the application of the Gaussian model are compatible
	with positivity, see text. Clearly, our approach respects
	this test of the positivity conditions.}
\end{figure}

In order to conduct a test we use the Gaussian Ansatz
(\ref{Eq:Gauss-f1}, \ref{Eq:Gauss-h1}, \ref{Eq:Gauss-f1Tperp},
\ref{Eq:Gauss-h1perp}, \ref{eq:g1t}, \ref{eq:h1l_final}) for the
TMDs and integrate over $k_\perp$. The results are shown in
Fig.~\ref{h1l_pos}, where we plot the ``normalized inequalities''
defined as follows:
given an inequality $a-b-\dots \ge 0$, the normalized inequality
is defined as: $0 \le (a-b-\dots)/(|a|+|b|+\dots) \le 1$.

Figure~\ref{h1l_pos} shows that the results of our approach for the
``normalized inequalities'' for both TMDs lie between
zero and one, as it is dictated by positivity constraints.
This is an important consistency cross-check for our approach.

\section{Subleading-twist asymmetries in WW-type approximation}
\label{Sec-7:twist-3-and-WW}

WW-type approximations can be applied to all eight twist-3 asymmetries,
see Sec.~\ref{Sec-4.2:WW-twist-3}. In this section we discuss all of them,
starting with less complex cases and proceeding then to those structure
functions whose description in WW-type approximation is more involved.

\subsection{\boldmath Subleading-twist  $A_{LU}^{\sin\phi_h}$}
\label{Sec-7.1:FLU}

We start our discussion with the structure function $F_{LU}^{\sin\phi_h}$,
Eq.~(\ref{FLUsinphi}), containing four terms:
two are proportional to the pure twist-3 fragmentation functions
$\tilde{G}^{\perp a}$ and $\tilde{E}^a$; 
the other two
terms are proportional to the twist-3 TMDs $e^a$ and $g^{\perp a}$, which
also turn out to be given in terms of pure twist-3 terms upon the
inspection of (\ref{Eq:WW-type-1},~\ref{Eq:WW-type-gperp}).
Hence, after consequently applying the WW-type approximation, we are left
with no term. Our approximation predicts this structure function to be zero.

In this asymmetry we encounter the generic limitation of the
WW-type approximation in most extreme form. As discussed in
Sec.~\ref{Sec-3.8:limitations}, if we have a function
$=\la\bar{q}q\ra + \la\bar{q}gq\ra$ the necessary condition for
the approximation to work is that $\la\bar{q}q\ra \neq 0$ and the
sufficient condition would be $|\la\bar{q}q\ra|\gg|\la\bar{q}gq\ra|$.
Remarkably, none of the twist-3 TMDs or FFs entering this structure
function satisfy even the necessary condition. In this situation we
do not expect the WW-type approximation to work.

Indeed, data from COMPASS, HERMES, and JLab show a clearly non-zero
asymmetry $A_{LU}^{\sin\phi_h} = F_{LU}^{\sin\phi_h}/F_{UU}$ of the order of 2\,$\%$
\cite{Avakian:2003pk,Airapetian:2006rx,Gohn:2009zz,
Aghasyan:2011ha,Adolph:2014pwc,Gohn:2014zbz} (which includes
the $1/Q$ factor intrinsic in twist-3 observables; without this factor
the asymmetry would reach $10\%$ at large $x$ at COMPASS).
This observable provides a unique opportunity to learn about the
physics of $\bar{q}gq$-terms, but is beyond the applicability of
the WW-type approximation.

With the numerator of the asymmetry proportional to $\bar{q}gq$--terms
and the denominator given in terms of $\bar{q} q$--terms, one could be
tempted to interpret this observation as
\be\label{Eq:ALU-small}
    	A_{LU}^{\sin\phi_h}
	\;\; \propto \;\;\frac{\la\bar{q}gq\ra}{\la\bar{q}q\ra}
    	\;\;\stackrel{\rm exp}{\sim} \;\;
	{\cal O}(2\,\%)
    	\;.
\ee
Thus, in some sense the observed effect hints at the smallness of the
involved $\bar{q}gq$--terms. While in principle a correct observation,
one should keep in mind several reservations. First,
the experimental result (\ref{Eq:ALU-small})
contains kinematic prefactors, which help to reduce the value.
Second, the denominator contains $f_1^a$ and $D_1^a$, which are the
largest TMD and FF because of positivity constraints. Third, the
numerator is a sum of four terms, so its overall smallness could result
from cancellation of different terms, rather than indicating that
each single $\bar{q}gq$--term is small. Fourth, some asymmetries
predicted to be non-zero in WW-approximation are not larger and
in some cases even smaller than the result in (\ref{Eq:ALU-small}).

To conclude, the WW-type approximation predicts $A_{LU}^{\sin\phi}\approx 0$
in contradiction to experiment. The size of the observed effect seems
in line with the WW-type approximation, as
$A_{LU}^{\sin\phi_h}\sim\la\bar{q}gq\ra/\la\bar{q}q\ra \sim {\cal O}(2\,\%)$
is not large, although this interpretation has reservations.
$F_{LU}^{\sin\phi}$ is the only twist-3 SIDIS structure function not
``contaminated'' by leading twist. Attempts to describe this observable
and relevant model studies have been reported
\cite{Efremov:2002qh,Schweitzer:2003uy,Ohnishi:2003mf,Cebulla:2007ej,
Efremov:2002ut,Afanasev:2003ze,Yuan:2003gu,Gamberg:2003pz,Metz:2004je,Afanasev:2006gw,Mao:2012dk,Mao:2013waa,Mao:2014dva,Lorce:2014hxa,Courtoy:2014ixa,Yang:2016mxl, Pasquini:2018oyz}.
But more phenomenological work and dedicated studies on the basis of models
of $\bar{q}gq$ terms are needed to fully understand this asymmetry.

\subsection{\boldmath Subleading-twist  $A_{LT}^{\cos\phi_S}$}
\label{Sec-7.2:FLTcosphiS}

In the WW-type approximation the structure function
$F_{LT}^{\cos\phi_S}$ arises from $g_T^a(x,\kperp^{2})$ and $D_1^a(z,\pperp^{2})$,
whose collinear counterparts are more or less known, see
Secs.~\ref{Sec-3.4:WW-classic-experiment} and \ref{Sec-5.1:FUU-basis}.
We assume the Gaussian Ansatz for $g_T^a(x,\kperp^{2})$, shown in
(\ref{eq:gtnew}) of App.~\ref{App-B:Gauss-Ansatz-non-basis-TMDs},
with ${\avkperp_{g_T}}={\avkperp_{g_1}}$. We then determine $g^{q}_{T}(x)$
according to (\ref{Eq:WW-original1}), which is a well-tested
approximation in DIS, see Sec.~\ref{Sec-3.4:WW-classic-experiment}.
In this way we obtain for $F_{LT}^{\cos\phi_S}$ the result
\begin{subequations}\ba
	F_{LT}^{\cos\phi_S}(x,z,\Phperp)
	&=& -\frac{2M_N}{Q}\; x^2 \sum_q e_q^2\,g_T^q(x)\,D_1^q(z)\;
	{ \cal G}(\Phperp)\, , \label{Eq:FLTcosphiS-Gauss}\\
	F_{LT}^{\cos\phi_S}(x,z)
	&=& -\frac{2M_N}{Q}\; x^2 \sum_q e_q^2\, g_T^q(x)\,D_1^q(z)\, ,
	\label{Eq:FLTcosphiS-Gauss-b}
\ea\end{subequations}
with the width $\lambda= z^2 \avkperp_{g_T} + \avpperp_{D_1}$
in the Gaussian ${\cal G}(\Phperp)$.
\begin{figure}[b!]
\centering
\begin{tabular}{ccc} \ \hspace{-8mm}
\includegraphics[height=3.3cm]{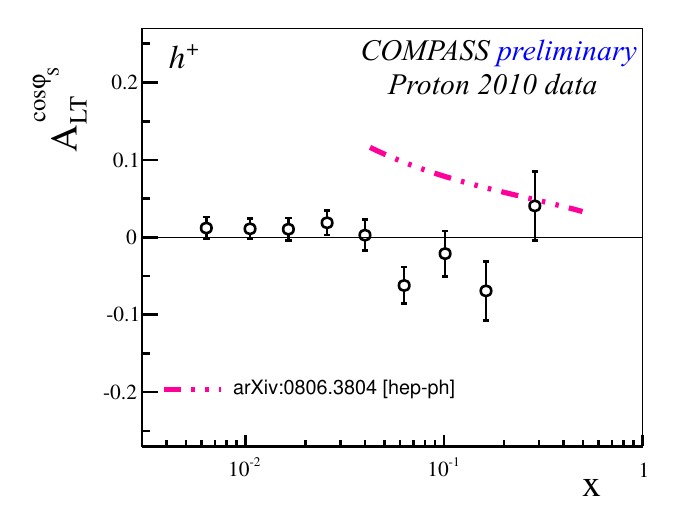}&
\includegraphics[height=3.3cm]{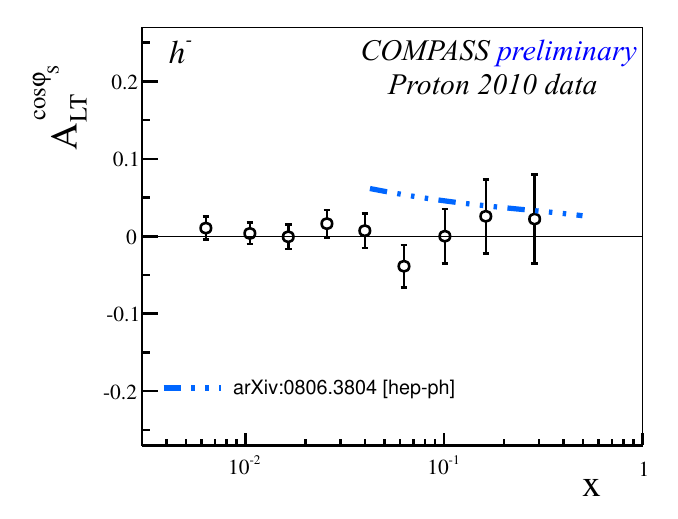}&
\includegraphics[height=3.1cm]{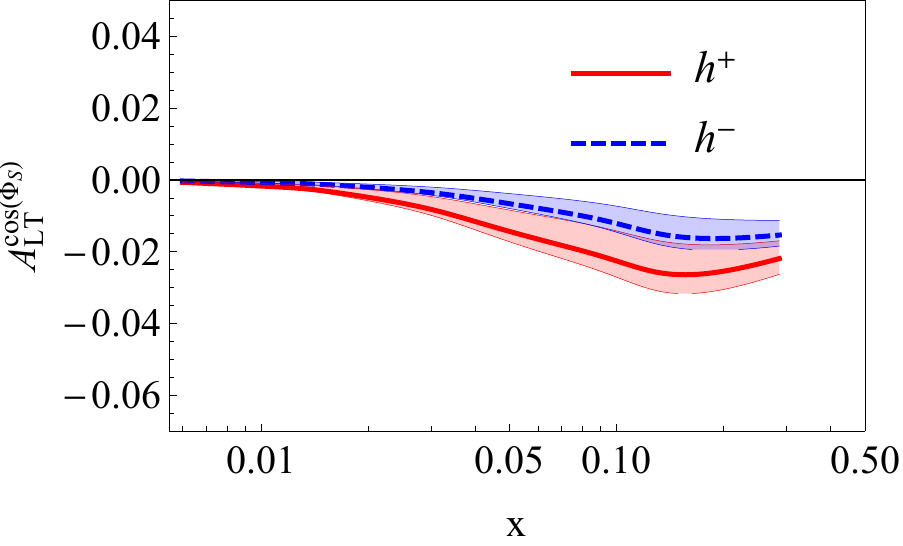}\\
{\tiny(a)}& {\tiny(b)}& {\tiny(c)}
\end{tabular}
\caption{\label{altcosphis} Subleading-twist asymmetry $A_{LT}^{\cos\phi_S}$
	as function of $x$ from scattering of 160 GeV
	longitudinally polarized muons off a transversely polarized
	proton target \cite{Parsamyan:2013fia} (a,b),  and our 
	calculation for COMPASS kinematics (c) where a different 
	scale is chosen to better visualize the theory curves.}
\end{figure}

Notice that we followed here the scheme explained in
Sec.~\ref{Sec-4.4:evaluation}: first assume Gaussian Ansatz, then apply
WW-type approximation. For some structure functions the order of
these steps is not relevant, but here it is. It is instructive to
discuss what the opposite order of these steps would yield.
Using first the WW-type approximation (\ref{Eq:WW-type-gT})
in the convolution integral (\ref{Eq:WW-original1}) and then
using the Gaussian Ansatz yields a bulkier
analytical expression than (\ref{Eq:FLTcosphiS-Gauss}) though
a numerically similar result. But there are two critical issues
with that. First, the WW-type approximation (\ref{Eq:WW-type-gT})
relates $xg^{q}_{T}(x,\kperp^2)=g^{\perp(1)q}_{1T}(x,\kperp^2)$ which
would imply $g^{q}_{T}(x, \kperp^2) \to 0$ for $\kperp\to0$
due to the weight $\kperp^2/(2M_N^2)$ in the (1)-moment,
which is not supported by model calculations \cite{Avakian:2010br}.
Second, the more economic (because
less bulky) expression in (\ref{Eq:FLTcosphiS-Gauss}) automatically
yields (\ref{Eq:FLTcosphiS-Gauss-b}), which is the correct collinear
result for this SIDIS structure function in (\ref{Eq:FLT-collinear})
in WW-type approximation. This technical remark confirms the
consistency of the scheme suggested in Sec.~\ref{Sec-4.4:evaluation}.

Figure~\ref{altcosphis} shows our predictions
for $A_{LT}^{\cos\phi_S}$ in comparison to preliminary
COMPASS data~\cite{Parsamyan:2013fia}. The predicted
asymmetry is small and compatible with the COMPASS data within
uncertainties.
Preliminary HERMES data~\cite{Pappalardo:2012zz} confirm a small asymmetry.
More precise data are necessary to judge how
well the WW-type approximation works in this case. Such data
could come from the JLab12 experiments.
For further model studies of this asymmetry we refer to
Refs.~\cite{Mao:2014fma,Wang:2016dti}.

\subsection{\boldmath Subleading-twist  $A_{LT}^{\cos(2\phi_h - \phi_S)}$}
\label{Sec-7.3:FLTcos2phi-phiS}

In the WW-type approximation this asymmetry is expressed in terms of
$g_T^{\perp a}(x,\kperp^{2})$, for which we assume a Gaussian Ansatz according to
(\ref{eq:gtperpnew}) in App.~\ref{App-B:Gauss-Ansatz-non-basis-TMDs},
and use the WW-type approximation (\ref{Eq:WW-type-gTperp}) as
\be
	xg_T^{\perp(2)q}(x) = \frac{\la\kperp^2\ra_{g_{1T}^\perp}}{M_N^2}\;
	g_{1T}^{\perp (1)q}(x)\,,
\ee
where we finally express $g_{1T}^{\perp (1)q}(x)$ in terms of $g_1^q(x)$
according to Eq.~(\ref{Eq:WW-approx-g1T}). For the Gaussian widths
we assume $\avkperp_{g_{T}^\perp}=\avkperp_{g_{1T}^\perp}=\avkperp_{g_1}$.
This yields for the structure function
\begin{subequations}\ba
	F_{LT}^{\cos(2\phi_h - \phi_S)}(x,z,\Phperp)
	&=& -\frac{2M_N}{Q} x \sum_q e_q^2\,x\,g_{T}^{\perp (2) q}(x)\,D_1^q(z)\;
	b^{(2)}_{C}\,\biggl(\frac{z \Phperp} {\lambda}\biggr)^{\!2}\!
	{ \cal G}(\Phperp), \quad \\
	F_{LT}^{\cos(2\phi_h - \phi_S)}(x,z,\la\Phperp\ra)
	&=& -\frac{2M_N}{Q} x \sum_q e_q^2\,x\,g_{T}^{\perp (2) q}(x)\,D_1^q(z)\;
	c^{(2)}_{C}\,\biggl(\frac{z} {\lambda^{1/2}}\biggr)^{\!2}\,
	\label{eq:asymmetry_auu_cos2phi-phiS}
\ea\end{subequations}
where $\lambda=z^2 \avkperp_{g_{T}^\perp} + \avpperp_{D_1}$ and
$b^{(2)}_{\rm C}=c^{(2)}_{\rm C} = M_N^2$,
see App.~\ref{App:convol-details} for details.

\begin{figure}[b!]
\centering
\begin{tabular}{ccc} \ \hspace{-8mm}
\includegraphics[width=0.30\textwidth]{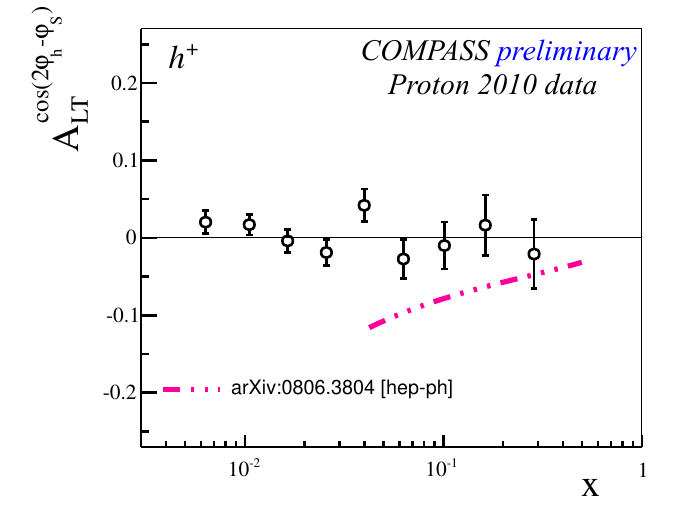}&
\includegraphics[width=0.30\textwidth]{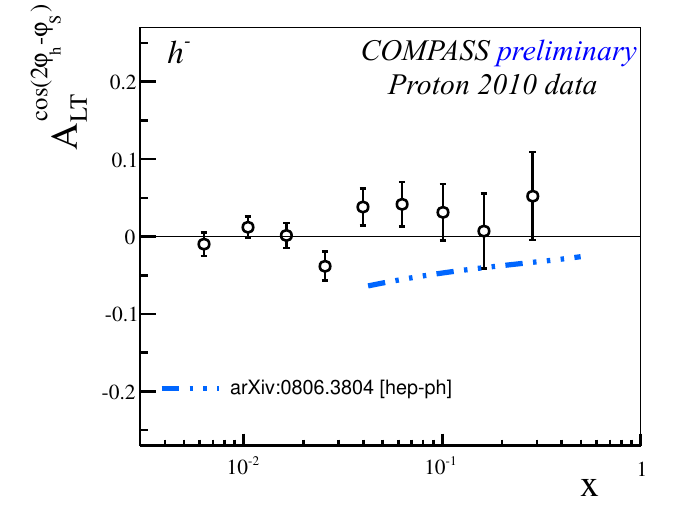}&
\includegraphics[width=0.36\textwidth]{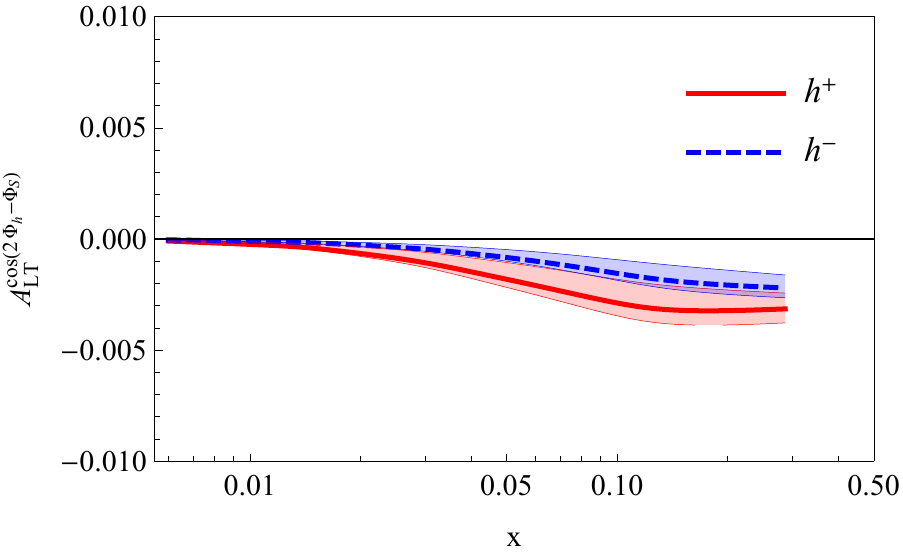}\\
{\tiny(a)}& {\tiny(b)}& {\tiny(c)}
\end{tabular}
\caption{\label{altcos2phihphis} $A_{LT}^{\cos(2\phi_h - \phi_S)}$
	as a function of $ x $ from a proton target: 
	preliminary COMPASS data \cite{Parsamyan:2013fia} (a,b), 
	and our calculation for COMPASS kinematics (c)
	shown separately for reasons explained in the
	caption of Fig.~\ref{aut_h1tp_jlab}.}
\end{figure}

Our predictions for the asymmetry
$A_{LT}^{\cos(2\phi_h -\phi_S)}=F_{LT}^{\cos(2\phi_h -\phi_S)}/F_{UU}$ as a function
of $x$ are displayed in Fig.~\ref{altcos2phihphis} in
addition to preliminary COMPASS data from Ref.~\cite{Parsamyan:2013fia}.
The asymmetry is very small, so at the current stage one may
conclude that the WW-type approximation for the asymmetry
$A_{LT}^{\cos(2\phi_h -\phi_S)}$ is compatible with available data. In view
of the smallness of the effect (cf.~Fig.~\ref{altcos2phihphis}),
it might be difficult to obtain more quantitative insights
in the near future.

For completeness, Fig.~\ref{altcos2phihphis} shows also results from a
quark-diquark model calculation, where a more sizable asymmetry was predicted
\cite{Kotzinian:2008fe}. Let us remark that the $A_{LT}^{\cos(2\phi_h -\phi_S)}$
asymmetry was also studied in a different spectator model in
Ref.~\cite{Mao:2014fma} predicting asymmetries of the same
sign but of smaller magnitudes of ${\cal O}(1\,\%)$.

\subsection{\boldmath   Subleading-twist  $A_{LL}^{\cos\phi_h}$}
\label{Sec-7.3:FLLcosphi}

In WW-type approximation the only contribution to $F_{LL}^{\cos\phi_h}$
is due to $g_{L}^{\perp q}(x,k_\perp^{2})$, which we assume to have a
Gaussian $k_\perp$--behavior according to (\ref{eq:gLperp})
in App.~\ref{App-B:Gauss-Ansatz-non-basis-TMDs} with
the Gaussian width $\avkperp_{g_{L}^\perp}=\avkperp_{g_1}$.
The structure function $F_{LL}^{\cos\phi_h}$ reads
\begin{subequations}\ba
	F_{LL}^{\cos \phi_h}(x,z,\Phperp)
	&=& -\,\frac{2M_N}{Q}\;x\sum_q e_q^2\,x\,g_{L}^{\perp (1) q}(x)\,D_1^{q}(z)\;
	b^{(1)}_{\rm B}\,\biggl(\frac{z \Phperp} {\lambda}\biggr)\,
	{ \cal G}(\Phperp ) \, , \\
	F_{LL}^{\cos\phi_h}(x,z,\la\Phperp\ra)
	&=& -\,\frac{2M_N}{Q}\;x\sum_q e_q^2\,x\,g_{L}^{\perp (1) q}(x)\,D_1^{q}(z)\;
	c^{(1)}_{\rm B}\,\biggl(\frac{z} {\lambda^{1/2}}\biggr)\, ,
\ea\end{subequations}
where $\lambda=z^2 \avkperp_{g_{L}^\perp} + \avpperp_{D_1}$, $b^{(1)}_{\rm B}=2M_N$,
$c^{(1)}_{\rm B} = \sqrt{\pi}\,M_N$, see App.~\ref{App:convol-details}. Finally
we explore the WW-type approximation (\ref{Eq:WW-type-gLperp}) to relate
$x\,g_L^{\perp(1) a}(x) = \frac{\la \kperp^2\ra_{g_1}}{2\,M_N^2}\,g_1^a(x)$.

\begin{figure}[t!]
\centering
\begin{tabular}{cc} \ \hspace{-8mm}
\includegraphics[width=0.6\textwidth]{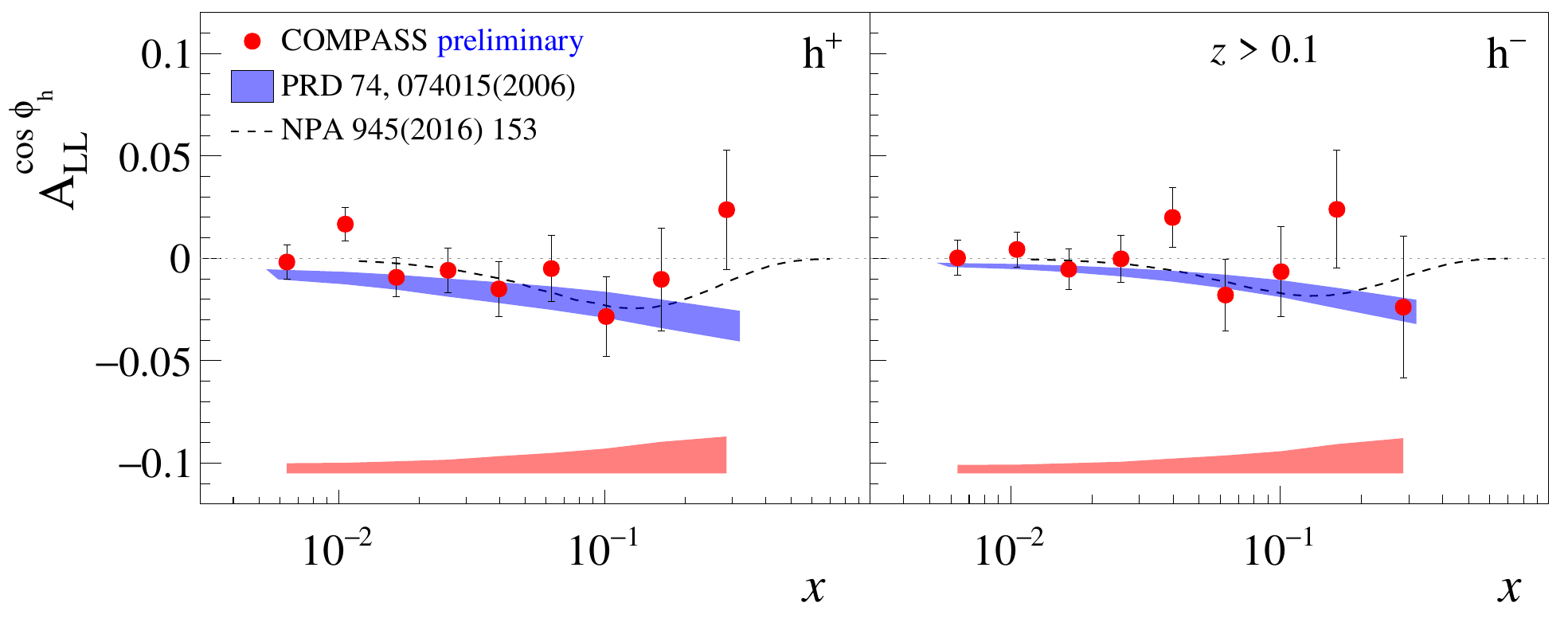}&
\includegraphics[width=0.38\textwidth]{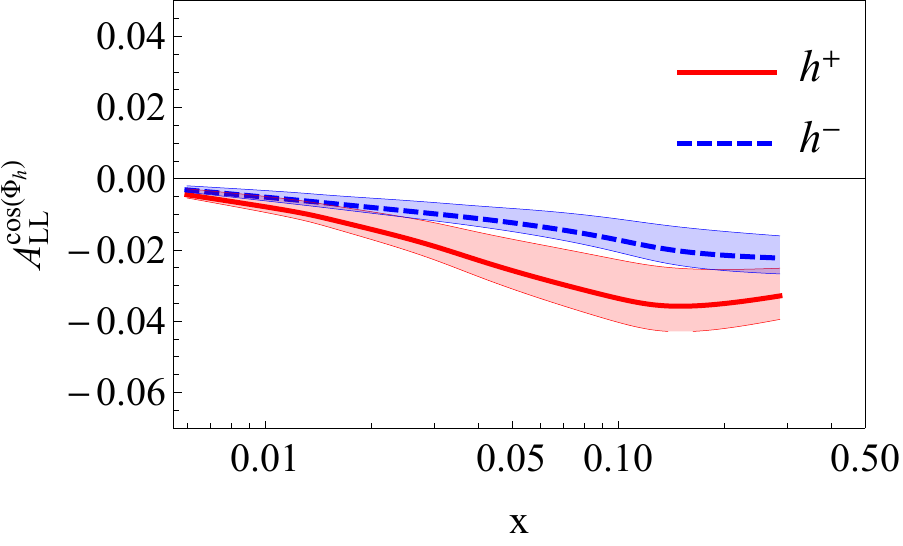}\\
{\tiny (a) \hspace{3cm} (b)}&{\tiny (c)}
\end{tabular}
	\caption{\label{allcosphi_jlab} $A_{LL}^{\cos(\phi_h)}$
	as a function of $ x $: preliminary COMPASS data
	\cite{Parsamyan:2018ovx,Parsamyan:2018evv} (a,b),  
	and our calculation for COMPASS kinematics in 
	WW-type approximation (c)
	shown separately for reasons explained in the
	caption of Fig.~\ref{aut_h1tp_jlab}.}
\end{figure}

The asymmetry $A_{LL}^{\cos \phi_h}=F_{LL}^{\cos \phi_h}/F_{UU}$ predicted by the
WW-type approximation in this case is compatible with preliminary COMPASS
\cite{Parsamyan:2018ovx,Parsamyan:2018evv} (see Fig.~\ref{allcosphi_jlab}) 
and HERMES \cite{Airapetian:2018rlq} data.
We remark that previously this asymmetry was studied in basically
WW-type approximation in \cite{Anselmino:2006yc} and more recently
also in a model study \cite{Mao:2016hdi}. The predictions from
both works are included in Fig.~\ref{allcosphi_jlab} for
comparison.

\

\

\subsection{\boldmath Subleading-twist $A_{UL}^{\sin\phi_h}$ }
\label{Sec-7.4:FULsinphi}

This was historically the first measurement of a single-spin
asymmetry in SIDIS, by HERMES \cite{Airapetian:1999tv,Airapetian:2001eg},
and consequently subject to numerous phenomenological and model studies
\cite{DeSanctis:2000fh,Oganessian:2000um,Efremov:2001ia,Efremov:2002td,
Efremov:2002sd,Ma:2001ie,Ma:2002ns,Schweitzer:2003yr}. A more recent
model study was reported in \cite{Lu:2014fva}.

In WW-type approximation $A_{UL}^{\sin\phi_h}$ is described by
$h_L^q(x,\kperp^{2})$, for which we assume the Gaussian Ansatz
(\ref{eq:hLnew}) in App.~\ref{App-B:Gauss-Ansatz-non-basis-TMDs}
with $\avkperp_{h_L}=\avkperp_{h_1}$. We explore (\ref{Eq:WW-type-6})
to estimate $x\,h_L^q(x) = -2 h_{1L}^{\perp(1)q}(x)$ and express
$h_{1L}^{\perp(1)q}(x)$ through $h_1^a(x)$ according to
(\ref{Eq:WW-approx-h1L}). This yields
\begin{subequations}\ba
	F_{UL}^{\sin\phi_h}(x,z,\Phperp)
	&=& \frac{2M_N}{Q}\,x\sum_q e_q^2\,x\,h_{L}^{q}(x)\,H_1^{\perp(1)q}(z)\;
	b^{(1)}_{\rm A}\,\biggl(\frac{z \Phperp} {\lambda}\biggr)\,
	{ \cal G}(\Phperp ) \, , \\
	F_{UL}^{\sin\phi_h}(x,z,\la\Phperp\ra)
	&=& \frac{2M_N}{Q}\,x\sum_q e_q^2\,x\,h_{L}^{q}(x)\,H_1^{\perp(1)q}(z)\;
	c^{(1)}_{\rm A}\,\biggl(\frac{z} {\lambda^{1/2}}\biggr)\,
\ea\end{subequations}
where $\lambda=z^2 \avkperp_{h_L} + \avpperp_{H_1^\perp}$ and
$b^{(1)}_{\rm A}=2\mh$ and $c^{(1)}_{\rm A} = \sqrt{\pi}\,\mh$,
see App.~\ref{App:convol-details} for details.

\begin{figure}[t!]
\centering
\begin{tabular}{cc} \ \hspace{-8mm}
\includegraphics[width=0.38\textwidth]{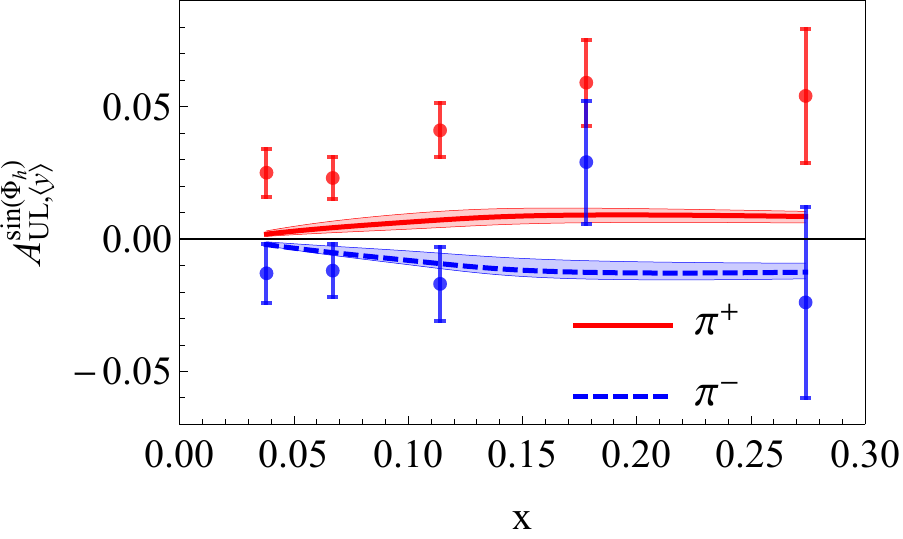} &
\includegraphics[width=0.38\textwidth]{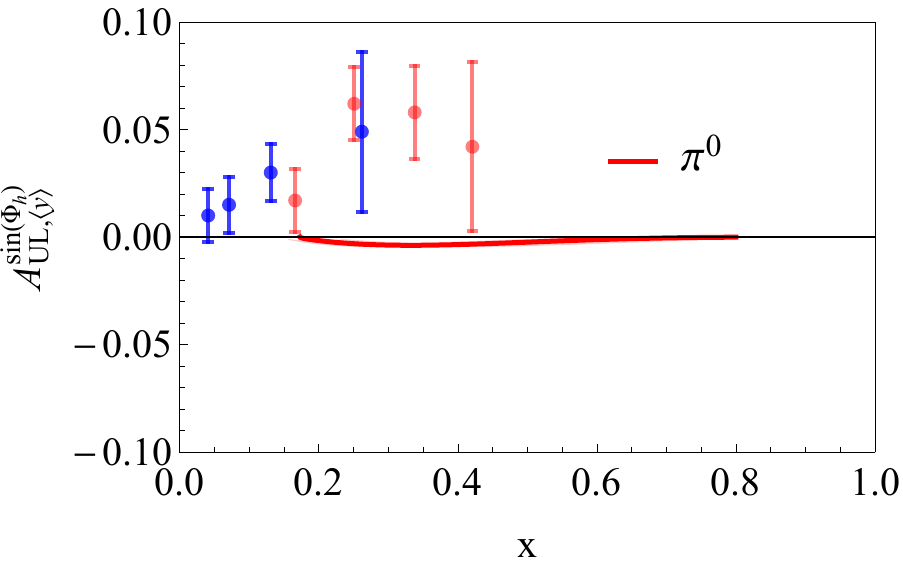} \\
{\tiny (a) }&{\tiny (b)} \vspace{5mm}\\
\ \hspace{-8mm}
\includegraphics[width=0.6\textwidth]{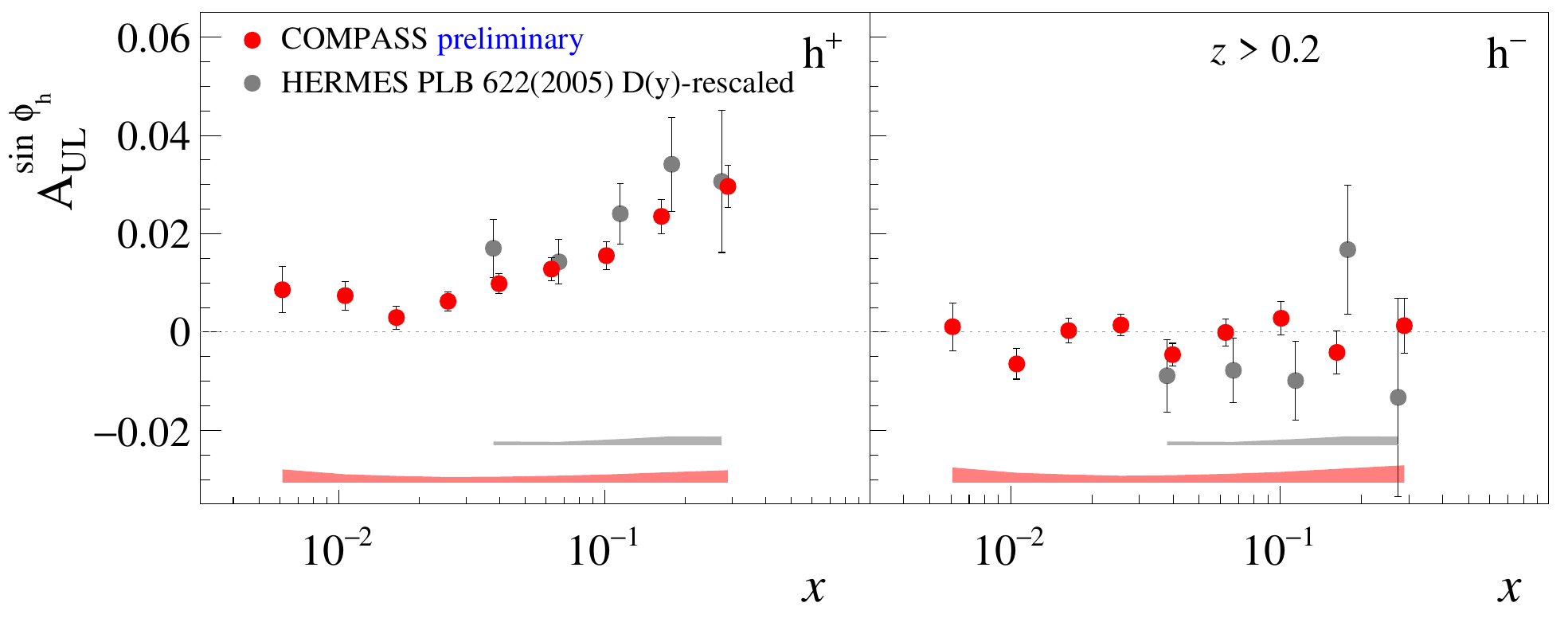}&
\includegraphics[width=0.38\textwidth]{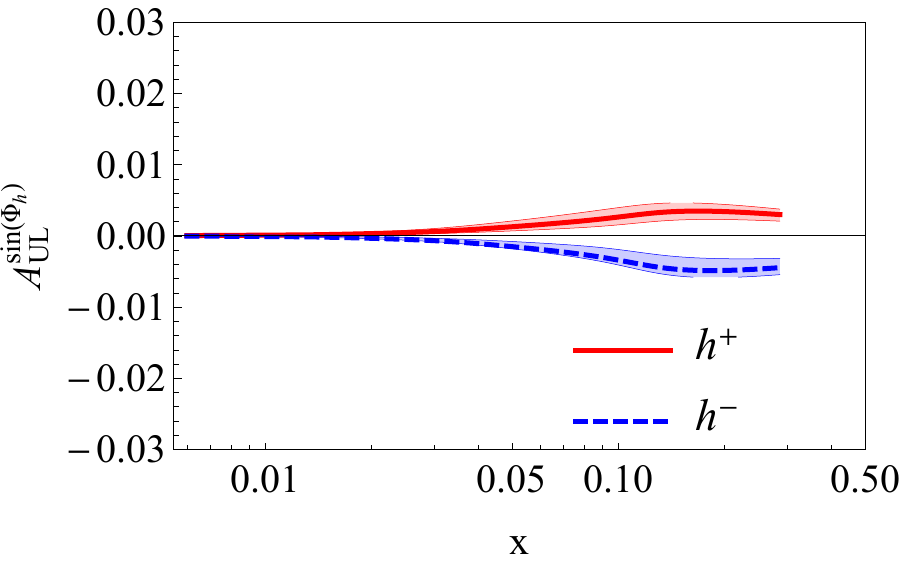}\\
{\tiny (c) \hspace{3cm} (d)}&{\tiny (e)} \vspace{5mm}
\end{tabular}
	\caption{\label{aulsinphi_jlab} 
	$A_{UL}^{\sin\phi_h}$ for proton vs.~$x$ from WW-type
	approximation and comparison to the data.
	$\pi^\pm$ from HERMES \cite{Airapetian:2005jc} (a), $\pi^0$ from HERMES (blue) \cite{Airapetian:2001eg} and JLab (red) \cite{Jawalkar:2017ube} (b), 
	preliminary COMPASS data
	\cite{Parsamyan:2018ovx,Parsamyan:2018evv} (c,d),  
	and our calculation for COMPASS kinematics in 
	WW-type approximation (e)
	shown separately for reasons explained in the
	caption of Fig.~\ref{aut_h1tp_jlab}.}
\end{figure}

The asymmetries $A_{UL}^{\sin\phi_h}=F_{UL}^{\sin\phi_h}/F_{UU}$ are compared
to HERMES, JLab, and preliminary COMPASS data in Fig.~\ref{aulsinphi_jlab}.
The WW-type approximation seems not incompatible with data for negative
pions and hadrons, but underestimates the magnitude of the asymmetry for
positive pions and hadrons at HERMES and COMPASS. For neutral pions the
approximation predicts a negligibly small effect (due to cancelling
contributions from favored und unfavored Collins fragmentation function)
and is not able to explain the large effect
observed at HERMES and JLab for $\pi^0$ in the large--$x$ region
$0.1< x < 0.5$ in Fig.~\ref{aulsinphi_jlab}.
This indicates that in this asymmetry the
tilde-terms are not negligible, and have a characteristic flavor
dependence that is distinct from that of the Collins effect.

\subsection{\boldmath Subleading-twist  $A_{UT}^{\sin\phi_S}$}
\label{Sec-7.6:FUTsinphiS}

In this structure function some interesting new features occur.
The first feature is that after applying the WW-type approximation,
not one but three terms are left, cf.\
Eqs.~(\ref{FUTsinphiS},~\ref{Eq:WW-type-FUTsinphiS}):
two terms proportional to $h_T^{\perp q}(x,\kperp^{2})$ and $h_T^q(x,\kperp^{2})$, respectively, 
and one term proportional to $f_T^q(x,\kperp^{2})$, which is associated with
the second interesting feature. This T-odd TMD must satisfy the sum rule
$\int d^2\kperp \,f_T^q(x,\kperp^{2})=0$, see (\ref{Eq:sum-rules-T-odd})
and Sec.~\ref{Sec-3.8:limitations}.
This could be implemented in two ways. One could describe it with a
superposition of Gaussians, see App.~\ref{App-B:comment-Todd-twist-3}.
But at this point we have no guidance from phenomenology or theory to
fix additional parameters. So we choose an alternative
and pragmatic solution, namely to neglect the contribution of
$f_T^q(x,\kperp^{2})$.\footnote{\label{Footnote:fT-single-Gauss} This
	corresponds to using a ``single Gaussian'' as
	$f_T^q(x,\kperp^{2}) = f_T^q(x)\;
	\frac{\exp\left(-\kperp^2/\la\kperp^2\ra_{f_T}^{ }\right)}
	{\pi\la\kperp^2\ra_{f_T}^{ }}$
	with the ``coefficient'' $f_T^q(x)=0$ as dictated
	by the sum rule (\ref{Eq:sum-rules-T-odd}).}
Assuming Gaussian Ansatz (\ref{eq:hTperpnew},~\ref{eq:hTnew})
for $h_T^{\perp q}(x,\kperp^{2})$, $h_T^q(x,\kperp^{2})$ in
App.~\ref{App-B:Gauss-Ansatz-non-basis-TMDs} and relating
them to transversity via the WW-type approximations
(\ref{Eq:WW-type-7},~\ref{Eq:WW-type-8}), the expression
for $F_{UT}^{\sin\phi_S}$ is given in terms of a single term
\begin{subequations}\begin{alignat}{1}
	F_{UT}^{\sin\phi_S}(x,z,\Phperp)
	&= \,\frac{2M_N}{Q}\;x\sum_q e_q^2\,
	h_1^{(1)q}(x)\,H_1^{\perp(1)q}(z)\; \frac{4z^2 \mh\,M_N}{\lambda}
	\left(1-\frac{\Phperp^2}{\lambda}\right) {\cal G}(\Phperp)
	\label{Eq:FUTsinphiS-final-PhT}\\
  	F_{UT}^{\sin\phi_S}(x,z)
	&= 0 \, .	\label{Eq:FUTsinphiS-final}
\end{alignat}\end{subequations}
with $\lambda=z^2\la\kperp^2\ra_{h_T^\perp}+\la\pperp^2\ra_{H_1^\perp}$ and
$\la\kperp^2\ra_{h_T^\perp}=\la\kperp^2\ra_{h_T^{ }}=\la\kperp^2\ra_{h_1^{ }}$.
The third interesting feature is the occurrence of a term that drops
out upon integrating the structure function over $\Phperp$, cf.\
(\ref{Eq:FUTsinphiS-final-PhT}) vs.\ (\ref{Eq:FUTsinphiS-final}).
This is a property of the weight $\omega^{\{2\}}_{\rm B}$, see
(\ref{Eq:wi}) and App.~\ref{App:factor} (which appears also
in $F_{LT}^{\cos\phi_S}$, (\ref{FLTcosphiS}), where it drops
out in WW-type approximation). This property can help
to discriminate experimentally the terms associated with this weight.

\begin{figure}
\centering
\begin{tabular}{ccc} \ \hspace{-8mm}
\includegraphics[height=3.3cm]{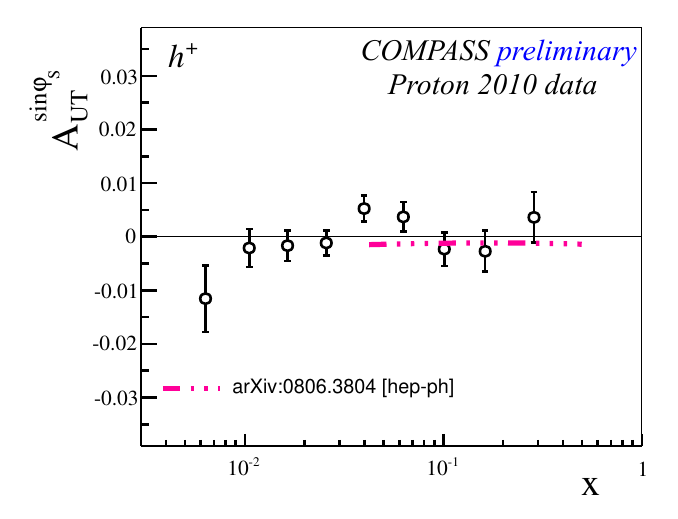}&
\includegraphics[height=3.3cm]{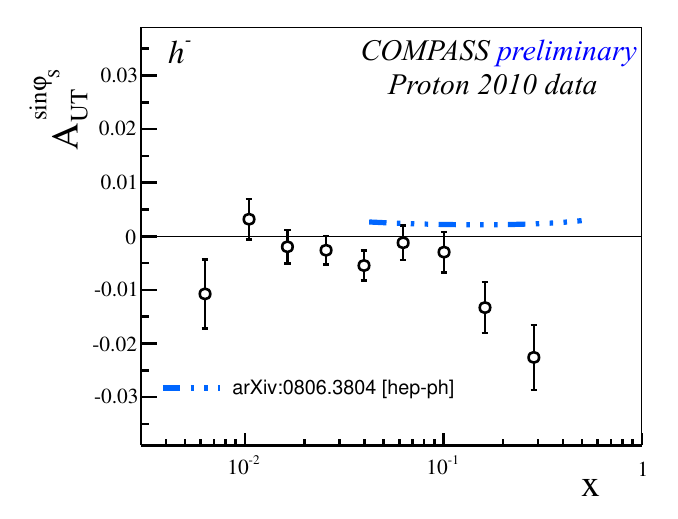}&
\includegraphics[height=3.3cm]{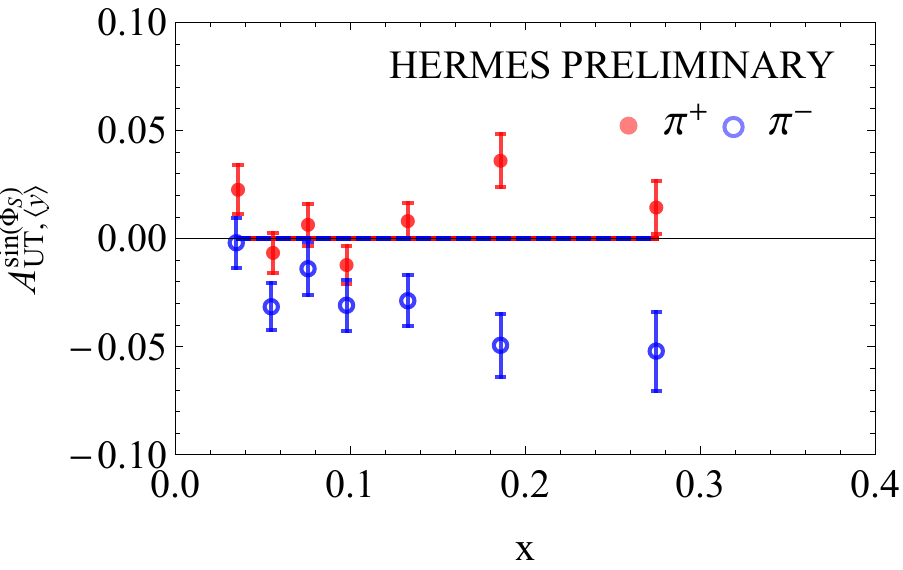}\\
{\tiny (a)}&{\tiny (b)}&{\tiny (c)}
\end{tabular}
\caption{\label{autsinphis}
 	Subleading-twist asymmetry $A_{UT}^{\sin\phi_S}(x)$, which
	is predicted to vanish in the WW-type approximation: 
	preliminary COMPASS~\cite{Parsamyan:2013fia} (a,b), 
	and HERMES data \cite{Schnell:2010zza} (c).}
\end{figure}

The final result in (\ref{Eq:FUTsinphiS-final}) is the consistent
result for $F_{UT}^{\sin\phi_S}(x,z)$ in WW-type
approximation, see (\ref{Eq:FUT-collinear}). Our prediction is
therefore $A_{UT}^{\sin\phi_S}(x)=0$. Preliminary
COMPASS \cite{Parsamyan:2013fia} (Fig.~\ref{autsinphis} top) and
HERMES \cite{Schnell:2010zza} (Fig.~\ref{autsinphis} bottom) data
indicate that at $x\gtrsim 0.1$ the signal is clearly non-zero and
thus inconsistent with this WW-type prediction. In order to
describe the data, it is therefore necessary to explicitly model
the involved tilde-functions. For a recent model study we refer
to Ref.~\cite{Mao:2014aoa}.

\newpage
\subsection{\boldmath Subleading-twist  $A_{UT}^{\sin(2\phi_h-\phi_S)}$ }
\label{Sec-7.8:FUTsin2phi-phiS}

In this asymmetry, two terms survive the WW-type approximation.
Using the Gaussian Ansatz for
$f_T^{\perp q}(x,\kperp^{2})$, $h_T^{\perp q}(x,\kperp^{2})$, $h_T^q(x,\kperp^{2})$,
according to (\ref{eq:hTperpnew},~\ref{eq:hTnew},~\ref{eq:ftperpnew})
in App.~\ref{App-B:Gauss-Ansatz-non-basis-TMDs}, with
$\la\kperp^2\ra_{h_T^\perp}=\la\kperp^2\ra_{h_T^{ }}=\la\kperp^2\ra_{h_{1T}^\perp}$
and $\la\kperp^2\ra_{f_T^\perp}=\la\kperp^2\ra_{f_{1T}^\perp}$
yields the expressions 
\begin{subequations}\begin{alignat}{1}
	F_{UT}^{\sin(2\phi_h -\phi_S)}(x,z,\Phperp) \;
	=
	\frac{2M_N}{Q}\;x\sum_q e_q^2\,\Biggl[
	x \, f_T^{\perp(2)q}(x) \; D_1^q(z) \;b^{(2)}_{\rm C}\;
	\left(\frac{z\Phperp}{\lambda}\right)^{\!\!2}\,{ \cal G}(\Phperp)&
	\nonumber\\
	+ \;
	\frac{x}{2}\left(h_T^{(1)q}(x)+h_T^{\perp(1)q}(x)\right)H_1^{\perp(1)q}(z)\,
	b^{(2)}_{\rm AB}
	\left(\frac{z\Phperp}{\lambda}\right)^{\!\!2}\,{\cal G}(\Phperp)&\Biggr] \, ,
	\label{Eq:FUTsin2phi-phiS-Gauss-PhT}\\
{ }
	F_{UT}^{\sin(2\phi_h -\phi_S)}(x,z,\la\Phperp\ra) \;
	= \;
	\frac{2M_N}{Q}\;x\sum_q e_q^2\,\Biggl[
	x \; f_T^{\perp(2)q}(x) \; D_1^q(z) \; c^{(2)}_{\rm C} \;
	\biggl(\frac{z}{\lambda^{1/2}}\biggr)^{\!\!2} &
	\nonumber\\
	+ \;
	\frac{x}{2}\left(h_T^{(1)q}(x)+h_T^{\perp(1)q}(x)\right)H_1^{\perp(1)q}(z)\,
	c^{(2)}_{\rm AB}
	\biggl(\frac{z}{\lambda^{1/2}}\biggr)^{\!\!2} & \Biggr] \, ,
	\label{Eq:FUTsin2phi-phiS-Gauss}
\end{alignat}\end{subequations}
with respectively
$\lambda=z^2\la\kperp^2\ra_{f_T^\perp}+\la\pperp^2\ra_{D_1}$ in the first, and
$\lambda=z^2\la\kperp^2\ra_{h_T^\perp}+\la\pperp^2\ra_{H_1^\perp}$ in the second
terms in (\ref{Eq:FUTsin2phi-phiS-Gauss-PhT},~\ref{Eq:FUTsin2phi-phiS-Gauss}).
The coefficients
$b^{(2)}_i$ and $c^{(2)}_i$ are defined in App.~\ref{App:convol-details}.
In the next step we explore the WW-type approximations
(\ref{Eq:WW-type-7},~\ref{Eq:WW-type-8},~\ref{Eq:WW-type-fTperp}) to
relate
$x \, f_T^{\perp(2)q}(x) =
\frac{\la\kperp^2\ra_{f_{1T}^\perp}}{M_N^2}\,f_{1T}^{\perp (1)q}(x)$ and
$-\,\frac12\,x \left(h_T^{(1)q}(x) + h_T^{\perp(1)q}(x)\right)
= h_{1T}^{\perp(2)q}(x)$.

\begin{figure}[t!]
\centering
\includegraphics[height=3.1cm]{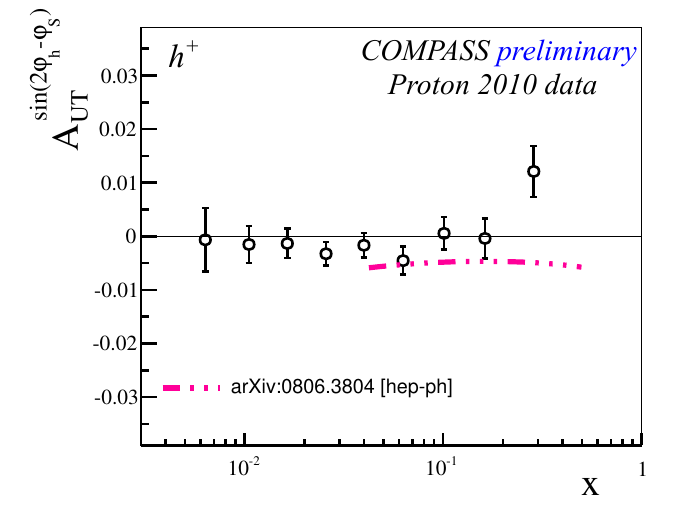} {\tiny (a)}%
\includegraphics[height=3.1cm]{\FigPath/UT_Pr10_sin2phihphis_pip} {\tiny (b)}

\includegraphics[height=2.4cm]{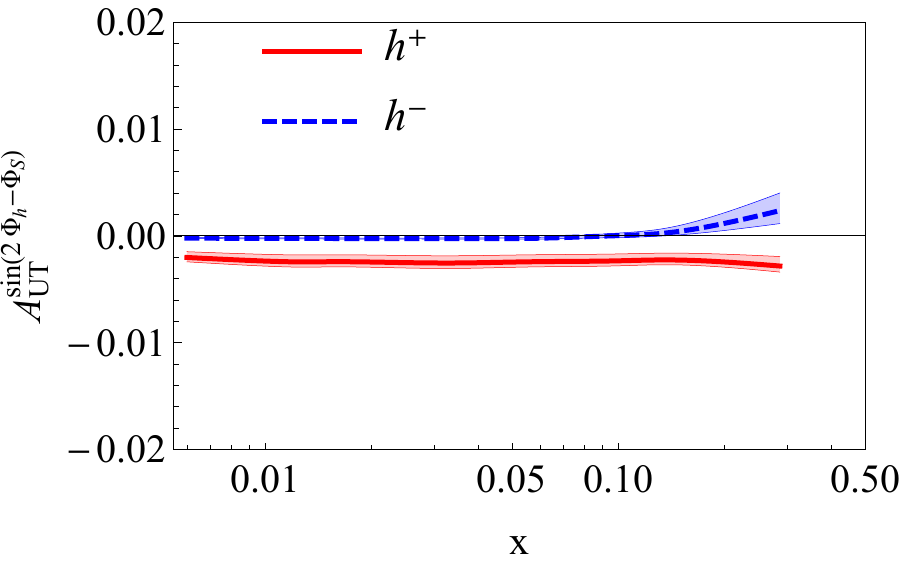} {\tiny (c)}
\includegraphics[height=2.4cm]{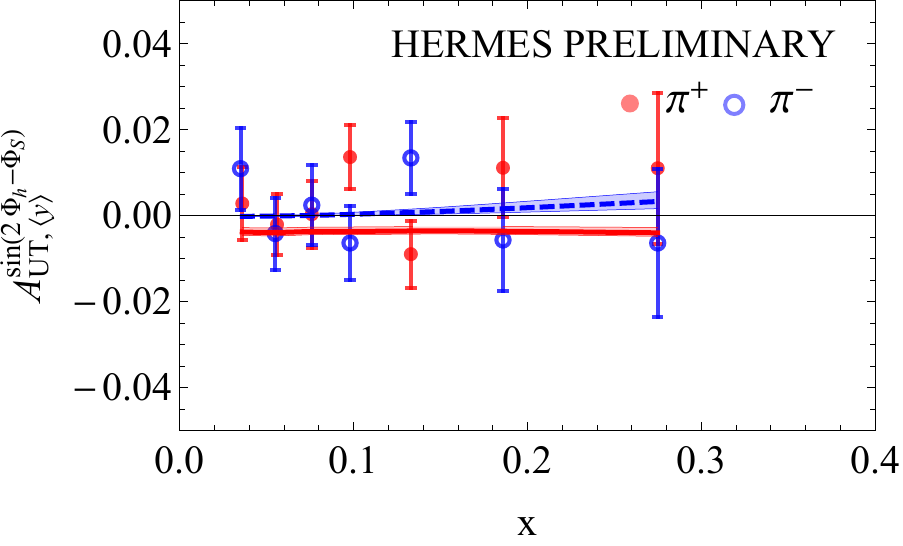} {\tiny (d)}

\vspace{-2mm}

\caption{\label{autsin2phihphis}
	$A_{UT}^{\sin(2 \phi_h - \phi_S)}(x)$: 
	preliminary COMPASS data~\cite{Parsamyan:2013fia} (a,b)
  and our calculation for COMPASS kinematics in WW-type 
      approximation (c) shown separately for reasons explained 
      in the caption of Fig.~\ref{aut_h1tp_jlab}.
	Similarly $A_{UT, \langle y \rangle}^{\sin(2\phi_h-\phi_S)}(x)$ vs
	preliminary HERMES data~\cite{Schnell:2010zza} (d).
	}
\end{figure}

The asymmetries $A_{UT}^{\sin (2 \phi_h-\phi_S)}=F_{UT}^{\sin (2 \phi_h-\phi_S)}/F_{UU}$
are plotted in Fig.~\ref{autsin2phihphis} in comparison
to preliminary COMPASS \cite{Parsamyan:2013fia} and
HERMES \cite{Schnell:2010zza} data. The predicted
asymmetry is small and compatible with the data that are consistent
with a zero effect within uncertainties.
For comparison, Fig.~\ref{altcos2phihphis} shows also the predictions
from the quark-diquark model of Ref.~\cite{Kotzinian:2008fe}.
More recently the $A_{UT}^{\sin(2\phi_h -\phi_S)}$ was also studied in the
model approach of Ref.~\cite{Mao:2014aoa}.

\subsection{\boldmath Subleading-twist  $A_{UU}^{\cos\phi_h}$ }
\label{Sec-7.7:FUUcosphi}

\begin{figure}[t!]
\centering
\includegraphics[width=0.38\textwidth]{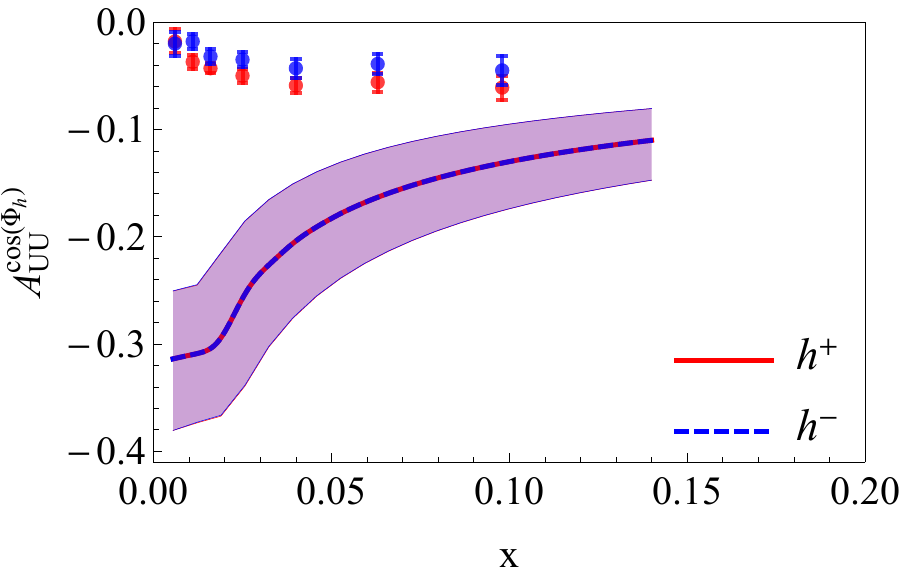}
\includegraphics[width=0.38\textwidth]{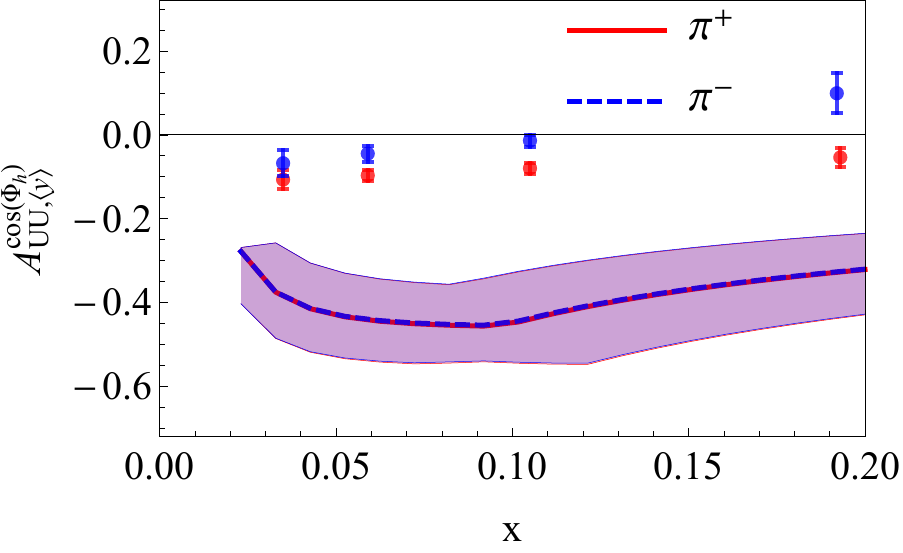}
\caption{\label{auucosphi_jlab}
	Left panel: asymmetry $A_{UU}^{\cos\phi_h}$ for positive and negative
	hadrons at COMPASS for a proton target \cite{Adolph:2014pwc}. Right
	panel: the corresponding $A_{UU, \langle y \rangle}^{\cos\phi_h}$ for 
	$\pi^\pm$ from HERMES \cite{Airapetian:2012yg}.}
\end{figure}

Historically this was the earliest azimuthal SIDIS asymmetry to be
discussed in literature: the first prediction for this asymmetry from
intrinsic $k_\perp$ was made in \cite{Cheng:1972sy,Cahn:1978se}, a
first measurement
was reported in \cite{Aubert:1983cz}.\footnote{First hints
	\cite{Dakin:1972db} of azimuthal modulations in SIDIS
	date back to the early 1970s, i.e., 10 years before
	the CERN measurements, but (unfortunately) were
	discarded by the authors.}
This structure function contains after the WW-type approximation initially
two contributions, proportional to $f^\perp(x,\kperp^{2})$ and $h^{q}(x,\kperp^{2})$.
The latter is T-odd and obeys the sum rule (\ref{Eq:sum-rules-T-odd}).
We treat $h^{q}(x,\kperp^{2})$ exactly as $f_T^q(x,\kperp^{2})$ in
Sec.~\ref{Sec-7.6:FUTsinphiS}.
Using the Gaussian Ansatz for $f^\perp(x,\kperp^{2})$ in (\ref{eq:fperpnew})
of App.~\ref{App-B:Gauss-Ansatz-non-basis-TMDs}, we obtain
\begin{subequations}\begin{alignat}{1}
	F_{UU}^{\cos\phi_h}(x,z,\Phperp)
	= \frac{2M_N}{Q}\;x\sum_q e_q^2 & \Biggl[
	- x\,f^{\perp(1)q}(x)\,D_1^q(z)\;b^{(1)}_{\rm B}\,
	  \biggl(\frac{z \Phperp} {\lambda}\biggr)\, { \cal G}(\Phperp )
	\Biggr] \, ,  \label{Eq:XXXa}\\
	F_{UU}^{\cos\phi_h}(x,z,\la\Phperp\ra)
	= \frac{2M_N}{Q}\;x\sum_q e_q^2 & \Biggl[
	-x\,f^{\perp(1)q}(x)\,D_1^q(z)\;c^{(1)}_{\rm B}\;
	  \biggl(\frac{z}{\lambda^{1/2}}\biggr)
	\Biggr] \, , \label{Eq:XXXb}
\end{alignat}\end{subequations}
with $\lambda=z^2\la\kperp^2\ra_{f^\perp}+\la\pperp^2\ra_{D_1}$. The coefficients
$b^{(1)}_i$ and $c^{(1)}_i$ are defined in App.~\ref{App:convol-details}.
Note that (\ref{Eq:XXXa}) is valid in the ``scheme'' of
footnote~\ref{Footnote:fT-single-Gauss}, but (\ref{Eq:XXXa})
holds independent of how one implements the sum rule (\ref{Eq:sum-rules-T-odd})
(as in footnote~\ref{Footnote:fT-single-Gauss} or
App.~\ref{App-B:comment-Todd-twist-3}).

For $f^{\perp(1)}(x)$ we explore (\ref{Eq:WW-type-Cahn}) as
$xf^{\perp(1)q}(x) = \frac{\la\kperp^2\ra_{f_1}}{2M_N^2}\,f_{1}^q(x)$ and
assume for its Gaussian width $\la\kperp^2\ra_{f^\perp}=\la\kperp^2\ra_{f_1}$.
The latter means the Gaussian factors of
$F_{UU}^{\cos\phi_h}$ and $F_{UU}$ cancel out, i.e.\ at some point
for $\Phperp\gtrsim1\,$GeV we would obtain from (\ref{Eq:XXXa})
an asymmetry $A_{UU}^{\cos\phi_h}=F_{UU}^{\cos\phi_h}/F_{UU}$ exceeding
100$\,\%$ and violating unitarity. This is of course an artifact of our
approximations, and reminds us that Gaussian and WW-type approximations
as well as the entire TMD formalism must be applied to small $\Phperp\ll Q$.

The asymmetries $A_{UU}^{\cos\phi_h}$ were measured by
EMC \cite{Aubert:1983cz}, at JLab \cite{Osipenko:2008aa,Mkrtchyan:2007sr},
HERMES \cite{Airapetian:2012yg}, and COMPASS \cite{Adolph:2014pwc}.
In Fig.~\ref{auucosphi_jlab} we compare our predictions to the HERMES and
COMPASS data.
The WW-type approximation tends to overestimate the data at
COMPASS especially in the small-$x$ region. It is also not
compatible with the flavor dependence seen at HERMES.
However, both experiments seem not to agree for instance on the shape
of the asymmetry for negative pions or hadrons. More experimental and
theoretical work is needed to clarify whether this could be due to
power corrections.

\section{Conclusions}
\label{Sec-8:conclusions}

In this work a comprehensive and complete treatment of SIDIS
spin and azimuthal asymmetries was presented.
The theoretical and phenomenological understanding of most of the
leading-twist SIDIS structure functions for the production of
unpolarized hadrons is relatively advanced: factorization is
proven, and each structure functions is unambiguously expressed in
terms of one of eight twist-2 TMDs convoluted with one of two twist-2 FFs.
For subleading-twist SIDIS structure functions the situation
is far more complex for two reasons. First, factorization is not
proven and must be assumed.
Second, each of the subleading-twist structure functions
receives four or six contributions from various TMDs and FFs one of
which is twist-2 and the other twist-3. Clearly, to make
predictions for new experiments or interpret early data, an
organizing theoretical guideline is needed.

In this work we have explored the so-called WW-type approximation
as a candidate guideline for the description of SIDIS structure
functions. This approximation consists of a systematic neglect
of $\bar{q}gq$-terms in the
correlators defining the TMDs and FFs. We have shown that in such
an approach all twist-2 and twist-3 structure functions can be
described in terms of eight basis functions: six TMDs and two FFs,
which are each twist-2. All other TMDs and FFs, assuming this
approximation, are either related to the basis functions or
vanish.
We remark that the generalized parton model
approach of Ref.~\cite{Anselmino:2011ch} provides a description
that is largely equivalent to ours.

To make this work self-contained, we included a review of the
available phenomenological information on the basis functions,
which is given in terms of six SIDIS structure functions.
(Of course, one cannot extract eight basis
functions from six observables: the extraction of two basis functions,
the unpolarized TMD and FF $f_1^q(x,\kperp^{2})$ and $D_1^q(x,\pperp^{2})$,
makes also use of other experiments, most notably Drell--Yan
and hadron production in $e^+e^-$ annihilations.)

The WW-type approximation for TMDs and FFs is inspired by the
observation that the classical WW-approximation for the twist-3
DIS structure function $g_2(x)$ (or PDF $g_T^q(x)$) works well.
This was predicted in theoretical studies in the instanton model
of the QCD vacuum, and confirmed by data and lattice QCD studies.
The instanton vacuum model predicts an analogous WW approximation
for the chirally odd twist-3 PDF $h_L^q(x)$ to work similarly
well. This prediction remains to be tested in experiment.

In each case, $g_T^q(x)$ and $h_L^q(x)$, we deal with nucleon
matrix elements of different $\bar{q}gq$ correlators, which are
assumed to be small. Can one generalize these approximations to
TMDs? This is a key question, which has been addressed in the
past in literature in selected cases. This work is the first
systematic investigation of this question.
As in the unintegrated correlators one deals with different
classes of operators, we prefer to speak of the
WW-type approximation to distinguish from the collinear case.

We have studied in detail all SIDIS structure function in
this approximation. This includes a review of results from
lattice QCD calculations, effective theories and models.
We found that from theoretical point of view the WW-type
approximations receive certain support, though there is
less evidence than in the collinear case.
Most importantly, we have conducted systematic tests of
WW approximations with available published or
preliminary (and soon to be published) SIDIS data
from HERMES, COMPASS, and JLab.

We found the following results. The two leading-twist
structure functions amenable to WW-type approximations,
	$F_{LT}^{\cos(\phi_h -\phi_S)}$ and
	$F_{UL}^{\sin(2\phi_h)}$,
are well-described (the former) or at least compatible
(the latter) with the data in this approximation. For
$F_{UL}^{\sin(2\phi_h)}$ more precise data are needed,
but also in this case the trend is encouraging
especially thanks to the recent preliminary COMPASS data.
We have also shown that our approach satisfies positivity
inequalities, which is a non-trivial consistency check
considering the crude approximations (WW-type, Gaussian
Ansatz for TMDs) in our approach.

At subleading twist the WW-type approximation for the
structure functions
	$F_{LT}^{\cos\phi_S}$,
	$F_{LT}^{\cos(2\phi_h-\phi_S)}$,
	$F_{LL}^{\cos\phi_h}$,
	$F_{UT}^{\sin(2\phi_h-\phi_S)}$
is compatible with data, too. Some of these asymmetries are
predicted to be very small in the WW-type approximation,
sometimes smaller than a fraction of a percent. This is
compatible with the available data in the sense that the
data are consistent with zero within their statistical
accuracy. This cannot be considered a thorough evidence
for the applicability of the WW-type approximations, but
on the positive side we also observe no hints
that the WW-type approximations fail in these cases.

In the case of the subleading-twist structure functions
	$F_{UU}^{\cos\phi_h}$,
	$F_{UL}^{\sin\phi_h}$,
	$F_{LU}^{\sin\phi_h}$, and
	$F_{UT}^{\sin\phi_S}$
the situation is clearer and indicates that here the WW-type
approximations do not work.
Incidentally, these asymmetries include the very first non-zero azimuthal
asymmetry measured in unpolarized SIDIS ($F_{UU}^{\cos\phi_h}$),
the very first non-zero target single-spin asymmetry measured in SIDIS
($F_{UL}^{\sin\phi_h}$), and the beam-spin asymmetry ($F_{LU}^{\sin\phi_h}$).
The WW-type prediction for $F_{UU}^{\cos(\phi_h)}$ tends to overshoot
the data. In the case of $F_{UL}^{\sin\phi_h}$ the WW-type approximation
undershoots data by a factor of two or so.
Most interestingly, in the case of $F_{LU}^{\sin\phi_h}$ the WW-type
approximation predicts exactly a zero asymmetry, but experiments
see small but non-zero effect.

The non-applicability of the WW--type approximation in these 
cases should not be too surprising. After all it is a crude method
to model TMDs and FFs and an uncontrollable ``expansion''
(in nuclear physics the concept of 2--body, 3--body, etc operators
is well-justified and an effective expansion can be conducted; in
the case of TMDs, however, it is less appropriate to speak about a
systematic expansion in terms of $\bar{q}q$, $\bar{q}gq$, etc
correlators). It will be very interesting to learn whether, e.g.,
in $F_{UL}^{\sin\phi_h}$ or $F_{LU}^{\sin\phi_h}$ a single $\bar{q}gq$--term
is anomalously large, or whether it is an accumulative effect of
several small terms $\bar{q}gq$--terms adding up to the observed asymmetry.

Among all SIDIS structure functions $F_{LU}^{\sin\phi_h}$ emerges as
a particularly interesting case: this asymmetry is due to $\bar{q}gq$
only, without ``contamination'' from $\bar{q}q$ terms.
Thus $F_{LU}^{\sin\phi_h}$ offers a unique view on the physics of $\bar{q}gq$
correlators, worth exploring for its own sake.

The results presented in this work are of importance for several reasons.
To the best of our knowledge, it is the first complete study of all SIDIS
structure functions up to twist-3 
using systematically a common theoretical guideline.
The results are
useful for experiments prepared in the near term (JLab 12) or proposed
in the long term (Electron-Ion Collider), and provide helpful input for
Monte Carlo event generators \cite{Avakian:2015vha}.
Our predictions will help to pave the way
towards a better understanding of the quark-gluon structure
of the nucleon beyond leading twist.

\section{Acknowledgments}
First and foremost, the authors would like to thank their families
for patience and constant support throughout writing this paper.
The authors would like to thank Werner Vogelsang for many useful
discussions. This work was partially supported by the U.S.\
Department of Energy under Contract No.~DE-AC05-06OR23177 (A.P.)
and within the TMD Collaboration framework, and by the National
Science Foundation under Contract No.\ PHY-1623454 (A.P.) and
Contract No.\ PHY-1812423 (P.S.), and the Wilhelm Schuler Stiftung (P.S.). The work was supported by DOE grant no. DE-FG02-04ER41309.
We thank the COMPASS and HERMES Collaborations for the 
permissions to show their preliminary data on several figures.

\appendix

\section{\boldmath The ``minimal basis'' of TMDs and FFs}
\label{App:basis}

This Appendix describes the technical details of the parametrizations
used in this work.

\subsection{\boldmath Unpolarized functions $f_1^a(x,k_\perp^2)$
			and $D_1^a(z,P_\perp^2)$}
\label{App:basis-f1-D1}

In this work we use the leading-order parametrizations
from \cite{Martin:2009iq} for the unpolarized PDF $f_1^a(x)$ and
from \cite{deFlorian:2007aj} for the unpolarized FF $D_1^a(z)$.
If not otherwise stated the parametrizations are taken at the scale
$Q^2=2.4\,{\rm GeV}^2$ typical for many currently available SIDIS data.
These parametrizations were used in \cite{Anselmino:2005nn} and other
works whose extractions we adopt.

To describe the transverse momentum dependence of $f_1^a(x,k_\perp^2)$
and $D_1^a(z,P_\perp^2)$ we use the Gaussian Ansatz in
Eqs.~(\ref{Eq:Gauss-f1},~\ref{Eq:Gauss-D1}). All early
\cite{Anselmino:2005nn,Collins:2005ie,D'Alesio:2007jt,Schweitzer:2010tt}
and some recent \cite{Signori:2013mda} analyses employed flavor and
$x$-- or $z$--independent widths $\avkperp$ and $\avpperp$.
In the analysis \cite{Anselmino:2013lza}
of HERMES multiplicities flavor-independence of the widths was
assumed. On long run one may anticipate that new precision data will
require to relax these assumptions. However, one may also expect that the
Gaussian Ansatz will remain a useful {\it approximation} as long as one is
interested in describing data on transverse hadron momenta $\Phperp \ll Q $.

\begin{table}[b!]
\centering
\begin{tabular}{rlccc}
\hline
  study \ \ \ \
	& $\;\;\la Q^2\ra$, $\la x\ra$, $\la z\ra$
	& $\avkperp_{f_1}$
  	& $\avpperp_{D_1}$
	& $\avkperp_{g_1}$ \cr
  	& {\footnotesize $[{\rm GeV}^2]$}
  	& {\footnotesize $[{\rm GeV}^2]$}
  	& {\footnotesize $[{\rm GeV}^2]$}
  	& {\footnotesize $[{\rm GeV}^2]$} \cr
\hline
\ fit of \cite{Anselmino:2005nn} & 5.0, 0.1, 0.3
	& $\sim 0.25$
	& $\sim 0.2$
	& -- \cr
\ fit of \cite{Schweitzer:2010tt} & 2.5, 0.1, 0.4
	& $0.38\pm0.06$
	& $0.16\pm0.01$
	& -- \cr
\ fit of \cite{Anselmino:2013lza} & 2.4, 0.1, 0.3
	& $0.57\pm0.08$
	& $0.12\pm0.01$
	& -- \cr
\ fit of \cite{Signori:2013mda} & 2.4, 0.1, 0.5
	& $\sim0.3$
	& $\sim0.18$
	& -- \cr
\ lattice \cite{Hagler:2009mb}  & 4.0, -- , --
	& 0.14--0.15
	& --
	& 0.11-0.15 \cr
\hline
\end{tabular}
\caption{\label{Table:Gauss-paramaters}
  	Gaussian model parameters for $f_1^a(x,k_\perp^{2})$, $D_1^a(z,P_\perp^{2})$,
 	$g_{1}^a(x,k_\perp^{2})$ from phenomenological and lattice QCD studies.
  	The kinematics to which the phenomenological results and the
	renormalization scale of the lattice results are indicated.
	The range of lattice values indicates flavor dependence
        (first number refers to $u$--flavor, second number to $d$--flavor).}
\end{table}

The parameters resulting from calculations or extractions are presented in
Table~\ref{Table:Gauss-paramaters}.
As most extractions of TMDs that we will use are done with the choice of
$\avkperp_{f_1} = 0.25$ GeV$^2$, $\avpperp_{D_1}= 0.2$ GeV$^2$, for our numerical estimates in this work  we will use these fixed widths.

Some comments are in order.
In \cite{Anselmino:2005nn} no attempt was made to assign an uncertainty of the
best fit result. The uncertainty of the numbers from \cite{Schweitzer:2010tt}
includes only the statistical uncertainty, but no systematic uncertainty.
For comparison lattice results from \cite{Hagler:2009mb} are included
whose range indicates flavor-dependence
(first number $u$--flavor,
second number $d$--flavor).
Notice that this is the contribution of the flavor averaged
over contributions from the respective quarks and antiquarks.
Chiral theories predict significant differences in the $\kperp$--behavior
of sea and valence quarks \cite{Schweitzer:2012hh}.
We will comment more on the lattice results in the next section.
In view of the large (and partly underestimated) uncertainties and the
fact that those parameters are anti-correlated the numbers from the
different approaches quoted in Table~\ref{Table:Gauss-paramaters}
can be considered to be in good agreement.

\subsection{\boldmath Helicity distribution $g_1^a(x,k_\perp^2)$}
\label{App:basis-g1}

For the helicity PDF $g_1^a(x)=\int d^2k_\perp\,g_1^a(x,k_\perp^2)\equiv
\int d^2k_\perp\,g_{1L}^a(x,k_\perp^2)$ we use in this work the leading-order
parametrizations from \cite{Gluck:1998xa}.
If not otherwise stated the parametrizations are taken at the scale
$Q^2=2.4\,{\rm GeV}^2$.

In lack of phenomenological information on the $k_\perp$--dependence of
$g_1^a(x,k_\perp^2)$ we explore lattice QCD results from \cite{Hagler:2009mb}
to constrain the Gaussian width in Eq.~(\ref{Eq:Gauss-g1}).
On a lattice with pion and nucleon masses
$m_\pi\approx 500\,{\rm MeV}$ and $M_N=1.291(23)\,{\rm GeV}$
and with an axial coupling constant $g_A^{(3)}= 1.209(36)$ reasonably
close to its physical value $1.2695(29)$ the following results were
obtained for the mean square transverse parton momenta \cite{Hagler:2009mb}.
For the unpolarized TMDs it was found
$\langle \kperp^2 \rangle_{f_1^u} = (0.3741\,{\rm GeV})^2$ and
$\langle \kperp^2 \rangle_{f_1^d} = (0.3839\,{\rm GeV})^2$.
For the helicity TMDs it was found
$\avkperp_{g_1^u} = (0.327\,{\rm GeV})^2$ and
$\avkperp_{g_1^d} = (0.385\,{\rm GeV})^2$.
These values are quoted in Table~\ref{Table:Gauss-paramaters}.

The lattice values for $\langle \kperp^2 \rangle_{f_1}$ consistently
underestimate the phenomenological numbers, see
Table~\ref{Table:Gauss-paramaters}.
The exact reasons for that are unknown, but it is natural to think it
might be related to the fact that the lattice predictions \cite{Hagler:2009mb}
do not refer to TMDs entering in SIDIS (or Drell--Yan or other process)
because a different gauge link was chosen, see Sec.~\ref{Sec-3.5:WW-lattice}.
Still one may expect these results to bear considerable information
on QCD dynamics.\footnote{
	The results \cite{Hagler:2009mb} refer also to pion masses above the
	physical value. This caveat is presumably less critical and
	will be overcome as lattice QCD simulations are becoming feasible
	at physical pion masses.}
To make use of this information we shall assume that the lattice results
\cite{Hagler:2009mb} provide robust predictions for the {\it ratios}
$\langle \kperp^2 \rangle_{g_1^u}/\langle \kperp^2 \rangle_{f_1^u}\approx 0.76$.
With the phenomenological value $\langle \kperp^2 \rangle_{f_1} = 0.25$ GeV$^2$
we then obtain the estimate for the width of the helicity TMD
$\langle \kperp^2 \rangle_{g_1} = 0.19$ GeV$^2$.
In our phenomenological study we use this value for $u$--quarks and for
simplicity also for $d$--quarks. Even though the lattice values indicate
an interesting flavor dependence, see Table~\ref{Table:Gauss-paramaters},
for a proton target this is a very good approximation due to $u$--quark
dominance. When precision data for deuterium and especially for $^3$He
from JLab become available, it will be interesting to
re-investigate this point in detail.

\subsection{\boldmath Sivers function $f_{1T}^{\perp q}(x,k_\perp^{2})$}
\label{App:basis-f1Tperp}

The Sivers distribution function was studied in Refs.~\cite{Efremov:2004tp,Anselmino:2008sga,Anselmino:2011gs,Anselmino:2011ch, Aybat:2011ge,Gamberg:2013kla,Bacchetta:2011gx,Anselmino:2012aa,Sun:2013dya,Echevarria:2014xaa}.
We will use parametrizations from  Refs.~\cite{Anselmino:2008sga,Anselmino:2011gs,Anselmino:2011ch}:
\ba
 \avkperp_{f_{1T}^\perp} &\equiv& \frac{\avkperp M_1^2}{\avkperp + M_1^2} \, ,\\
f_{1T}^{\perp}(x, \kperp^2)& =& - \frac {M}{M_1} \sqrt{2e} \;
{\cal N}_q(x) \, f_{q/p} (x,Q) \, \frac{e^{-\kperp^2/{\avkperp_{f_{1T}^\perp}}}}{\pi \avkperp} ,
\label{fiTp}
\ea
where $M_1$ is a mass parameter, $M$ the proton mass and
\ba
{\cal N}_q(x)= N_q \, x^{\alpha} (1-x)^{\beta} \,
\frac{(\alpha + \beta)^{(\alpha +\beta)}}
{\alpha^{\alpha} \beta^{\beta}} \, .
 \ea
The first moment of the Sivers function is:
\ba
f_{1T}^{\perp (1) q}(x)  &=& -\frac{\sqrt{\frac{e}{2}} \ \avkperp M_1^3}{M (\avkperp + M_1^2)^2}  \ {\cal N}_q(x)  f_q(x, Q) \nonumber \\ 
&=&  -\sqrt{\frac{e}{2}} \frac{1}{M M_1}  \frac{\avkperp_{f_{1T}^\perp}^2}{\avkperp}    \ {\cal N}_q(x)  f_q(x, Q)
\label{siversm} \ .
\ea
We can rewrite the parametrizations of the Sivers functions as
\ba
f_{1T}^{\perp q}(x,\kperp^2) =  f_{1T}^{\perp (1) q}(x)   \frac{2 M^2}{\pi \avkperp_{f_{1T}^\perp}^2} e^{-\kperp^2/{\avkperp_{f_{1T}^\perp}}}
\label{sivers_new} \ .
\ea

The fit the HERMES proton and COMPASS deuteron data from
including only Sivers functions for $u$ and $d$ quarks was done in Ref.~\cite{Anselmino:2011gs},
corresponding to seven free parameters, which are shown in Table~\ref{tab:a}.

\begin{table}
\centering
\begin{tabular}{cccc}
\hline
$N_u=\phantom{-}0.40$ & $\alpha_u=0.35$ & $\beta_u=2.6\phantom{0}$ & \\
$N_d=-0.97$ & $\alpha_d=0.44$ & $\beta_d=0.90$ & $M_1^2=0.19\;\textrm{(GeV}^2$)\\
\hline
\end{tabular}
	\caption{Best values of the fit of the Sivers functions.
	Table from Ref.~\cite{Anselmino:2011gs}}
\label{tab:a}
\end{table}

\subsection{\boldmath Transversity $h_{1}^{q}(x,k_\perp^{2})$ and
Collins function $H_{1}^{\perp q}(x,P_\perp^{2})$}
\label{App:basis-h1-H1perp}

These functions were studied in
Refs.~\cite{Anselmino:2007fs,Anselmino:2008jk,Anselmino:2013vqa,
Kang:2014zza,Kang:2015msa,Anselmino:2015sxa}.
The following shape was assumed for parametrizations in
Refs.~\cite{Anselmino:2007fs,Anselmino:2008jk,Anselmino:2013vqa}:
 \ba
h_{1}^{q} (x, \kperp^2) &=&h_{1}^{q} (x)  \frac{e^{-{\kperp^2}/{\avkperp_{h_1} }}}{\pi \avkperp_{h_1}} \; ,\label{tr-funct}\\
h_{1}^{q} (x) &=&\frac{1}{2} {\cal N}^{\T}_q(x)\,
[f_{1}(x)+g_1(x)]\; ,\\
H_{1 h/q}^{\perp}(z, \pperp^2) = \frac{z \mh}{2 \pperp} \Delta^N \! D_{h/q^\uparrow}(z,\pperp^2) &=&  \frac{z \mh}{M_C} e^{-p_{\perp}^2/M_C^2} \sqrt{2 e} H_{1 h/q}^{\perp}(z) \,\frac{e^{-\pperp^2/{\avpperp}}}{\pi \avpperp}\,,
\label{coll-funct}
 \ea
 with $\mh$ the mass of the produced hadron and
 \ba
 {\cal N}^{\T}_q(x) &=& N^{\T}_q
\,x^{\alpha} (1-x)^{\beta} \, \frac{(\alpha + \beta)^{(\alpha
+\beta)}} {\alpha^{\alpha} \beta^{\beta}}\; ,
\\
H_{1 h/q}^{\perp}(z) &=&  {\cal N}^{\C}_q(z) D_{h/q}(z)\, , \\
{\cal N}^{\C}_q(z) &=& N^{\C}_q \, z^{\gamma} (1-z)^{\delta} \,
\frac{(\gamma + \delta)^{(\gamma +\delta)}}
{\gamma^{\gamma} \delta^{\delta}}\, ,
 \ea
and $-1\le N^{\T}_q\le 1$, $-1 \le N^{\C}_q \le 1$. The helicity distributions $g_1(x)$ are taken
from Ref.~\cite{Gluck:2000dy}, parton distribution and fragmentation functions are the GRV98LO PDF set~\cite{Gluck:1998xa} and the
DSS fragmentation function set~\cite{deFlorian:2007aj}. Notice that with these choices both
the transversity and the Collins function automatically obey their
proper positivity bounds. Note that as in Ref.~\cite{Anselmino:2013vqa} we use two
Collins fragmentation functions, {\it favored} and {\it disfavored} ones, see Ref.~\cite{Anselmino:2013vqa} for details on implementation, and corresponding parameters ${N}^{\C}_a$ are then  ${N}^{\C}_{fav}$ and ${N}^{\C}_{dis}$. For numerical estimates we use parameters extracted in Ref.~\cite{Anselmino:2013vqa}, see Table~\ref{fitpar}.

\begin{table}[t]
\centering
\renewcommand{\tabcolsep}{0.4pc} 
\renewcommand{\arraystretch}{1.2} 
\begin{tabular}{@{ }lll}
 \hline
 $N_{u}^{\T}$ = $0.46^{+0.20}_{-0.14}$ & $N_{d}^{\T}$ = $ -1.00^{+1.17}_{-0.00}$ \\
 $\alpha$ =  $1.11^{+0.89}_{-0.66}$ & $\beta$  = $3.64^{+5.80}_{-3.37}$ &
 $\avkperp_{h_1} = 0.25$ (GeV$^2$) \\
 \hline
 $N_{\rm fav}^{\C}$  = $0.49^{+0.20}_{-0.18}$ & $N_{\rm dis}^{\C}$  =
 $-1.00^{+0.38}_{-0.00}$ \\
 $\gamma$  = $1.06^{+0.45}_{-0.32}$  & $\delta$   = $0.07^{+0.42}_{-0.07}$  &
 $M^2_C = 1.50^{+2.00}_{-1.12}$ (GeV$^2$) \\
 \hline
\end{tabular}
	\caption{Best values of the 9 free parameters fixing
	the $u$ and $d$ quark transversity distribution functions and
	favored and disfavored Collins fragmentation functions.
	The table is from Ref.~\cite{Anselmino:2013vqa}.
\label{fitpar}}
\end{table}

According to Eq.~(\ref{eq:moments}) we obtain the following expression for the first moment of
Collins fragmentation function:
\ba
H_{1 h/q}^{\perp (1)}(z) = \frac{H_{1 h/q}^{\perp}(z) \sqrt{e/2}  \avpperp M_C^3}{z \mh  (M_C^2+\avpperp)^2}\; .
\ea
We also define the following variable:
\ba
\avpperp_{H_1^\perp} = \frac{\avpperp M_C^2 }{\avpperp + M_C^2} .
\ea

We can rewrite the parametrizations of Collins FF as

\ba
H_{1}^{\perp}(z,\pperp^2) =  H_{1}^{\perp (1)}(z)   \frac{2 z^2 \mh^2}{\pi \avpperp_{H_{1}^\perp}^2} e^{-\pperp^2/{\avpperp_{H_{1}^\perp}}}
\label{coll-funct_new} \, .
\ea

\subsection{\boldmath Boer--Mulders function $h_{1}^{\perp q}(x,k_\perp^{2})$}
\label{App:basis-h1perp}

The Boer--Mulders function $h_{1}^{\perp}$~\cite{Boer:1997nt} measures
the transverse polarization asymmetry of quarks inside an unpolarized
nucleon. The Boer--Mulders functions were studied phenomenologically in
Refs.~\cite{Barone:2009hw,Barone:2010gk,Barone:2015ksa},  we present the parameters of extractions of 
the Boer-Mulders function from Ref.~\cite{Barone:2010gk} in Tablel~\ref{fitparbm}.

Ref.~\cite{Barone:2010gk} used the parametrization in which the Boer--Mulders function is proportional to the Sivers functions, such that
\ba
\avkperp_{h_1^\perp} &=& \frac{\avkperp \, M^2_{BM}}{\avkperp + M^2_{BM}} \, , \\
h_{1}^{\perp}(x, \kperp^2) &= &
- \,\frac{M}{M_{BM}}
\sqrt{2e}\; N_{BM}^q N_q (x)
\, f_{q/p} (x, Q)\frac{e^{-\kperp^2/\avkperp_{h_{1}^{\perp}}}}{\pi\avkperp},
\label{BM-dist}
\ea
 where
\ba
{\cal N}_q(x)= N_q \, x^{\alpha} (1-x)^{\beta} \,
\frac{(\alpha + \beta)^{(\alpha +\beta)}}
{\alpha^{\alpha} \beta^{\beta}}\; .
 \ea

The first moment of the Boer--Mulders function is
\ba
h_{1}^{\perp (1) q}(x)  &=& -\frac{\sqrt{\frac{e}{2}} \ \avkperp M_{BM}^3}{M (\avkperp + M_{BM}^2)^2}  \ {N}_q f_q(x, Q)  \nonumber \\
&=& -\sqrt{\frac{e}{2}} \frac{1}{M M_{BM}}  \frac{\avkperp_{h_1^\perp}^2}{\avkperp}    \ {N}_q  f_q(x, Q)
\label{bm} \ .
\ea

We can rewrite the parametrizations of Boer--Mulders functions as
\ba
h_{1}^{\perp q}(x,\kperp^2) =  h_{1}^{\perp (1) q}(x)   \frac{2 M^2}{\pi \avkperp_{h_{1}^\perp}^2} e^{-\kperp^2/{\avkperp_{h_{1}^\perp}}}\label{bm_new} \ .
\ea

\begin{table}
\centering
\begin{tabular}{crll}
\hline
$N_{BM}^u=2.1\pm0.1$ & $N_{BM}^d=$&$-1.111\pm0.001$   \\
$N_u=0.35$ & $\alpha_u=0.73$ & $\beta_u=3.46$ \\
$N_d=-0.9$ & $\alpha_d=1.08$ & $\beta_d=3.46$
& $M_{BM}^2=0.34\; \textrm{(GeV}^2$)    \\
\hline
\end{tabular}
\caption{Fitted parameters (those with error bars)
and fixed parameters (those without error bars)
of the Boer--Mulders quark distributions from Ref.~\cite{Barone:2009hw}.}
\label{fitparbm}
\end{table}
%

\subsection{\boldmath Pretzelosity distribution $h_{1T}^{\perp q}(x,k_\perp^{2})$}
\label{App:basis-h1Tperp}

Pretzelosity distribution function
$h_{1T}^{\perp}$~\cite{Lefky:2014eia} describes transversely polarized quarks
inside a transversely polarized nucleon.
We use the following form of $h_{1T}^{\perp a}$ ~\cite{Lefky:2014eia}:
\ba
h_{1T}^{\perp a}(x,k_{\perp}^2) = \frac{M^2}{M_{TT}^2} e^{-\kperp^2/M_{TT}^2} h^{\perp a}_{1T}(x) \frac{e^{-{\kperp^2}/{\avkperp}}}{\pi \avkperp}=\frac{M^2}{M_T^2} h^{\perp a}_{1T}(x) \frac{e^{-{\kperp^2}/{\avkperp_{h_{1T}^\perp}}}}{\pi \avkperp}\,\;,
\label{eq:h1Tperp}
\ea
where
\ba
h^{\perp a}_{1T}(x) &=& e  \ {\cal N}^a(x) (f_{1}^{a}(x, Q) - g_{1}^{a}(x, Q)), \label{eq:hx_par}\\
{\cal N}^a(x) &=& N^{a} x^{\alpha} (1-x)^{\beta} \frac{(\alpha + \beta)^{\alpha + \beta}}{\alpha^{\alpha} \beta^{\beta}}\, ,  \\
\avkperp_{h_{1T}^\perp}  &=& \frac{\avkperp M_{TT}^2}{\avkperp + M_{TT}^2}\, ,
\ea
and where ${N}^a$, $\alpha$, $\beta$, and $M_T$ are parameters fitted to data and can be found in Table~\ref{fitparI}.

\begin{table}[t]
\centering
\begin{tabular}{l c l l c l l}
\hline
$\alpha$ &=& $2.5\pm1.5$ & $\beta$ &=& $2$ fixed \\
 $N_{u}$ &=& $1 \pm 1.4$ & $N_{d}$ &=& $-1 \pm 1.3$ &
 $M_{TT}^2=0.18 \pm  0.7$ GeV$^2$\\
\hline
\end{tabular}
\caption{Fitted parameters of the pretzelosity quark distributions. Table from Ref.~\cite{Lefky:2014eia}}
\label{fitparI}
\end{table}

We use Eq.~(\ref{eq:moments}) to obtain for the second moment
of $ h_{1T}^{\perp a}(x,\kperp^2)$ in (\ref{eq:h1Tperp}) the result
shown below, and use it to rewrite parametrization of pretzelosity
functions as
\ba
h_{1T}^{\perp (2) a}(x) =  \frac{h^{\perp a}_{1T}(x) \avkperp_{h_{1T}^\perp}^3}{2 M^2 M_{TT}^2 \avkperp} \, , \quad
h_{1T}^{\perp q}(x,\kperp^2) =  h_{1T}^{\perp (2) q}(x)   \frac{2 M^4}{\pi \avkperp_{h_{1T}^\perp}^3} e^{-\kperp^2/{\avkperp_{h_{1T}^\perp}}}
\label{pretzelosity_new} \ .
\ea

\section{Convolution integrals and expressions in Gaussian Ansatz}
\label{App:factor}

In this Appendix we explain the notation for convolution integrals
of TMDs and FFs and give the explicit results obtained assuming the
Gaussian Ansatz.

\subsection{Notation for convolution integrals \label{ApendixB1}}

Structure functions are expressed as convolutions of TMDs and FFs
in the Bjorken limit at tree level. For reference we quote the
convolution integrals in ``Amsterdam notation'' \cite{Bacchetta:2006tn}
\be
	{\cal C}\bigl[ w\slim f\slim D \bigr]
	=  x \, \sum_a e_a^2 \int d^2 {\bm p}_T \,  d^2 {\bm k}_T
	\, \delta^{(2)}\bigl({\bm p}_T - {\bm k}_T - {\bm P}_{h \perp}/z \bigr)
	\,w({\bm p}_T,{\bm k}_T)\,
	f^a( x ,p_T^2)\,D^a(z,z^2 k_T^2) , \label{eq:convolution-Amsterdam}
\ee
where all transverse momenta refer to the virtual photon-proton
center-of-mass frame and $\bfhp  ={\bm P}_{h \perp}/{P}_{h\perp}$.
Hereby ${\bm p}_T$ is
the transverse momentum of quark with respect to nucleon,
${\bm k}_T$ is the transverse momentum of the fragmenting quark
with respect to produced hadron. The notation is not unique.
The one chosen in this work, in comparison to other works, is
\begin{alignat}{3}
	\mbox{transverse momentum in TMD:}\;\;\;&
    	\left[\bfkperp\right]_{\rm our}
    &=& 	\left[{\bm k}_\perp\right]_{\mbox{\tiny Ref.\cite{Anselmino:2011ch}}} \;
    =  	\;\;\;\;\;\left[{\bm p}_T\right]_{\mbox{\tiny Ref.\cite{Bacchetta:2006tn}}}\,,
	\\
	\mbox{transverse momentum in \ FF:}\;\;\;&
	\left[\bfpperp\right]_{\rm our}
    &=& 	\left[{\bm p}_\perp\right]_{\mbox{\tiny Ref.\cite{Anselmino:2011ch}}} \;
    =  	-z\left[{\bm k}_T\right]_{\mbox{\tiny Ref.\cite{Bacchetta:2006tn}}}\,,
	\\
	\mbox{transverse hadron momenta:}\;\;\;&
    	\left[\bfPhperp\right]_{\rm our}
    &=&	\left[{\bm P}_T\right]_{\mbox{\tiny Ref.\cite{Anselmino:2011ch}}} \;
    =  	\;\;\;\left[{\bm P}_{h\perp}\right]_{\mbox{\tiny Ref.\cite{Bacchetta:2006tn}}}\,.
\end{alignat}
Notice that
$\left[\bfpperp\right]_{\rm our}=
-z\left[{\bm k}_T\right]_{\mbox{\tiny Ref.\cite{Bacchetta:2006tn}}}$
is the transverse momentum the hadron acquires in the fragmentation process.
The normalization for unpolarized fragmentation functions is
\be
	D_1^a(z)
	= \left[\,\int d^2{\bm P}_\perp D_1^a(z,P_\perp^2)\right]_{\rm our}
	= \left[z^2\int d^2{\bm k}_TD_1^a(z, z^2  k_T^2)
	  \right]_{\mbox{\tiny Ref.~\cite{Bacchetta:2006tn}}}
	\label{eq:D1}\;.
\ee
The ``Amsterdam'' convolution integral (\ref{eq:convolution-Amsterdam})
reads in our notation
\be
	{\cal C}\bigl[ w\slim f\slim D \bigr]
	=  x \,
	\sum_a e_a^2 \int d^2 \bfkperp\,  d^2 \bfpperp^{ }
	\, \delta^{(2)}\bigl(z\bfkperp + \bfpperp - \bfPhperp \bigr)
	\,w\left(\bfkperp,-\frac{\bfpperp^{ }}{z}\right)
	f^a( x ,\kperp^2)\,D^a(z,\pperp^2) . \label{eq:convolution_our}
\ee

\subsection{Gaussian Ansatz}

For a generic TMD and FF the Gaussian Ansatz is given by
\be
    f^a(x,\kperp^2)=
    f^a(x)\,\frac{\exp(-\kperp^2/\la\kperp^2\ra)}{\pi\la\kperp^2\ra}\,,\;\;\;
    D^a(z,\pperp^2)=
    D^a(z)\,\frac{\exp(-\pperp^2/\la\pperp^2\ra)}{\pi\la\pperp^2\ra}
\ee
where
$\la\kperp^2\ra$ could be $x$--dependent,
and $\la\pperp^2\ra$ $z$--dependent.
Both could be flavor-dependent.
The variable $\pperp$ is convenient because phenomenological experience
shows that $\pperp$ in $D_1^{q/h}(z,\pperp^2)$ exhibits a Gaussian distribution
with weakly $z$--dependent Gaussian width. The distribution of transverse
momenta in $\left[D^a(z,z^2 k_T^2)\right]_{\mbox{\tiny Ref.\cite{Bacchetta:2006tn}}}$
would require a strongly $z$--dependent Gaussian width. It is a matter of
taste which one prefers to use.

It is convenient to work with transverse moments of TMDs and FFs
which are defined, and in the Gaussian model given by
\begin{alignat}{6}
	f^{(n)}(x) &=& \int d^2\bfkperp
	\left(\frac{\kperp^2}{2M^2}\right)^{\!\!n}\; f(x, \kperp^2)
	\;&\stackrel{\rm Gauss}{=}&\;
	\frac{n!\,\la \kperp^2\ra^n}{ 2^n\,M_N^{2n}}\, f(x),\;\;\;\nonumber\\
	D^{(n)}(z) &=& \int d^2\bfpperp
	\left(\frac{\pperp^2}{2 z^2 \mh^2}\right)^{\!\!n}\; D(z, \pperp^2)
	\;&\stackrel{\rm Gauss}{=}&\;
	\frac{n!\,\la \pperp^2\ra^n}{ 2^n_{ }z^{2n}_{ }\,m_{h}^{2n}} D(z)\,. \;\;\;
	\label{eq:moments}
\end{alignat}
It is important to keep in mind that these objects are well-defined
in the Gaussian model. However, in QCD and even in simple models
\cite{Avakian:2010br,Schweitzer:2012hh} one faces issues with UV
divergences and has to carefully define how to deal with them.

In Eqs.~(\ref{eq:moments}) the Gaussian dependence is factorized
from $x$ or $z$ dependence and parametrizations are made with
respect to either $f(x)$ or $D(z)$. As we saw in Appendix
\ref{App:basis} some TMD functions are parametrized with higher
moments directly as operator product expansion of TMDs may start
from higher twist matrix element instead of the usual twist-2 one.
In those cases equivalent formulas to Eqs.~(\ref{eq:moments}) can
be easily derived.

\subsection{Gaussian Ansatz for the derived TMDs used in this work}
\label{App-B:Gauss-Ansatz-non-basis-TMDs}

Having discussed the Gaussian Ansatz for the 8 basis functions
in Eqs.~(\ref{Eq:Gauss-f1}--\ref{Eq:Gauss-h1Tperp}) of
Sec.~\ref{Sec-4.4:evaluation} and in App.~\ref{App:basis},
we list here the Ans\"atze for the following derived TMDs:
\begin{subequations}\begin{align}
	g_{1T}^{\perp q}(x,\kperp^2)
	  &=	g_{1T}^{\perp (1) q}(x)\,
		\frac{2 M_N^2}{\pi \avkperp_{g_{1T}^\perp}^2}\,
		e^{-\kperp^2/{\avkperp_{g_{1T}^\perp}}}  \,,
	  && 	\mbox{cf.\ Sec.~\ref{Sec-6.1:FLTcosphi-phiS},}
		\label{eq:g1t}\\
	h_{1L}^{\perp a}(x,\kperp^2)
	  &= 	h_{1L}^{\perp (1) a}(x)\,
		\frac{2 M_N^2}{\pi{\avkperp_{h_{1L}^\perp}^2}}\,
		e^{-\kperp^2/{\avkperp_{h_{1L}^\perp}}}
	  && 	\mbox{cf.\ Sec.~\ref{Sec-6.2:FULsin2phi},}	
		\label{eq:h1l_final}\\
	g^{q}_{T}(x,\kperp^2)
	  &=	g^{q}_{T}(x)\,\frac{1}{\pi \avkperp_{g_{T}}}\,
		e^{-\kperp^2/{\avkperp_{g_{T}}}}\;,
	  && 	\mbox{cf.\ Sec.~\ref{Sec-7.2:FLTcosphiS},}	
		\label{eq:gtnew} \\
	g^{\perp q}_{T}(x,\kperp^2)
	  &= 	g^{\perp (2) q}_{T}(x)\,\frac{2 M^4}{\pi \avkperp^3_{g_{T}^{\perp} }}\,
		e^{-\kperp^2/{\avkperp_{g_{T}^{\perp} }}}\;,
	  && 	\mbox{cf.\ Sec.~\ref{Sec-7.3:FLTcos2phi-phiS},}	
		\label{eq:gtperpnew} \\
	g_{L}^{\perp q}(x,\kperp^2)
	  &=	g_{L}^{\perp (1) q}(x)\,
		\frac{2 M_N^2}{\pi \avkperp_{g_{L}^\perp}^2}\,
		e^{-\kperp^2/{\avkperp_{g_{L}^\perp}}}  \,,
	  && 	\mbox{cf.\ Sec.~\ref{Sec-7.3:FLLcosphi},}
		\label{eq:gLperp} \\
	h_L^q(x,\kperp^2)
	  &= 	h_L^q(x) \,\frac{1}{\pi \avkperp_{h_{L}}}\,
		e^{-\kperp^2/{\avkperp_{h_{L}}}}\;,
	  && 	\mbox{cf.\ Sec.~\ref{Sec-7.4:FULsinphi},}
		\label{eq:hLnew}\\
	h_T^{\perp q}(x,\kperp^2)
	  &=	h_T^{\perp (1) q}(x)\,\frac{2 M^2}{\pi \avkperp_{h_{T}^\perp}^2}\,
	  	e^{-\kperp^2/{\avkperp_{h_{T}^\perp}}}\;,
	  && 	\mbox{cf.\ Sec.~\ref{Sec-7.6:FUTsinphiS},}
		\label{eq:hTperpnew}\\
	h_T^{q}(x,\kperp^2)
	  &=	h_T^{(1) q}(x) \,\frac{2 M^2}{\pi \avkperp_{h_{T}}^2}\,
		e^{-\kperp^2/{\avkperp_{h_{T}}}}\;,
	  && 	\mbox{cf.\ Sec.~\ref{Sec-7.6:FUTsinphiS},}
		\label{eq:hTnew}\\
	f^{\perp q}_{T}(x,\kperp^2)
	  &= 	f^{\perp (2) q}_{T}(x)\,\frac{2 M^4}{\pi \avkperp^3_{f_{T}^{\perp} }}\,
		e^{-\kperp^2/{\avkperp_{f_{T}^{\perp} }}}\;,
	  && 	\mbox{cf.\ Sec.~\ref{Sec-7.8:FUTsin2phi-phiS},}
		\label{eq:ftperpnew}\\
	f^{\perp q}(x,\kperp^2)
	  &= 	f^{\perp (1) q}(x) \,\frac{2 M^2}{\pi \avkperp_{f^\perp}^2}\,
		e^{-\kperp^2/{\avkperp_{f^\perp}}}\;,
	  && 	\mbox{cf.\ Sec.~\ref{Sec-7.7:FUUcosphi},}
		\label{eq:fperpnew}
\end{align}\end{subequations}

\subsection{Comment on TMDs subject to the sum rules (\ref{Eq:sum-rules-T-odd})}
\label{App-B:comment-Todd-twist-3}

In this section we comment on the twist-3 TMDs
$f_T^q(x,\kperp^{2})$,  $h^q(x,\kperp^{2})$, $e_L^q(x,\kperp^{2})$,
which are T-odd, appear in the decompositions of the correlator with no
explicit $k_\perp^j$--prefactors, and would have collinear PDF counterparts.
But T-odd PDFs are forbidden by time-reversal and parity invariance of
strong interactions, which dictate the sum rules (\ref{Eq:sum-rules-T-odd}),
see Sec.~\ref{Sec-3.8:limitations}.
Such TMDs could be described by functions with a node in
$k_\perp$\footnote{The possibility of TMDs with nodes is not unrealistic.
	For instance in the covariant parton model the helicity TMDs
	exhibit nodes for the $u$-- and $d$--flavor \cite{Efremov:2010mt}.
	We will have to revise our description of $g_1^q(x,\kperp^{2})$
	in Eq.~(\ref{Eq:Gauss-g1}) and App.~\ref{App:basis-g1} to something
	of the type (\ref{Eq:multiple-Gauss}), if that prediction is
	confirmed experimentally. }
such that they can integrate to zero in Eq.~(\ref{Eq:sum-rules-T-odd}).
A single Gaussian has no node and is not adequate for that.
However, one could work with a superposition of Gaussians
with different widths,
\begin{alignat}{1}
	x \, f_T^q(x,\kperp^{2}) =  - \, f_{1T}^{\perp (1)q}(x)\;
	&\sum\limits_{i=1}^{n} a_i\;
	\frac{\exp(-\kperp^2/\la\kperp^2\ra_i^{ })}{\pi\la\kperp^2\ra_i^{ }}\,,
	\label{Eq:multiple-Gauss}\\
	&\sum\limits_{i=1}^n a_i = 0\,, \;\;
	\la\kperp^2\ra_i^{ }\neq\la\kperp^2\ra_j^{ }\;\;\forall\;i\neq j,
	\; 1\le i,\,j\le n,\;n\ge 2.\nonumber
\end{alignat}
Notice that in (\ref{Eq:multiple-Gauss}) we cannot write ``$f_T^q(x)$'',
which would be zero according to (\ref{Eq:sum-rules-T-odd}), and we
explore here the WW-type approximation (\ref{Eq:WW-type-fT}).
The minimal choice would be $n=2$ with $a_1=-a_2=1$ and
$\la\kperp^2\ra_1^{ } = \la\kperp^2\ra_{f_{1T}^\perp}$ to make use
of the theoretical guidance provided by the WW-type approximation
(\ref{Eq:WW-type-fT}).
The second Gaussian width $\la\kperp^2\ra_2^{ }$ could be chosen
very large $\la\kperp^2\ra_2^{ } \gg \la\kperp^2\ra_{f_1^\perp}$ to
model the Gaussian description of $f_T^{q}(x,\kperp^{2})$ similar to
that of $f_{1T}^{\perp(1)q}(x,\kperp^2)$ at intermediate $\kperp$.
A very large parameter $\la\kperp^2\ra_2^{ }$ could be thought of as
a relict which enters in the sum rule (\ref{Eq:sum-rules-T-odd})
where the $\kperp$--integration formally extends up to infinity
where the TMD description does not apply. The theoretical
understanding of higher--twist TMDs is too limited at
the present stage, but in principle this could be a
pragmatic way of modeling the TMD $f_T^q(x,\kperp^{2})$
and analogously $h^q(x,\kperp^{2})$, $e_L^q(x,\kperp^{2})$.

\subsection{Convolution integrals in Gaussian Ansatz}
\label{App:convol-details}

Solving the convolution integrals relevant for SIDIS in the
Gaussian Ansatz yields
\begin{subequations}\label{Eq:Gaussian-integrals-RAW}\ba
  {\cal C}\bigl[\;\omega^{\{0\}}\, f \, D \bigr]
    &=&	\phantom{-}\,u \;{\cal G}(\Phperp)\\
  {\cal C}\bigl[ \omega^{\{1\}}_{\rm A} \, f \, D \bigr]
    &=&	\phantom{-}\,u \;{\cal G}(\Phperp)\;
	\biggl(\frac{z\Phperp}{\mh}\biggr)\;\frac{\la\pperp^2\ra}{z^2\lambda}\\
  {\cal C}\bigl[\omega^{\{1\}}_{\rm B} \, f \, D \bigr]
    &=&	- \,u \;{\cal G}(\Phperp) \;
	\biggl(\frac{z\Phperp}{M_N}\biggr)\;\frac{\la\kperp^2\ra}{\lambda}\\
  {\cal C}\bigl[ \omega^{\{2\}}_{\rm A} \, f \, D \bigr]
    &=&	\phantom{-}\,u \; {\cal G}(\Phperp)\;
	\frac{\la\kperp^2\ra\la\pperp^2\ra}{\lambda M_N \mh}
	\left(-1+\frac{2\Phperp^2}{\lambda}\right)\\
  {\cal C}\bigl[ \omega^{\{2\}}_{\rm B} \, f \, D \bigr]
    &=&	\phantom{-}\,u \; {\cal G}(\Phperp)\;
	\frac{\la\kperp^2\ra\la\pperp^2\ra}{\lambda M_N \mh}
	\left(1-\frac{\Phperp^2}{\lambda}\right)\\
  {\cal C}\bigl[ \, \omega^{\{2\}}_{\rm AB} \: f \, D \bigr]
    &=&	\phantom{-}\,u \; {\cal G}(\Phperp)\;
	\biggl(\frac{z\Phperp}{M_N}\biggr)
	\biggl(\frac{z\Phperp}{\mh}\biggr)
	\;\frac{\la\kperp^2\ra}{\lambda}
	\;\frac{\la\pperp^2\ra}{z^2\lambda}\\
  {\cal C}\bigl[ \, \omega^{\{2\}}_{\rm C} \: f \, D \bigr]
    &=&	\phantom{-}\,\frac{u}{2} \; {\cal G}(\Phperp)\;
	\biggl(\frac{z\Phperp}{M_N}\biggr)
	\biggl(\frac{z\Phperp}{M_N}\biggr)
	\;\frac{\la\kperp^2\ra}{\lambda}
	\;\frac{\la\kperp^2\ra}{\lambda}\\
  {\cal C}\bigl[ \, \omega^{\{3\}}_{\rm { }} \: f \, D \bigr]
    &=&	\phantom{-}\,\frac{u}{2} \; {\cal G}(\Phperp)\;
	\biggl(\frac{z\Phperp}{M_N}\biggr)
	\biggl(\frac{z\Phperp}{M_N}\biggr)
	\biggl(\frac{z\Phperp}{\mh}\biggr)
	\;\frac{\la\kperp^2\ra}{\lambda}
	\;\frac{\la\kperp^2\ra}{\lambda}
	\;\frac{\la\pperp^2\ra}{z^2\lambda}
\ea\end{subequations}
with the $\omega^{\{n\}}_i$ as defined in Eq.~(\ref{Eq:wi}),
and we introduced the abbreviations
\be
	u=x\,\sum_a e_a^2 f^a(x)D^a(z) \,, \;\;\;
	{\cal G}(\Phperp)=\frac{\exp(-\Phperp^2/\lambda)}{\pi\lambda}\,,\;\;\;
	\lambda = z^2\la\kperp^2\ra+\la\pperp^2\ra\,,
\ee
with the normalization $\int d^2\Phperp\,{\cal G}(\Phperp) = 1$.
It is important to keep in mind that strictly speaking
${\cal G}(\Phperp) = {\cal G}(\Phperp,x,z)$ {\it also} depends
on $x$ and $z$.
The ``non-compact'' notation in Eqs.~(\ref{Eq:Gaussian-integrals-RAW})
was chosen to display the pattern. The masses $M_N$ or $\mh$ in the
denominators of the $\Phperp$ indicate the ``origins'' of the
contributions: due to intrinsic $\kperp$ from target, due to
transverse momenta $\pperp$ acquired during fragmentation, or both.
The weight $\omega^{\{2\}}_{\rm B}$ is the only which enters
cross sections and does not have a homogeneous scaling in $\Phperp$.

For practical application it is convenient to absorb as many
(Gaussian model) parameters as possible into expressions that can
be more easily fitted to data. One way to achieve this is to make
use of the transverse moments (\ref{eq:moments}).
We introduce the following abbreviations
\begin{alignat}{6}
u^{\{1\}}_{\rm A} 	&=& \;x\,\sum_a e_a^2 f^{   a}(x)D^{(1)a}(z) \,, &\hspace{1cm}&
u^{\{1\}}_{\rm B} 	&=& \;x\,\sum_a e_a^2 f^{(1)a}(x)D^{   a}(z) \,, \\
u^{\{2\}}_{\rm AB}	&=& \;x\,\sum_a e_a^2 f^{(1)a}(x)D^{(1)a}(z) \,, &\hspace{1cm}&
u^{\{2\}}_{\rm C}	&=& \;x\,\sum_a e_a^2 f^{(2)a}(x)D^{   a}(z) \,, \\
u^{\{3\}}_{\rm C}	&=& \;x\,\sum_a e_a^2 f^{(2)a}(x)D^{(1)a}(z) \,. &\hspace{1cm}& &&
\end{alignat}
In this notation the results in Eqs.~(\ref{Eq:Gaussian-integrals-RAW})
read
\begin{subequations}\label{Eq:Gaussian-integrals-working}\ba
  {\cal C}\bigl[ \omega^{\{1\}}_{\rm A} \, f \, D \bigr]
    &=&	\phantom{-}\,u^{(1)}_{\rm A} \;{\cal G}(\Phperp)\;
	\biggl(\frac{z\Phperp}{\mh}\biggr)\;\frac{2\mh^2}{\lambda}\\
  {\cal C}\bigl[ \omega^{\{1\}}_{\rm B} \, f \, D \bigr]
    &=&	- \,u^{(1)}_{\rm B} \;{\cal G}(\Phperp) \;
	\biggl(\frac{z\Phperp}{M_N}\biggr)\;\frac{2M_N^2}{\lambda}\\
  {\cal C}\bigl[ \omega^{\{2\}}_{\rm B} \, f \, D \bigr]
    &=&	\phantom{-}\, u^{(2)}_{\rm B}\; {\cal G}(\Phperp)\;
	\frac{4z^2 \mh\,M_N}{\lambda}
	\left(1-\frac{\Phperp^2}{\lambda}\right)\\
  {\cal C}\bigl[ \, \omega^{\{2\}}_{\rm AB} \: f \, D \bigr]
    &=&	\phantom{-}\,u^{(2)}_{\rm AB} \; {\cal G}(\Phperp)\;
	\biggl(\frac{z\Phperp}{M_N}\biggr)
	\biggl(\frac{z\Phperp}{\mh}\biggr)
	\;\frac{2M_N^2}{\lambda}
	\;\frac{2\mh^2}{\lambda}\\
  {\cal C}\bigl[ \, \omega^{\{2\}}_{\rm C} \: f \, D \bigr]
    &=&	\phantom{-}\,\frac{u^{(2)}_{\rm C}}{2} \; {\cal G}(\Phperp)\;
	\biggl(\frac{z\Phperp}{M_N}\biggr)
	\biggl(\frac{z\Phperp}{M_N}\biggr)
	\;\frac{2M_N^2}{\lambda}
	\;\frac{2M_N^2}{\lambda}\\
  {\cal C}\bigl[ \, \omega^{\{3\}}_{\rm { }} \: f \, D \bigr]
    &=&	\phantom{-}\,\frac{u^{(3)}_{\rm  }}{2} \; {\cal G}(\Phperp)\;
	\biggl(\frac{z\Phperp}{M_N}\biggr)
	\biggl(\frac{z\Phperp}{M_N}\biggr)
	\biggl(\frac{z\Phperp}{\mh}\biggr)
	\;\frac{2M_N^2}{\lambda}
	\;\frac{2M_N^2}{\lambda}
	\;\frac{2\mh^2}{\lambda}
\ea\end{subequations}
In this notation the results in Eqs.~(\ref{Eq:Gaussian-integrals-RAW})
read
\be
	{\cal C}\bigl[ \omega^{\{n\}}_{i} \, f \, D \bigr]
	= u^{(n)}_{i} \;{\cal G}(\Phperp)\;\times
	\left[
	\delta_{n2}\,\delta_{i\rm B}\;a^{(2)}_{B} + b^{(n)}_{i}
	\left(\frac{z\Phperp}{\lambda}\right)^{\!\!n}\;\right]
\ee
with
\begin{alignat}{7}
	b^{(0)}_{\rm  } 	&= 1 \, , \\
	b^{(1)}_{\rm A} 	&= 2\mh	\, , &
	b^{(1)}_{\rm B} 	&= 2M_N	\, , \\
	a^{(2)}_{\rm B}	&= 4M_N\mh\lambda^{-1}\,z^2  \, , \;\;\;\;\;\; &
	b^{(2)}_{\rm AB} 	&= -\,b^{(2)}_{\rm B}
			 = 4M_N\mh 	\, , \;\;\;\;\;\; &
	b^{(2)}_{\rm C} 	&= M_N^2  \, ,\\
	b^{(3)}_{\rm  } 	&= 2M_N^2\mh \, .
	\label{Eq:Gaussian-integrals-working-III}
\end{alignat}
Finally, integrating out transverse hadron momenta yields
\be
	\int d^2\Phperp\;
	{\cal C}\bigl[\,\omega^{\{n\}}_{i} \, f \, D \bigr]
	= u^{(n)}_{i} \;c^{(n)}_{i}\;\biggl(\frac{z}{\lambda^{1/2}}\biggr)^{\!\!n}
\ee
with
\begin{alignat}{7}
	c^{(0)}_{\rm  } 	&= 1 \, , \\
	c^{(1)}_{\rm A} 	&= \sqrt{\pi}\,\mh	\, , &
	c^{(1)}_{\rm B} 	&= \sqrt{\pi}\,M_N	\, , \\
	c^{(2)}_{\rm AB} 	&= 4M_N\mh		\, , \;\;\;\;\;\; &
	c^{(2)}_{\rm C} 	&= M_N^2 		\, , \;\;\;\;\;\;\;\;\;\; &
	c^{(2)}_{\rm B}	&= 0 			\, ,\\
	c^{(3)}_{\rm  } 	&= {\textstyle\frac32}\sqrt{\pi} \,M_N^2\mh
	\, . \;\;\;     &
			&
	\label{Eq:Gaussian-integrals-working-IV}
\end{alignat}


\providecommand{\href}[2]{#2}\begingroup\raggedright\endgroup

\end{document}